\documentclass[letterpaper,11pt]{article}
\pdfoutput=1

\usepackage{jheppub}

\usepackage{multirow}

\usepackage{subfig}
\usepackage{xspace}
\usepackage[countmax]{subfloat}

\def\nslash{n\hspace{-2mm}\slash}

\DeclareRobustCommand{\Sec}[1]{Sec.~\ref{#1}}

\DeclareRobustCommand{\App}[1]{App.~\ref{#1}}
\DeclareRobustCommand{\Tab}[1]{Table~\ref{#1}}

\DeclareRobustCommand{\Fig}[1]{Fig.~\ref{#1}}
\DeclareRobustCommand{\Figs}[2]{Figs.~\ref{#1} and \ref{#2}}
\DeclareRobustCommand{\Eq}[1]{Eq.~(\ref{#1})}
\DeclareRobustCommand{\Eqs}[2]{Eqs.~(\ref{#1}) and (\ref{#2})}
\DeclareRobustCommand{\Ref}[1]{Ref.~\cite{#1}}
\DeclareRobustCommand{\Refs}[1]{Refs.~\cite{#1}}

\newcommand{\pythia}[1]{\textsc{Pythia\xspace #1}}

\newcommand{\fastjet}[1]{\textsc{FastJet\xspace #1}}

\newcommand{\mathematica}{\texttt{Mathematica }}

\preprint{MIT--CTP 4525}

\title{Toward Multi-Differential Cross Sections:\\
Measuring Two Angularities on a Single Jet
}

\author{Andrew J. Larkoski,}
\author{Ian Moult,}
\author{and Duff Neill}

\affiliation{Center for Theoretical Physics, Massachusetts Institute of Technology, Cambridge, MA 02139, USA}

\emailAdd{larkoski@mit.edu}
\emailAdd{ianmoult@mit.edu}
\emailAdd{dneill@mit.edu}

\abstract{
The analytic study of differential cross sections in QCD has typically focused on individual observables, such as mass or thrust, to great success. Here, we present a first study of double differential jet cross sections considering two recoil-free angularities measured on a single jet. By analyzing the phase space defined by the two angularities and using methods from soft-collinear effective theory, we prove that the double differential cross section factorizes at the boundaries of the phase space. We also show that the cross section in the bulk of the phase space cannot be factorized using only soft and collinear modes, excluding the possibility of a global factorization theorem in soft-collinear effective theory. Nevertheless, we are able to define a simple interpolation procedure that smoothly connects the factorization theorem at one boundary to the other. We present an explicit example of this at next-to-leading logarithmic accuracy and show that the interpolation is unique up to $\alpha_s^4$ order in the exponent of the cross section, under reasonable assumptions. This is evidence that the interpolation is sufficiently robust to account for all logarithms in the bulk of phase space to the accuracy of the boundary factorization theorem. We compare our analytic calculation of the double differential cross section to Monte Carlo simulation and find qualitative agreement. Because our arguments rely on general structures of the phase space, we expect that much of our analysis would be relevant for the study of phenomenologically well-motivated observables, such as $N$-subjettiness, energy correlation functions, and planar flow.
}

\begin{document} 
\maketitle

\section{Introduction}

Historically, there has been significant effort devoted to understanding and computing the all-orders distributions of jet observables in QCD \cite{Catani:1991kz,Catani:1991bd,Catani:1992jc,Seymour:1997kj,Dokshitzer:1998kz,Berger:2003iw,Banfi:2004yd,Schwartz:2007ib,Becher:2008cf,Ellis:2010rwa,Abbate:2010xh,Chiu:2012ir,Feige:2012vc,Becher:2012qc,Chien:2012ur,Larkoski:2012eh,Dasgupta:2012hg,Jouttenus:2013hs,Dasgupta:2013ihk,broadening}.  This has led to incredibly precise predictions for the differential cross sections of these observables which have been used, for example, to determine the strong coupling $\alpha_s$ to high precision \cite{Becher:2008cf,Davison:2008vx,Abbate:2010xh}.  For all their successes, though, this program can only answer questions about individual observables.  In this paper, we move beyond this paradigm of single differential cross sections, to exploring the full phase space of multi-differential cross sections analyzed on a single jet.\footnote{A first example (to our knowledge) of joint resummation of any form was in \Ref{Laenen:2000ij}.}  Multi-differential cross-sections have been studied before in an SCET context, but these are also multi-jet cross-sections as well, where each jet sector receives at most one measurement \cite{Ellis:2010rwa,Kelley:2011ng,Jouttenus:2011wh,Hornig:2011iu}. The closest in spirit to our current study was the construction of SCET$_+$ \cite{Bauer:2011uc}. Using angularities as a case study, we find that factorization methods are confined to the boundaries on the physical phase space regions, and we propose an interpolation method to connect all the various forms of factorization possible.   

There exist multiple motivations for why one might want to study such multi-differential cross sections.  
 Aside from purely formal interest in connecting different effective field theory regimes, we focus on two motivations here: for studying the correlations between different observables and for understanding the properties of observables formed from the ratio of two infrared and collinear (IRC) safe observables. Phenomenologically one would want to know the correlations between different observables so as to determine the extent to which they probe identical physics.  However, this cannot be done by studying the differential cross sections of individual observables alone.  Correlations are encoded in the multi-differential cross section of the observables and so to understand the correlations between two observables we must study their double differential cross section.  

Studying the correlations between two observables is not necessary to make highly precise predictions for QCD. However, with the advent and boom of the jet substructure program \cite{Abdesselam:2010pt,Altheimer:2012mn,Altheimer:2013yza} increasingly detailed questions about the dynamics of QCD jets are being asked  and probed by experiment \cite{ATLAS-CONF-2011-073,Miller:2011qg,ATLAS-CONF-2011-053,ATLAS:2012am,ATLAS:2012xna,Aad:2012meb,Aad:2012raa,ATLAS:2012dp,ATLAS-CONF-2012-066,ATLAS-CONF-2012-065,Aad:2013fba,Aad:2013gja,TheATLAScollaboration:2013pia,TheATLAScollaboration:2013qia,TheATLAScollaboration:2013ria,TheATLAScollaboration:2013sia,TheATLAScollaboration:2013tia,CMS-PAS-QCD-10-041,CMS-PAS-JME-10-013,CMS:2011bqa,Chatrchyan:2012mec,Chatrchyan:2012tt,Chatrchyan:2012sn,CMS:2013kfa,CMS:2013wea,CMS:2013vea,CMS:2013uea}.  In particular, one of the goals of jet substructure is to design highly efficient observables and procedures for discriminating QCD jets from boosted heavy particle decays.  Many of the proposed procedures for doing so involve the measurement of several observables on the jet and making appropriate cuts.  Thus, to determine if a QCD jet can fake looking like a boosted $W$, $Z$, $H$ or top quark requires a thorough analysis of the correlations of the observables that go into the discrimination procedure.

Several of the most powerful discrimination observables are formed from the ratio of two IRC safe observables. This includes $N$-subjettiness \cite{Thaler:2010tr,Thaler:2011gf}, energy correlation functions \cite{Banfi:2004yd,Larkoski:2013eya}, planar flow \cite{Thaler:2008ju,Almeida:2008yp} and angular structure functions \cite{Jankowiak:2011qa,Jankowiak:2012na}.  While it might seem like the ratio of two IRC safe observables is still IRC safe and so calculable order-by-order in perturbation theory, it was shown in \Ref{Soyez:2012hv} that ratio-type observables are actually IRC unsafe, if the denominator observable can become arbitrarily small.  Na\"ively, this is an insurmountable barrier to understanding these observables in perturbative QCD.  Indeed, this is true with the standard procedure of computing single differential cross sections which require IRC safety to be well-defined in perturbation theory.

However, in \Ref{Larkoski:2013paa} it was shown that ratio observables can actually be made well-defined, if all-orders effects are taken into account.  There, the simple observable formed from the ratio of two angularities \cite{Berger:2003iw,Almeida:2008yp,Ellis:2010rwa} measured on a single jet was studied, where the angularity $e_\alpha$ is\footnote{We normalize to the jet radius so that when comparing two angularities with different angular exponents, the jet radius is not relevant.}
\begin{equation}\label{eq:ang_def}
e_\alpha = \frac{1}{E_J} \sum_{i\in J} E_i \frac{\sin\theta_i \tan^{\alpha-1}\frac{\theta_i}{2}}{\sin R_0 \tan^{\alpha-1}\frac{R_0}{2}} \approx \frac{1}{E_J} \sum_{i\in J} E_i \left(\frac{\theta_i}{R_0}\right)^\alpha\ .
\end{equation}
$E_J$ is the jet energy, $R_0$ is the jet radius, $\theta_i$ is the angle between particle $i$ and an appropriately defined jet axis, and $\alpha>0$ for IRC safety.  The approximation is accurate for $R_0\ll 1$, which we assume throughout this paper.   In practice, we will take $R_0\simeq 0.4$, which is not strictly much smaller than 1; however, it has been shown that finite jet radius corrections are small \cite{Dasgupta:2007wa,Dasgupta:2012hg}.

The differential cross section of the ratio $r$ of two angularities $e_\alpha$ and $e_\beta$ can be found by marginalizing the double differential cross section of the two angularities:
\begin{equation}
\frac{d\sigma}{dr} \equiv \int de_\alpha de_\beta \, \frac{d^2\sigma}{de_\alpha \, de_\beta}\, \delta\left( r - \frac{e_\alpha}{e_\beta}  \right) \ .
\end{equation}
\Ref{Larkoski:2013paa} showed that, while the ratio observable is not IRC safe and so cannot be computed order-by-order in $\alpha_s$, by resumming the large logarithms present in the double differential cross section to all orders, the differential cross section for the ratio $r$ is well-defined and calculable.  This property was called ``Sudakov safety'' because the calculability of the cross section of $r$ relied on the fact that small values of the angularities $e_\alpha$ and $e_\beta$ are exponentially suppressed by the Sudakov factor.  
The calculation of the double differential cross section of angularities was done to leading logarithmic (LL) accuracy in \Ref{Larkoski:2013paa} with no robust predictions about what happens at higher logarithmic orders.  In particular, Sudakov safety was only exhibited to LL accuracy, and some important and subtle physics might arise at higher orders that could change the story.  

Given these motivations, the double differential cross section of two angularities measured on a jet provides a laboratory for understanding multi-differential cross sections. To have adequate control over large logarithmic corrections, we need to prove a factorization theorem which would provide an order-by-order recipe for resumming to arbitrary accuracy. We will find that establishing such a factorization theorem for all of phase space in the double differential cross section is not possible with identified soft and collinear modes. In particular, a subtlety in the resummation of double differential cross sections is that the two measured observables do not define a unique set of scales for soft and collinear radiation in the jet.

Nevertheless, we will show that there do exist factorization theorems on the boundaries of phase space for the double differential cross section of two angularities using soft-collinear effective theory (SCET) \cite{Bauer:2000yr,Bauer:2000ew,Bauer:2001yt,Bauer:2001ct,Bauer:2002nz}.  Single differential cross sections factorize when the observable is sufficiently small, when one can say that soft and collinear dynamics dominate the structure of the jet.  For the case of the double differential cross section of angularities $e_\alpha$ and $e_\beta$, small values of the angularities does mean that soft and collinear dynamics dominate the jet.  However, the physical phase space for the double differential cross section is two dimensional, and the precise scaling of $e_\alpha$ and $e_\beta$ with respect to one another emphasizes soft over collinear physics, or vise-versa.  Strictly speaking, only on the boundaries of the phase space are the soft and collinear modes on-shell, where the factorization theorems hold.\footnote{Chris Lee has humorously referred to this as a ``holographic factorization theorem''.}

The boundaries of phase space are defined by the requirements of energy conservation and clustering of emissions into the jet of radius $R_0$.  Energy conservation corresponds to the boundary\footnote{This is true to logarithmic accuracy in the double differential cross section.  Power-suppressed corrections deform this boundary, but for most of this paper we will ignore these effects.} where $e_\alpha^\beta = e_\beta^\alpha$ and the jet radius requirement is the boundary line $e_\alpha = e_\beta$.  The physical phase space lies in between.   We will show that, at these boundaries, the double differential cross section of the angularities $e_\alpha$ and $e_\beta$ reduces to the single differential cross section for one of the angularities times a $\delta$-function for the other angularity, depending on the boundary, plus terms that integrate to 0.  For example, near the boundary $e_\alpha^\beta = e_\beta^\alpha$, the double differential cross section reduces to
\begin{align}\label{eq:fact_def}
\left.\frac{d^2\sigma}{de_\alpha\, de_\beta}\right|_{e_\alpha^\beta\sim e_\beta^\alpha}&\simeq \sigma_0\, H\times J(e_\alpha,e_\beta)\otimes S(e_\alpha)\nonumber \\
& = \frac{d\sigma}{de_\alpha}\,\delta(e_\beta) +\frac{1}{e_\alpha^{1+\frac{\beta}{\alpha}}}f_+^{\alpha}\left(\frac{e_\beta}{e_\alpha^{\beta/\alpha}}   \right)\ ,
\end{align} 
where $\simeq$ denotes the relationship that follows from the factorization theorem and $f_+^{\alpha}$ is a function that integrates to zero on $e_\beta \in [0,e_\alpha^{\beta/\alpha}]$.  $H$ represents the hard function, $J(e_\alpha,e_\beta)$ is the double differential jet function and $S(e_\alpha)$ is the soft function for $e_\alpha$ alone.  

Importantly, this relationship captures the effect of canonical resummation on this boundary as predicted by the factorization theorem and the only non-trivial dependence on $e_\beta$ exists in the ratio $e_\beta/e_\alpha^{\beta/\alpha}$. The fact that the soft function is independent of $e_\beta$ implies that the ultraviolet singular structure of the cross section exists on the line $e_\beta=0$, as enforced by $\delta(e_\beta)$. This is a non-trivial statement of the factorization theorem on this boundary that to all orders the RG evolution does not generate a non-zero value for $e_\beta$. On the other boundary of phase space, where $e_\alpha = e_\beta$, we find a similar relationship for the singular terms, with the single differential cross section of $e_\beta$ times $\delta(e_\alpha)$:
\begin{align}\label{eq:fact_def_II}
\left.\frac{d^2\sigma}{de_\alpha\, de_\beta}\right|_{e_\alpha \sim e_\beta}&\simeq \sigma_0\, H\times J(e_\beta)\otimes S(e_\alpha,e_\beta)\nonumber \\
& = \frac{d\sigma}{de_\beta}\,\delta(e_\alpha) +\frac{1}{e_\beta^{2}}f_+^{\beta}\left(\frac{e_\alpha}{e_\beta}   \right)\ .
\end{align} 
Non-trivial dependence on $e_\alpha$ only exists in the ratio $e_\alpha/e_\beta$.

\begin{figure}
\begin{center}
\includegraphics[width=7cm]{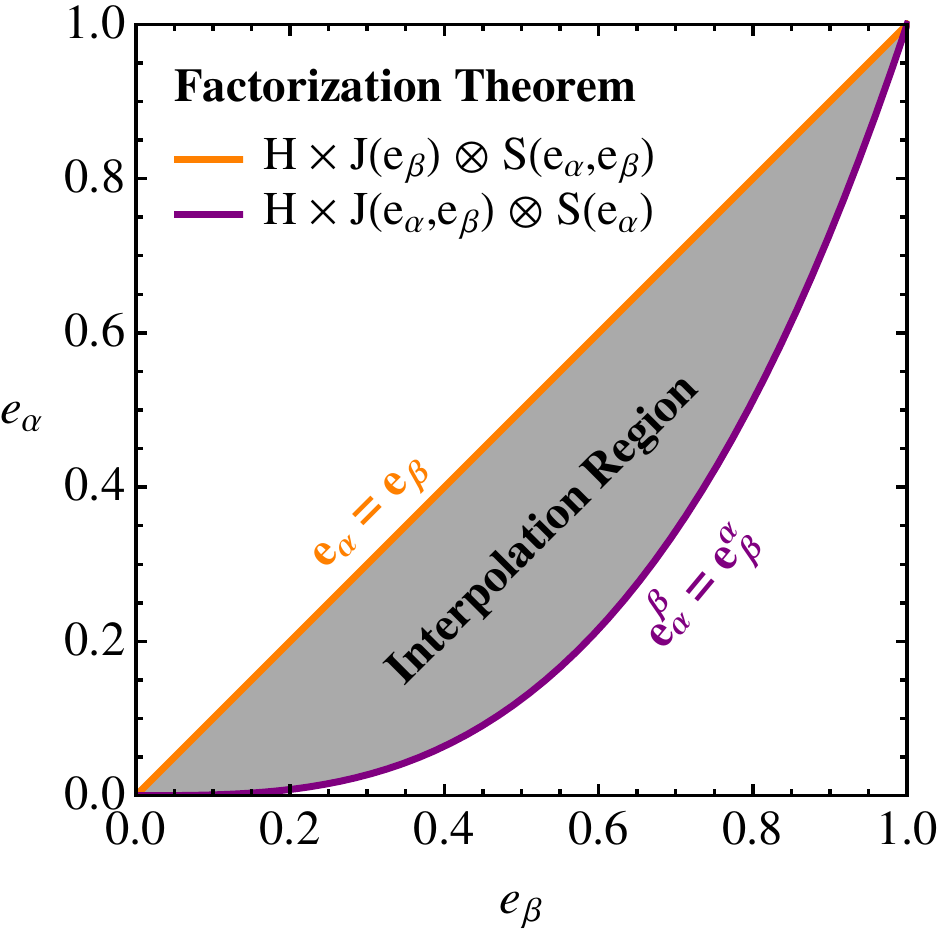}
\end{center}
\caption{
Summary of the results of the factorization theorem of the double differential cross section of angularities.  The factorization theorems exists near the boundaries of the allowed phase space where the double differential cross section reduces to the appropriate single differential cross section plus terms that integrate to 0.  The bulk of the phase space is described by an interpolating function.
}
\label{fig:ps_fact_sum}
\end{figure}

Because the factorization theorem only applies near the boundaries of the phase space, we cannot formally claim any logarithmic accuracy of the double differential cross section in the bulk of the phase space. However, we are able to determine a function that interpolates into the bulk of the phase space between the boundaries; crucial to this is the existence of factorization theorems at the boundaries. The interpolation between the boundary factorization theorems can be determined most simply by appropriately setting scales in the logarithms and by adding terms that are subleading at the boundaries. This conjectured double differential cross section must satisfy several consistency conditions, such as correctly reproducing the single differential cross section of one of the angularities.  Thus, while we are unable to fully demonstrate formal logarithmic accuracy in all of the phase space, we will present a conjecture for the double differential cross section to next-to-leading-logarithmic accuracy (NLL) which satisfies all consistency conditions.\footnote{In this paper, we will only consider the resummation of global logarithms.  A study of non-global \cite{Dasgupta:2001sh} and clustering \cite{Banfi:2005gj} logarithms from the jet algorithm restriction in the double differential cross section will be left to future work.}  The summary of this factorization theorem discussion is illustrated in \Fig{fig:ps_fact_sum}.

The structure of the cross section as found from the interpolation procedure is fascinating and manifests the barrier to proving a factorization theorem in the bulk of the phase space.  The interpolation procedure defines a double cumulative cross section containing the following logarithms:
\begin{equation}
\Sigma(e_\alpha,e_\beta) \supset \log e_\alpha,\ \log e_\beta^{1/\beta},\ \log e_\alpha^{\frac{1-\beta}{\alpha-\beta}}e_\beta^{\frac{\alpha-1}{\alpha-\beta}} \ .
\end{equation}
$\log e_\alpha$ and $\log e_\beta^{1/\beta}$ can naturally be understood as arising from soft and collinear divergences, respectively, and so correspond to the modes that are identified in SCET.  The other logarithms, which we refer to as ``$k_T$'',\footnote{For a single emission, this new logarithm reduces to the relative transverse momentum of the emitted parton.} are novel, arising neither from soft nor collinear modes over all of the phase space of $e_\alpha$ and $e_\beta$.  Indeed, the fact that there are three logarithmic structures in the bulk of the phase space suggests that there must be three modes in a factorization theorem of the double differential cross section that would be valid everywhere.\footnote{It might seem that the case when $\alpha = 1$ or $\beta = 1$ is special, where the logarithms degenerate, which may suggest that the number of modes that contribute to these cases is reduced.  However, as in the case of traditional broadening, just because the contributions from different modes to the observable degenerate does not mean that the number of modes that contribute changes.  As was observed with recoil-free angularities in \Ref{broadening}, we expect that there is smooth behavior through $\alpha =1$ and $\beta = 1$.}  At the boundaries of phase space, the $k_T$ logarithms degenerate to soft or collinear logarithms, which is why SCET factorization applies there.  This situation is very different than, for example, the recoil convolution in the broadening factorization theorem \cite{Becher:2011pf,Chiu:2012ir}. In that case, the relevant modes were still only soft and collinear.  Any factorization theorem of the double differential cross section must be super-SCET.

A possible complaint with the interpolation procedure\footnote{We thank Jesse Thaler for extensive discussions of this point.} is that it is not unique and therefore there is no control over the logarithms that appear in the bulk of the phase space in the double differential cross section.  This is an especially valid point because there is no factorization theorem in the bulk of the phase space and therefore no formal accuracy of the interpolation conjecture in this region is guaranteed.  However, we will show that (under reasonable assumptions on the double differential cross section) to NLL accuracy, the boundary conditions are sufficiently robust to forbid all logarithms that are not generated by our procedure for interpolation up to ${\cal O}(\alpha_s^4)$ in the exponent of the double cumulative cross section.  This is strong evidence that our interpolation procedure of setting scales can capture all logarithms that exist in the double differential cross section of two angularities to NLL accuracy over all of the phase space.

The outline of this paper is as follows.  In \Sec{sec:ps} we discuss the relevant phase space for the double differential cross section in the two angularities $e_\alpha$ and $e_\beta$.  This will also necessitate a discussion of the definition of the axis about which the angularities are measured.  To remove sensitivity to recoil from soft wide angle emissions, we measure angularities about the broadening axis of a jet \cite{Larkoski:2013eya,Larkoski:2013paa,broadening}.  In \Sec{sec:fo} we compute the double differential cross section at fixed-order.  This will illustrate some of the subtleties of resummation of the double differential cross section.  In \Sec{sec:ft} we present the factorization theorem of the double differential cross section.  We first discuss what can be learned simply from the phase space, then turn to the relevant SCET modes that contribute to the two angularities, and finally explicitly show that the double differential cross section factorizes near the boundaries of the phase space. Because the factorization theorem contains unfamiliar double differential jet and soft functions, we discuss the structure of these objects in \Sec{sec:ddiffunc} from constraints of power counting and consistency of the factorization. In \Sec{sec:int} we suggest a simple procedure for interpolating the double differential cross section from the boundaries into the bulk of the phase space.  We will show that this interpolating conjecture for the double differential cross section satisfies non-trivial consistency conditions and provide evidence that it captures all logarithms to NLL accuracy.  In \Sec{sec:mc} we compare our expression for the double differential cross section to Monte Carlo simulation and find good qualitative agreement.  Finally we close in \Sec{sec:conc} and suggest several future directions and applications for studying double differential cross sections.

\section{Angularities Phase Space}\label{sec:ps}

We begin with a discussion of the phase space of the differential cross section of two angularities.  From the introduction, we define the angularity $e_\alpha$ measured on a narrow jet as
\begin{equation}
e_\alpha=\frac{1}{E_J}\sum_{i\in J}E_i \left(\frac{\theta_i}{R_0}\right)^\alpha \ ,
\end{equation}
where $E_J$ is the jet energy, $R_0\ll 1$ is the jet radius, and the sum runs over all constituents in the jet.  For IRC safety, $\alpha>0$.  $\theta_i$ is the angle between particle $i$ and an appropriately chosen axis.  Historically, this has been chosen to be the jet axis, defined as the sum of three-momenta of all particles in the jet.  However, recently \cite{Larkoski:2013eya} it has been emphasized that this choice of axis is sensitive to recoil effects from the emission of soft, wide angle particles.  At small values of the angular exponent $\alpha$, the effect of recoil dominates the value of the angularity, significantly reducing its power for quark versus gluon jet discrimination, for example.

Instead, one can define an axis that is insensitive to these recoil effects and one example of this is the broadening axis \cite{Larkoski:2013eya,Larkoski:2013paa}.\footnote{It should be noted that the broadening axis is one definition that results in recoil-free observables.  Other recoil-free examples include energy correlation function observables \cite{Banfi:2004yd,Jankowiak:2011qa,Larkoski:2013eya} and the axis defined by the winner-take-all jet algorithm recombination scheme \cite{Bertolini:2013iqa,broadening}.  To the accuracy that we work in this paper, all of the recoil-free choices are identical.}  The broadening axis is defined as the axis in the jet that minimizes the $\beta=1$ measure of $N$-subjettiness \cite{Thaler:2010tr,Thaler:2011gf}; equivalently, the broadening axis is defined as the axis that minimizes the jet broadening \cite{Rakow:1981qn,Ellis:1986ig,Catani:1992jc}.  That is, the broadening axis $\hat{b}$ corresponds to the axis that minimizes the scalar sum of momentum transverse to it:
\begin{equation}
\min_{\hat{b}} \left[ \sum_{i \in J} E_i \theta_{i\hat{b}}  \right] \ .
\end{equation}
For a jet with two constituents, the broadening axis aligns with the hardest particle.  In general, the broadening axis typically aligns with the direction of the hard core of energy in the jet.  We also define the broadening axis to be the center of the jet so that all particles in the jet are closer than the jet radius $R_0$ to the broadening axis.

With this set-up, now consider the allowed phase space of the double differential cross section of two angularities $e_\alpha$ and $e_\beta$.  We will assume that $\alpha>\beta$ and so, because all angles between particles and the broadening axis are less than $R_0$, $e_\alpha < e_\beta$.  This implies that as $e_\beta\to0$, then $e_\alpha\to0$.  Also, because the angularities $e_\alpha$ and $e_\beta$ are first non-zero at the same order in perturbation theory, then $e_\alpha\to0$ implies that $e_\beta\to0$.  Therefore, in addition to the upper bound on the phase space, there must also be a lower bound on the phase space for two angularities $e_\alpha$ and $e_\beta$.  This lower bound of the phase space follows from energy conservation.

\begin{figure}
\begin{center}
\includegraphics[width=7cm]{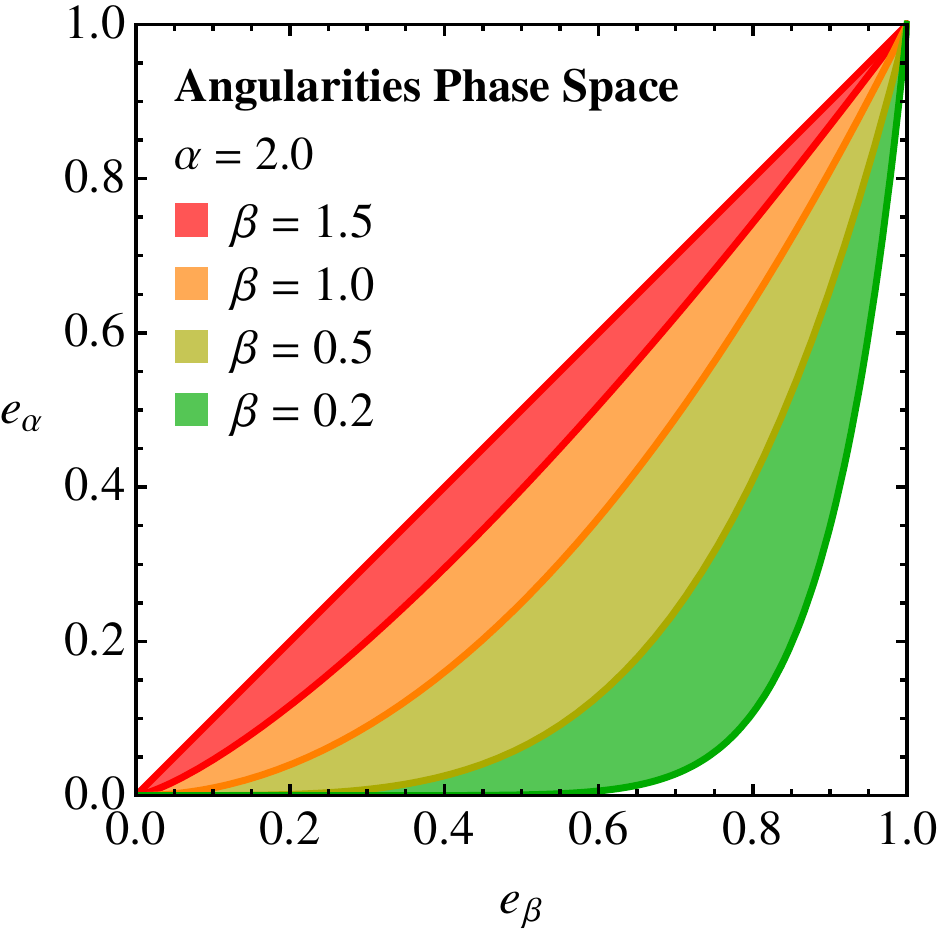}
\end{center}
\caption{
The allowed phase space of the double differential cross section of angularities $e_\alpha$ and $e_\beta$.  The angular exponent $\alpha$ is fixed to be $2$ and $\beta$ is varied.  For a given value of $\beta$, the phase space consists of the respective shaded region and all shaded regions above. 
}
\label{fig:ps_ddiff}
\end{figure}

These properties can be seen explicitly in a jet with two constituents.  The phase space can be described by the splitting angle $\theta$ and the energy fraction $z$ of the emission.  For the emission to be in the jet, $\theta<R_0$ and for energy to be conserved $z<1$.  The matrix element then necessarily contains the restrictions
\begin{equation}
\Theta(1-z)\Theta(R_0-\theta) \ .
\end{equation}
In these phase space coordinates, in the soft emission limit, the recoil-free angularity $e_\alpha$ is\footnote{Strictly speaking, the recoil-free angularities to this order in $\alpha_s$ are $$e_\alpha = \min[z,1-z] \frac{\theta^\alpha}{R_0^\alpha} \ .$$  Throughout this paper, we will only consider logarithmically-enhanced contributions to the angularities.  Therefore, to the accuracy that we consider, the definition of the angularities in \Eq{eq:ang_LO} is sufficient.
}
\begin{equation}\label{eq:ang_LO}
e_\alpha = z \left(\frac{\theta}{R_0}\right)^\alpha \ ,
\end{equation}
which ranges from $0$ to $1$.  The phase space coordinates $z$ and $\theta$ can be rewritten in terms of the two angularities $e_\alpha$ and $e_\beta$ as
\begin{equation}\label{eq:ps_irc}
z=e_\alpha^{-\frac{\beta}{\alpha-\beta}}e_\beta^{\frac{\alpha}{\alpha-\beta}} \ , \qquad \frac{\theta}{R_0} =  e_\alpha^{\frac{1}{\alpha-\beta}}e_\beta^{-\frac{1}{\alpha-\beta}} \ .
\end{equation}
The phase space restrictions written in terms of $e_\alpha$ and $e_\beta$ are then
\begin{equation}\label{eq:ps_consts}
\Theta(1-z)\Theta\left(R_0-\theta\right)\quad \Rightarrow \quad \Theta\left( e_\alpha^\beta - e_\beta^\alpha \right) \Theta\left( e_\beta - e_\alpha  \right) \ ,
\end{equation}
where the first $\Theta$-function follows from energy conservation and the second $\Theta$-function follows from demanding that the emission is in the jet.  The allowed phase space in the $(e_\alpha,e_\beta)$ plane is illustrated in \Fig{fig:ps_ddiff}, setting $\alpha = 2$ and scanning over a range of values for $\beta$.\footnote{This phase space has been discussed previously in \Ref{Larkoski:2013paa}.}

\section{Fixed-Order Cross Section}\label{sec:fo}

In this section, we will explicitly compute the double differential cross section of two jet angularities at ${\cal O}(\alpha_s)$.  The process we will consider is gluon emission from a quark and we will use the soft emission form of the angularities from \Eq{eq:ang_LO}.
For simplicity, we will just use the QCD splitting function as representative of the
matrix element, but this only differs from the full QCD matrix element at ${\cal O}(\alpha_s)$ by non-singular terms. 
To the accuracy that we consider, the (normalized) cumulative distribution of two angularities can be computed from
\begin{align}
\Sigma(e_\alpha,e_\beta) &= 1+\frac{\alpha_s}{\pi} \int_0^{R_0} \frac{d\theta}{\theta} \int_0^1 dz\, P_q(z) \left[  \Theta\left( e_\alpha - z\frac{\theta^\alpha}{R_0^\alpha}  \right)\Theta\left( e_\beta - z\frac{\theta^\beta}{R_0^\beta}  \right)-1 \right] \nonumber \\
&=1-\frac{\alpha_s}{\pi} \int_0^{1} \frac{d\theta}{\theta} \int_0^1 dz\, P_q(z) \left[ \Theta\left( z\theta^\beta -e_\beta \right) +\Theta\left( z\theta^\alpha-e_\alpha  \right)\Theta\left( e_\beta - z\theta^\beta  \right) \right] \ , 
\end{align}
where $P_q(z)$ is the quark splitting function given by
\begin{equation}
P_q(z) = C_F \frac{1+(1-z)^2}{z} \ .
\end{equation}
  The $-1$ in the first line is the subtraction of the virtual contribution which, by unitarity, we can assume is defined on the same phase space as the real contribution.  On the physical phase space defined by $e_\beta > e_\alpha$ and $e_\alpha^\beta>e_\beta^\alpha$, we find\footnote{Note that we have ignored contributions to the cumulative cross section that affect the total cross section at ${\cal O}(\alpha_s)$.}
\begin{align}
\Sigma(e_\alpha,e_\beta) &= 1+ \frac{\alpha_s}{\pi} C_F \Theta\left( e_\beta - e_\alpha  \right)\Theta\left( e_\alpha^\beta - e_\beta^\alpha  \right) \left\{  \ -\frac{7}{4\beta} - \frac{3}{2}\frac{\log e_\beta}{\beta} -\frac{\log^2 e_\beta}{\beta} \right. \nonumber\\
&\qquad\qquad+ \left. \frac{2}{\alpha}e_\alpha - \frac{e_\alpha^2}{4\alpha}  +  \frac{2(\alpha-\beta)}{\alpha \beta} e_\alpha^{-\frac{\beta}{\alpha-\beta}}e_\beta^{\frac{\alpha}{\alpha-\beta}} -\frac{\alpha-\beta}{4\alpha\beta}e_\alpha^{-\frac{2\beta}{\alpha-\beta}}e_\beta^{\frac{2\alpha}{\alpha-\beta}} -\frac{\log^2\frac{e_\alpha}{e_\beta}}{\alpha-\beta}   \right\} \ . \nonumber \\
\end{align}

From the double cumulative cross section, the double differential cross section is found by differentiating with respect to $e_\alpha$ and $e_\beta$.  Away from the boundaries of the phase space, we find
\begin{align}
\frac{d^2\sigma}{de_\alpha\, de_\beta} &\equiv \frac{\partial}{\partial e_\alpha}\frac{\partial}{\partial e_\beta} \Sigma(e_\alpha,e_\beta) \nonumber \\
&=2 \frac{\alpha_s}{\pi}\frac{C_F}{\alpha-\beta}\Theta\left( e_\beta - e_\alpha  \right)\Theta\left( e_\alpha^\beta - e_\beta^\alpha  \right) \left(
-e_\alpha^{-\frac{\alpha}{\alpha-\beta}}e_\beta^{\frac{\beta}{\alpha-\beta}} + \frac{e_\alpha^{-\frac{\alpha+\beta}{\alpha-\beta}}  e_\beta^{\frac{\alpha+\beta}{\alpha-\beta}}}{2}  +\frac{1}{e_\alpha e_\beta}
\right) \ .
\end{align}

The structures of the cumulative cross section and the differential cross section have some surprising distinctions.  In the cumulative distribution, there are several terms which appear power-suppressed in the physical phase space region.  For example, consider the term $$\frac{2(\alpha-\beta)}{\alpha \beta} e_\alpha^{-\frac{\beta}{\alpha-\beta}}e_\beta^{\frac{\alpha}{\alpha-\beta}} \ .$$  Because $e_\beta^\alpha < e_\alpha^\beta$ in the physical phase space, this term is suppressed by powers of the angularities.  Specifically, it is constant on the curve $e_\alpha^\beta = e_\beta^\alpha$, but otherwise vanishes in the physical phase space as $e_\alpha,e_\beta\to 0$.  However, in the double differential cross section, this term produces $$-2\frac{e_\alpha^{-\frac{\alpha}{\alpha-\beta}}e_\beta^{\frac{\beta}{\alpha-\beta}}}{\alpha-\beta} \ .$$
Because $e_\alpha < e_\beta$ on the physical phase space, this term actually diverges as $e_\alpha,e_\beta\to 0$.  Clearly, this term is integrable so one would not necessarily think that it needs to be resummed.

This term, however, is actually vital to reproduce the single differential cross section of one angularity to single logarithmic accuracy.  By marginalizing over one of the angularities, we have
\begin{align}\label{eq:sing_diff_marg}
\frac{d\sigma}{de_\beta} &= \int_0^1 de_\alpha \frac{d^2\sigma}{de_\alpha\, de_\beta} = \int_0^1 de_\alpha \frac{\partial}{\partial e_\alpha}\frac{\partial}{\partial e_\beta} \Sigma(e_\alpha,e_\beta) \nonumber \\
&=\left.\frac{\partial}{\partial e_\beta}\Sigma(e_\alpha,e_\beta)\right|_{e_\alpha = e_\beta}-\left.\frac{\partial}{\partial e_\beta}\Sigma(e_\alpha,e_\beta)\right|_{e_\alpha = e_\beta^{\alpha/\beta}} \ .
\end{align}
Note that the first term is evaluated at the upper limit of the phase space.  This means that in this term, $e_\alpha$ has been integrated over its entire physical range and so by itself, this term must be the differential cross section of $e_\beta$.  That is,
\begin{equation}
\frac{d\sigma}{de_\beta} =\left. \frac{\partial}{\partial e_\beta} \Sigma(e_\alpha,e_\beta)\right|_{e_\alpha= e_\beta} \ .
\end{equation}
The second term on the second line of \Eq{eq:sing_diff_marg} therefore must be zero to reproduce the correct cross section.  

This can be checked explicitly.  The derivative of the cumulative distribution with respect to $e_\beta$ is
\begin{align}
\frac{\partial}{\partial e_\beta}\Sigma(e_\alpha,e_\beta) &= \frac{\alpha_s}{\pi}  C_F \Theta\left(e_\alpha^\beta - e_\beta^\alpha\right) \Theta(e_\beta - e_\alpha) \left( -\frac{3}{2\beta}\frac{1}{e_\beta}  -\frac{2}{\beta}\frac{\log e_\beta}{e_\beta} \right. \nonumber \\
&\left.\qquad\qquad+ \ \frac{2}{\beta}e_\alpha^{-\frac{\beta}{\alpha-\beta}}e_\beta^{\frac{\beta}{\alpha-\beta}}- \frac{e_\alpha^{-\frac{2\beta}{\alpha-\beta}} e_\beta^{\frac{\alpha+\beta}{\alpha-\beta}}}{2\beta}   -\frac{2}{\alpha-\beta}\frac{\log\frac{e_\beta}{e_\alpha}}{e_\beta}    \right) \ .
\end{align}
For $e_\alpha = e_\beta$, this produces
\begin{equation}
\left. \frac{\partial}{\partial e_\beta} \Sigma(e_\alpha,e_\beta)\right|_{e_\alpha= e_\beta} =  \frac{\alpha_s}{\pi}  C_F\left(-\frac{3}{2\beta}\frac{1}{e_\beta}  -\frac{2}{\beta}\frac{\log e_\beta}{e_\beta}+\frac{2}{\beta}-\frac{e_\beta}{2\beta} \right) \ ,
\end{equation}
which is correct to this accuracy.  For $e_\alpha = e_\beta^{\alpha/\beta}$, it indeed vanishes.  However, there is a delicate cancelation of terms that is necessary for this term to vanish.  Note that if the na\"ively power-suppressed terms in the double cumulative cross section are removed, the derivative becomes
\begin{align}
\frac{\partial}{\partial e_\beta}\Sigma(e_\alpha,e_\beta)_{\log} &= \frac{\alpha_s}{\pi}  C_F \Theta\left(e_\alpha^\beta - e_\beta^\alpha\right) \Theta(e_\beta - e_\alpha) \left( -\frac{3}{2\beta}\frac{1}{e_\beta}  -\frac{2}{\beta}\frac{\log e_\beta}{e_\beta}   -\frac{2}{\alpha-\beta}\frac{\log\frac{e_\beta}{e_\alpha}}{e_\beta}    \right) \ .
\end{align}
At the boundary where $e_\alpha = e_\beta^{\alpha/\beta}$, we find
\begin{equation}
\left.\frac{\partial}{\partial e_\beta}\Sigma(e_\alpha,e_\beta)_{\log}\right|_{e_\alpha = e_\beta^{\alpha/\beta}} = -\frac{\alpha_s}{\pi}  C_F \Theta\left(e_\alpha^\beta - e_\beta^\alpha\right) \Theta(e_\beta - e_\alpha)\frac{3}{2\beta}\frac{1}{e_\beta}  \ ,
\end{equation}
which is clearly non-zero.  Therefore, to guarantee that the double differential cross section is accurate and consistent to single logarithmic accuracy requires that the cumulative cross section contains terms that are na\"ively power-suppressed with respect to logarithmic terms.  This is unfamiliar from the calculation of resummed single differential cross sections because there is no analogous consistency condition and will be important in the following sections.  Note that the double logarithms are correct, even when all power-suppressed terms are removed. 

\section{Factorization Theorem}\label{sec:ft}

Having discussed the phase space and fixed-order calculation of the double differential cross section, we now turn to studying its all-orders properties.  We present the factorization theorem for the double differential cross section of angularities measured on a single jet.  This section consists of three parts ordered in increasing technical detail, but only the first part is necessary to understand the remainder of this paper.  First, we return to discussing the phase space of the double differential cross section.  Nearly all of the conclusions from this section follow from simple geometric arguments about the limiting behavior of the double cumulative distribution at the boundaries of the phase space.  We then discuss the relevant on-shell SCET modes which contribute to the angularities.  We will show that there are two soft modes with different invariant mass which are relevant in the bulk of the phase space of the angularities.  This will be an obstacle to factorization on the full phase space but, by an appropriate partition, we are able to prove factorization of the cross section at the boundaries of the phase space.  The form of the factorization theorem will result in non-trivial identities between the cross sections at the two boundaries and we will use this in the following section to interpolate the cross section from the two boundaries into the bulk region.

\subsection{A Study of the Phase Space}\label{sec:ps_fact}

\begin{figure}
\begin{center}
\subfloat[]{\label{fig:cum_a}
\includegraphics[width=7cm]{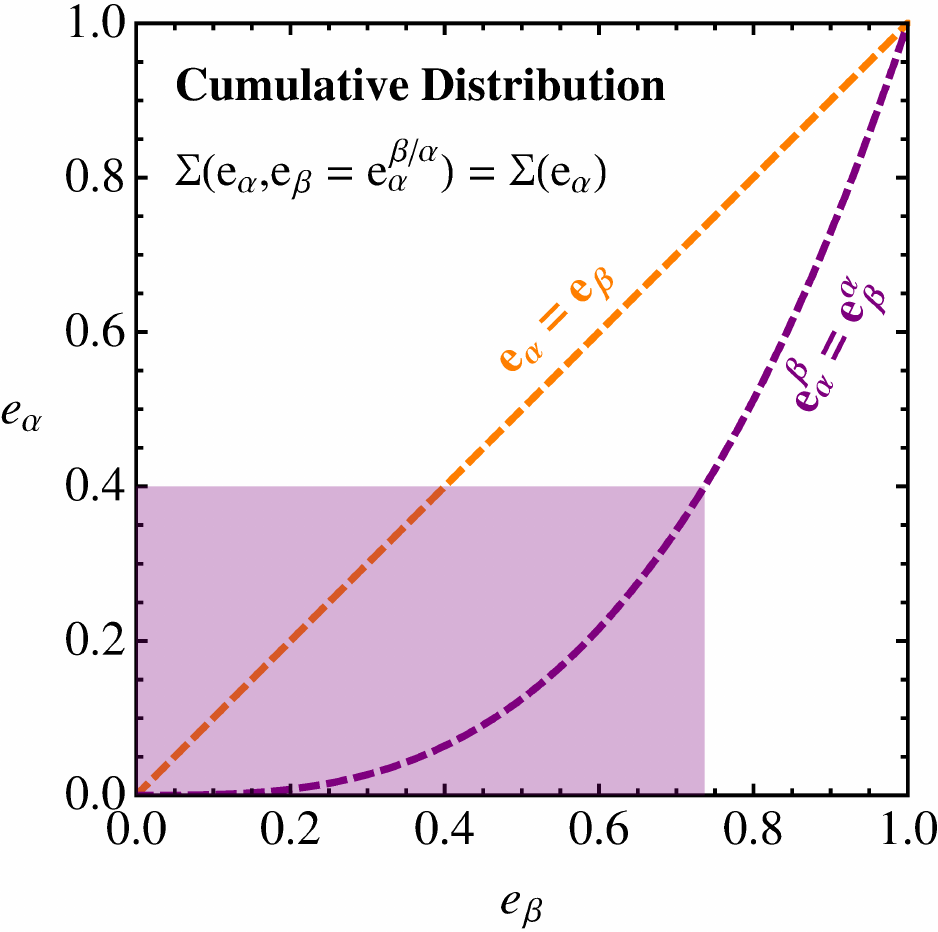}
}
$\qquad$
\subfloat[]{\label{fig:cum_b} 
\includegraphics[width=7cm]{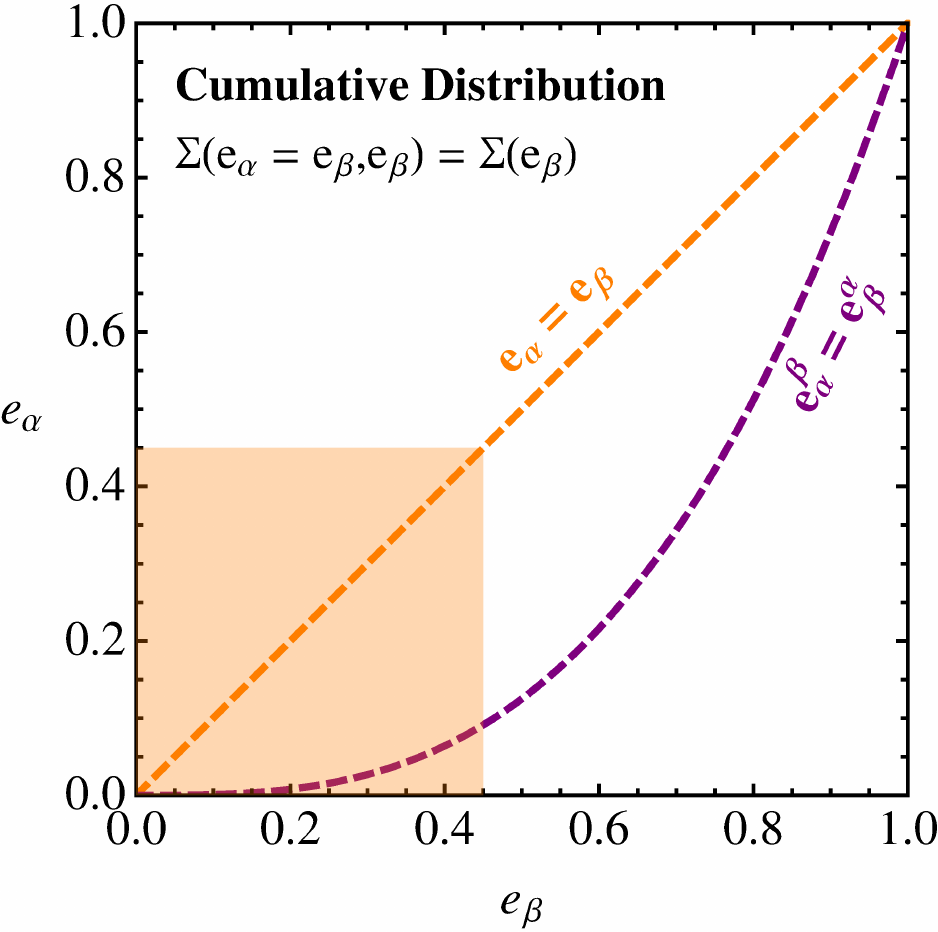}
}
\end{center}
\caption{
Illustration of the double cumulative distribution evaluated on the boundaries of phase space.  Left: Evaluated on the boundary $e_\alpha^\beta = e_\beta^\alpha$  which reduces the double cumulative distribution to $\Sigma(e_\alpha)$.  Right: Evaluated on the boundary $e_\alpha = e_\beta$  which reduces the double cumulative distribution to $\Sigma(e_\beta)$.
}
\label{fig:cum_bound}
\end{figure}

Consider again the allowed phase space for the two angularities $e_\alpha$ and $e_\beta$, with $\alpha>\beta$.  The double cumulative distribution $\Sigma(e_\alpha,e_\beta)$ is the integral of the double differential cross section over a rectangle that includes the origin of the phase space.  In particular, the double cumulative distribution can be evaluated at one of the boundaries of the phase space, illustrated in \Fig{fig:cum_bound}.  For example, if the double cumulative distribution is evaluated at the boundary where $e_\alpha^\beta = e_\beta^\alpha$, note that $e_\beta$ has been integrated over its entire allowed range: from $e_\beta = e_\alpha$ to $e_\beta = e_\alpha^{\beta/\alpha}$.  Therefore, on this boundary, the cumulative distribution can only depend on $e_\alpha$:
\begin{equation}
\Sigma(e_\alpha,e_\beta=e_\alpha^{\beta/\alpha}) = \Sigma(e_\alpha) \ ,
\end{equation}
where $\Sigma(e_\alpha)$ is the cumulative distribution for $e_\alpha$ alone.  A similar relationship exists on the other boundary, where
\begin{equation}
\Sigma(e_\alpha=e_\beta,e_\beta) = \Sigma(e_\beta)  \ .
\end{equation}

To determine the differential cross section, we differentiate the double cumulative distribution.  The boundary behavior of the cumulative distribution implies that
\begin{equation}
\left.\frac{\partial}{\partial e_\alpha}\Sigma(e_\alpha,e_\beta)\right|_{e_\beta = e_\alpha^{\beta/\alpha}} = \frac{\partial}{\partial e_\alpha} \Sigma(e_\alpha) = \frac{d\sigma}{de_\alpha} \ ,
\end{equation}
which is the single differential cross section for $e_\alpha$.  This can be related to the double differential cross section by integration:
\begin{equation}
\left.\frac{\partial}{\partial e_\alpha}\Sigma(e_\alpha,e_\beta)\right|_{e_\beta = e_\alpha^{\beta/\alpha}} = \int^{e_\alpha^{\beta/\alpha}} de_\beta \,\frac{d^2\sigma}{de_\alpha de_\beta} = \frac{d\sigma}{de_\alpha} \ ,
\end{equation}
which holds for all values of $e_\alpha^{\beta/\alpha}>0$. For this relationship to be true at this boundary, the double differential cross section should be expressable as\footnote{All that is necessary is that the function that multiplies the differential cross section of $e_\alpha$ integrates to 1 and the remainder function integrates to 0.  Using distributions, a function that integrates to 1 can always be expressed as an appropriate $\delta$-function plus a distribution that integrates to 0.  Therefore, the expression in \Eq{eq:general_form_of_FO_result} is not unique, but will be justisfied with the factorization theorem in the following sections.}
\begin{equation}\label{eq:general_form_of_FO_result}
\left.\frac{d^2\sigma}{de_\alpha de_\beta}\right|_{e_\beta \sim e_\alpha^{\beta/\alpha}} = \frac{d\sigma}{de_\alpha} \delta(e_\beta) + f_+(e_\alpha,e_\beta) \ ,
\end{equation}
where the $f_+$ function integrates to zero on $e_\beta\in[0,e_{\alpha}^{\beta/\alpha}]$.  A similar relationship holds for the other boundary, where $e_\alpha = e_\beta$. 

Then, the statement of the boundary factorization theorem is: the double differential cross section simplifies at the boundaries:
\begin{align}\label{eq:statement_of_analytic_structure}
\left.\frac{d^2\sigma}{de_\alpha de_\beta}\right|_{e_\beta \sim e_\alpha^{\beta/\alpha}} &\simeq \frac{d\sigma}{de_\alpha} \delta(e_\beta) + \frac{1}{e_{\alpha}^{1+\frac{\beta}{\alpha}}}f_{+}^{\alpha}\Bigg(\frac{e_{\beta}}{e_\alpha^{\beta/\alpha}}\Bigg)\ , \nonumber \\
\left.\frac{d^2\sigma}{de_\alpha de_\beta}\right|_{e_\alpha \sim e_\beta} &\simeq \frac{d\sigma}{de_\beta} \delta(e_\alpha) + \frac{1}{e_{\beta}^{2}}f_{+}^{\beta}\Bigg(\frac{e_\alpha}{e_{\beta}}\Bigg)\ .
\end{align}
We have made use of the fact that the arguments of the non-trivial functions assume a very specific form dictated by the factorization, as discussed in detail in \Sec{sec:ddiffunc}. Factorization theorems for differential cross sections of individual angularities are well-known \cite{Berger:2003iw,Ellis:2010rwa,Chiu:2012ir,broadening}; therefore, the double differential cross section factorizes at the boundaries.   In the following sections, we argue for factorization by studying the on-shell modes of the double differential cross section in detail. This relationship between the single and double differential cross section is quite remarkable, and can be understood as a precise statement of the UV structure of the effective theory. The $e_\alpha=0$ and $e_\beta=0$ lines are where all UV divergences are localized. This is consistent with the fact that these lines are parametrically far away from the boundary where the factorization theorem is valid, since this is a statement about the UV structure of the theory. 

\subsection{Modes of the Double Differential Cross Section}\label{sec:modes}

We now turn to studying the on-shell SCET modes that contribute to the double differential cross section.  For small values of a jet angularity $e_\alpha$, the dominant contributions to $e_\alpha$ come from collinear and soft radiation in the jet.  In general, the contribution to $e_\alpha$ from collinear modes scales like $\theta^\alpha$, where $\theta$ is the characteristic angular size of the collinear splittings.  Soft modes, by contrast, contribute an amount that scales like their energy.  Therefore, for the soft and collinear modes to be on-shell and contribute comparably to the angularity $e_\alpha$, their momenta must scale like\footnote{The literature is not in agreement whether to assign the $\alpha$ dependence to the soft or collinear mode, since it is only their relative invariant mass that is physical.  If we instead put the $\alpha$ dependence in the collinear mode, we would find two jet modes (instead of two soft modes) in \Tab{tab:scaling}.  However, the form of the factorization theorem would be identical.
}
\begin{align}\label{eq:mode_scale}
p_C &\sim Q(1,\lambda^2,\lambda) \nonumber \\
p_S &\sim Q(\lambda^\alpha,\lambda^\alpha,\lambda^\alpha) \ ,
\end{align}
in the $-$, $+$ and $\perp$ lightcone coordinates, respectively.  $\lambda$ is a small parameter which sets the size of the angularity; here $\lambda\sim e_\alpha^{1/\alpha}$.

For the double differential cross section, this analysis can be extended to the two angularities, $e_\alpha$ and $e_\beta$.  We will only consider on-shell modes, which for small values of $e_\alpha$ and $e_\beta$ are only soft and collinear radiation.  For on-shell collinear modes, the scaling of their momenta must be the same as for a single observable, from \Eq{eq:mode_scale}.  Because the angular scaling of the angularities $e_\alpha$ and $e_\beta$ is different, the collinear modes contribute an amount of order $\lambda^\alpha$ to $e_\alpha$ and $\lambda^\beta$ to $e_\beta$.  Soft modes are more subtle.  Now, because there are two angularities, there are two possible scalings of the soft modes.  Either the momenta of the soft modes scale like $\lambda^\alpha$ or they scale like $\lambda^\beta$.  Any other scaling would either be off-shell or would not be consistent with the collinear modes.  Therefore, while there is a single collinear mode that contributes to the double differential cross section of $e_\alpha$ and $e_\beta$, there are two soft modes whose scalings are set by the angular exponents of the angularities.

This is shown in \Tab{tab:scaling} where the scaling in the small parameter $\lambda$ of the collinear and soft modes is given in terms of their energy fraction $z$ and their splitting angle $\theta$.  Also, we show the contribution to the two angularities from each mode.  The collinear mode contributes a different amount to each angularity, depending on the angular exponent.  By contrast, each soft mode contributes the same amount to the two angularities, because the angularities are linear in the energy of the modes.  Thus, in addition to having to deal with two soft modes on a single jet, the scaling of the angularities will be unfamiliar from the single differential cross section.

\begin{table}[t]
\begin{center}
\begin{tabular}{c|c|c|c|c}
&$z$ & $\theta$ & $e_\alpha$ & $e_\beta$\\ 
\hline
$C$ & 1 & $\lambda$ & $\lambda^\alpha$ & $\lambda^\beta$\\
$S_\alpha$ & $\lambda^\alpha$ & 1 &  $\lambda^\alpha$ &  $\lambda^\alpha$\\
$S_\beta$ & $\lambda^\beta$ & 1 &  $\lambda^\beta$ &  $\lambda^\beta$
\end{tabular}
\end{center}
\caption{Scaling of the on-shell collinear ($C$) and soft ($S_\alpha,S_\beta$) modes of the double differential cross section.  $z$ is the energy fraction of the mode and $\theta$ is the angle of the mode from the jet axis.}
\label{tab:scaling}
\end{table}

We will prove in the following section that with this scaling of the modes, the double differential cross section does factorize.  Here, we will give a heuristic argument for the factorization of the cross section.  If we assume that $\alpha > \beta$, we can determine the dominant modes that contribute at leading power in $\lambda$ to the cross section.  For now, we will assume that the cross section can be written in the factorized form:
\begin{equation}\label{eq:toy_factorization}
\frac{1}{\sigma_0}\frac{d^2\sigma}{de_\alpha \, de_\beta} = H \times J(e_\alpha,e_\beta)\otimes S(e_\alpha,e_\beta) \ ,
\end{equation}
where $\sigma_0$ is the Born-level cross section, $H$ is the hard function, $J(e_\alpha,e_\beta)$ is the jet function describing the collinear modes' contribution and $S(e_\alpha,e_\beta)$ is the soft function describing the soft modes' contribution.  The $\otimes$ symbol denotes convolution. This form of the cross section is suggestive, but must be expanded in powers of $\lambda$ to ensure that the divergences in the hard, jet and soft functions cancel consistently at leading power in $\lambda$.

This expansion can be done depending on the chosen scaling of the soft modes.  By choosing a particular scaling of the soft modes, we restrict ourselves to a small region of the full angularities phase space, described in \Sec{sec:ps}, where those soft modes are on-shell.  
If we first choose the $S_\alpha$ soft modes, then the soft and collinear contributions to the angularity $e_\alpha$ both scale like $\lambda^\alpha$.  Therefore, they both will appear in the leading-power cross section.  However, for this choice of soft mode scaling, the contribution from soft modes to $e_\beta$, which scale like $\lambda^\alpha$, is power-suppressed with respect to the contribution from collinear modes, which scale like $\lambda^\beta$.
Explicitly, this is the limit in which $e_\alpha \ll e_\beta$, corresponding to a region of phase space far from the boundary $e_\alpha = e_\beta$.
Therefore with this choice of scaling of the soft modes, the leading-power factorized cross section has the form:
\begin{equation}
\frac{1}{\sigma_0}\frac{d^2\sigma^\alpha}{de_\alpha \, de_\beta} = H \times J(e_\alpha,e_\beta)\otimes S(e_\alpha) \ ,
\end{equation}
where the superscript $\alpha$ denotes the scaling of the soft modes.  Note that both angularities appear in the jet function and so the angle of the splitting is dominating the double differential cross section.  Thus, this form of the cross section is valid in the region of phase space controlled by collinear emissions, near the boundary where $e_\alpha^\beta = e_\beta^\alpha$.  By similar arguments, choosing the other scaling of the soft modes produces the factorized cross section
\begin{equation}\label{beta_bound_fact}
\frac{1}{\sigma_0}\frac{d^2\sigma^\beta}{de_\alpha \, de_\beta} = H \times J(e_\beta)\otimes S(e_\alpha,e_\beta) \ ,
\end{equation}
which corresponds to an expansion with $e_\beta \ll e_\alpha^{\beta/\alpha}$, which is far from the boundary where $e_\beta^\alpha = e_\alpha^\beta$.
Therefore, the factorization theorem of \Eq{beta_bound_fact} is valid near the boundary dominated by soft emissions, where $e_\alpha = e_\beta$.

Therefore, the double differential cross section factorizes near the boundaries of the phase space. The form of the factorization is quite interesting.  Near the $\alpha$ boundary, consistency of the renormalization group implies that
\begin{equation}
\gamma_H + \gamma_J(e_\alpha,e_\beta)+\gamma_S(e_\alpha) = 0 \ ,
\end{equation}
where $\gamma_F$ denotes the anomalous dimensions of the appropriate function $F$.  The hard function has no dependence on the observable and the soft anomalous dimension only depends on the angularity $e_\alpha$.  Thus, near this boundary of phase space the anomalous dimension of the jet function can only have non-trivial dependence on $e_\alpha$.  Now we are in a position to see that the analytic forms of the double differential factorization theorems given in \Eq{eq:statement_of_analytic_structure} capture the UV structure of each factorization. The $\delta$-function term (which multiplies the single differential cross-section) contains all of the divergences of the factorization, and hence dictates its canonical resummation. It must be the case that the UV-divergence structure is localized by these $\delta$-functions, since each factorization contains a single differential function, and between all sectors divergences cancel. Put simply, on a boundary the UV structure of the factorization reduces to that of the single differential cross-section.

\subsection{Proof of Boundary Factorization Theorem}

We now present a proof  in SCET that the double differential cross section factorizes near the boundaries of the phase space. In this proof, we will implicitly use many results from \Ref{Ellis:2010rwa} which discussed the factorization of jet observables for the first time, and so, here, will only focus on the novel aspects of the factorization theorem of the double differential cross section.  Also, our analysis will focus on jets in $e^+e^- \to q \bar{q}$ events, but using \Ref{Ellis:2010rwa}, this can be extended to jets in $e^+e^-$ collision events with any number of well-separated jets.

We begin with the double differential cross section in QCD for $e^+e^-\to q\bar{q}$:
\begin{equation}\label{eq:QCD_xsec}
\frac{d^2\sigma}{de_\alpha de_\beta}=L_{\mu\nu}\int d^dx\, \langle 0|J^{\mu}(x)\delta(\hat{e}_{\alpha}-e_{\alpha})\delta(\hat{e}_{\beta}-e_{\beta}){\cal O}_{J}J^{\nu}(0)|0\rangle \ ,
\end{equation}
where $L_{\mu\nu}$ is the leptonic tensor and $J^\mu(x)$ is the QCD current at position $x$.  The two $\delta$-functions enforce the measured values of the angularities and the operator ${\cal O}_J$ is the jet algorithm restriction.  Here, we will mostly be agnostic to the form of this operator.  It is defined to return a jet in the event on which the angularities are measured.  To the order to which we work, the jet algorithm in $e^+e^-\to q\bar{q}$ events can be enforced by integrating over one hemisphere of the event and boosting to constrain radiation of the other hemisphere to exist in a cone of radius $R_0$.  This setup will be sufficient for our discussion here, with a more detailed discussion of jet algorithm factorization left to \Ref{Ellis:2010rwa}.  As mentioned in the introduction, we will only discuss the resummation of global logarithms.  Factorization-violating non-global and clustering logarithms will be left to future work.

To be able to factorize the cross section, we match the QCD current with the corresponding current written in terms of fields in SCET as
\begin{align}\label{eq:QCD_Current_Matching}
J^{\mu}(x)&=\sum_nC \bar{\chi}_{n}S^{\dagger}_n\Gamma^\mu S_{\bar{n}}\chi_{\bar{n}}(x) \ ,
\end{align}
where $\bar{n}$ ($n$) is the (anti-)quark light-cone direction, $\chi$ ($\bar{\chi}$) is the collinear (anti-)quark field, $S$ ($S^\dagger$) is a light-like Wilson line, and $C$ is the matching coefficient matrix.  Spinor indices have been suppressed for simplicity.  However, matching SCET to QCD is more subtle than for single differential cross sections.  For the case of the double differential cross section of two angularities, the form of the factorization depends not only on the fact that the angularities are small, but also the way in which they scale with respect to one another.  This is important because the current matching does not set the virtuality of the soft radiation.  For a single differential cross section, the measurement of the observable sets the virtuality of the soft emission, but for the double differential cross section, the soft radiation does not have a unique, well-defined virtuality, as discussed in \Sec{sec:modes}.  Only once the relative scaling of the angularities $e_\alpha$ and $e_\beta$ is specified does the soft radiation have a well-defined virtuality.

\begin{figure}
\begin{center}
\includegraphics[width=7cm]{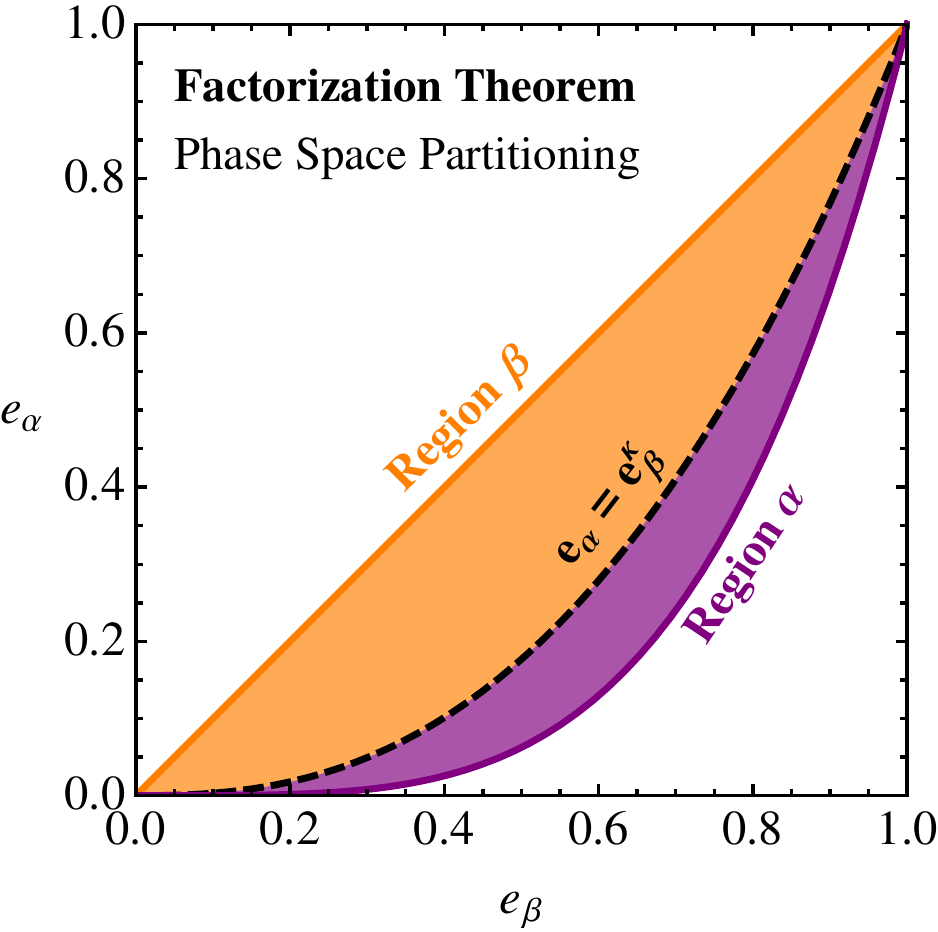}
\end{center}
\caption{
Angularities phase space divided into boundary regions $\alpha$ and $\beta$ in which different factorization theorems live, defined by the soft modes' virtuality.  The virtuality of the soft modes in region $\alpha$ ($\beta$) is $\lambda^{2\alpha}$ ($\lambda^{2\beta}$).  The dividing line of the regions is $e_\alpha = e_\beta^\kappa$, where $\kappa\in[1,\alpha/\beta]$.
}
\label{fig:ps_fact}
\end{figure}

To enforce a virtuality of the soft modes, we can restrict the measurement operator to only have support in the region of phase space where the relative scaling of the angularities $e_\alpha$ and $e_\beta$ produce a unique soft mode.  From \Sec{sec:modes} we found two on-shell soft modes, and so to do this, we can partition the phase space into two regions: one in which the soft modes have virtuality $\lambda^{2\alpha}$ and the other in which the soft modes have the virtuality $\lambda^{2\beta}$.  That is, the measurement operator can be written as
\begin{align}\label{eq:partition}
\delta(\hat{e}_{\alpha}-e_{\alpha})\delta(\hat{e}_{\beta}-e_{\beta})&=\delta(\hat{e}_{\alpha}-e_{\alpha})\delta(\hat{e}_{\beta}-e_{\beta})[\Theta(e_\alpha-e_\beta^\kappa)+\Theta(e_\beta^\kappa-e_\alpha)]  \ ,
\end{align}
where we have inserted the identity.  $\kappa$ controls the relative scaling of $e_\alpha$ with respect to $e_\beta$.  On the physical phase space, $\kappa\in[1,\alpha/\beta]$.  At this level, \Eq{eq:partition} is an operator identity, however, each $\Theta$-function constrains the angularities in a region of phase space with a unique, on-shell soft mode.  This partitioning is illustrated in \Fig{fig:ps_fact} where boundary region $\alpha$ corresponds to $e_\beta^\kappa > e_\alpha$ and boundary region $\beta$ corresponds to $e_\alpha > e_\beta^\kappa$.

Inserting \Eq{eq:partition} into the expression for the full QCD cross section, we have
\begin{align}\label{eq:partitionxsec}
\frac{d^2\sigma}{de_\alpha de_\beta}&=L_{\mu\nu}\int d^dx\, \langle 0|J^{\mu}(x)\delta(\hat{e}_{\alpha}-e_{\alpha})\delta(\hat{e}_{\beta}-e_{\beta}){\cal O}_{J}J^{\nu}(0)|0\rangle \nonumber \\
&=L_{\mu\nu}\int d^dx\,\Theta(e_\alpha-e_\beta^\kappa)\, \langle 0|J^{\mu}(x)\delta(\hat{e}_{\alpha}-e_{\alpha})\delta(\hat{e}_{\beta}-e_{\beta}){\cal O}_{J}J^{\nu}(0)|0\rangle \nonumber \\
& \qquad +L_{\mu\nu}\int d^dx\,\Theta(e_\beta^\kappa-e_\alpha)\, \langle 0|J^{\mu}(x)\delta(\hat{e}_{\alpha}-e_{\alpha})\delta(\hat{e}_{\beta}-e_{\beta}){\cal O}_{J}J^{\nu}(0)|0\rangle \ .
\end{align}
Note that the $\Theta$-functions commute with all operators because they are functions of pure numbers (i.e., the value of the angularities). 
Because each term after the second equal sign in \Eq{eq:partitionxsec} only has a single on-shell soft mode, each term separately can be factorized by matching currents as defined in \Eq{eq:QCD_Current_Matching}.  For single angularities measured with respect to the jet thrust axis, this was done in \Ref{Ellis:2010rwa} and for angularities measured with respect to the jet broadening axis this was done in \Ref{broadening}.  Because it is a relatively standard and familiar procedure, we do not present the details of the factorization of the QCD cross section into SCET operators.

Performing the factorization and expanding to leading power, we find the following form for the cross section for angularities $e_\alpha$ and $e_\beta$ measured on a single jet
\begin{equation}\label{eq:factordd}
\frac{1}{\sigma_0}\frac{d^2\sigma}{de_\alpha de_\beta}= \Theta(e_\beta^\kappa-e_\alpha)H\times J(e_{\alpha},e_{\beta})\otimes_\alpha S(e_\alpha)+ \Theta(e_\alpha-e_\beta^\kappa)H\times J(e_{\beta})\otimes_\beta S(e_\alpha,e_\beta) \ ,
\end{equation}
where the hard function $H$ is the absolute square of the matching coefficient matrix $C$, $H = C^\dagger C$.  The subscript on the symbol $\otimes$ denotes the appropriate convolution.  For example, in the first term, because the soft function is independent of $e_\beta$, the jet and soft functions are only convolved in $e_\alpha$.
The single differential functions are
\begin{align}\label{sing_diff_jet_soft_func}
J(e_{\beta})&=  \frac{(2\pi)^3}{N_c} \,  \langle 0 \vert \,    \bar{\chi}_{\bar n}   \,\delta( n \cdot \hat P - Q ) \, \delta(\hat{e}_{\beta}-e_{\beta}) \, {\cal O}_J \delta^{(2)}(\hat P_{\perp})\,\frac{\nslash}{2} \, \chi_{\bar n}    \, \vert 0\rangle \,,\nonumber \\
S(e_{\alpha})& = \frac{1}{N_c}{\rm tr} \, \langle 0 \vert \,  T\Big\{      S^{\dagger}_{\bar n} \, S_{n}  \Big\}\,  \delta(\hat{e}_{\alpha}-e_{\alpha})\, {\cal O}_J \, 
\bar{T}\Big\{   S^{\dagger}_{n} \, S_{\bar n} \Big\}  \, \vert 0\rangle \, ,
\end{align}
where the jet's $-$ component of momentum is $Q$ and we have assumed that the jet is in the $\bar{n}$ direction.
The double differential jet and soft functions are
\begin{align}\label{double_diff_jet_soft_func}
J(e_{\alpha},e_{\beta})&=  \frac{(2\pi)^3}{N_c} \,  \langle 0 \vert \,    \bar{\chi}_{\bar n}   \,\delta( n \cdot \hat P - Q ) \, \delta(\hat{e}_{\alpha}-e_{\alpha})\delta(\hat{e}_{\beta}-e_{\beta}) \, {\cal O}_J \, \delta^{(2)}(\hat P_{\perp})\,\frac{\nslash}{2} \, \chi_{\bar n}    \, \vert 0\rangle \,,\nonumber \\
S(e_{\alpha},e_{\beta})& = \frac{1}{N_c}{\rm tr} \, \langle 0 \vert \,T\Big\{      S^{\dagger}_{\bar n} \, S_{n}\Big\} \,   \delta(\hat{e}_{\alpha}-e_{\alpha})\delta(\hat{e}_{\beta}-e_{\beta}) \,{\cal O}_J \, 
  \bar{T}\Big\{  S^{\dagger}_{n} \, S_{\bar n}  \Big\} \, \vert 0\rangle \ .
\end{align}
The calculation of the double differential jet and soft functions will be discussed in \Sec{sec:ddiffunc}. The definitions of the various operators appearing in these functions can be found \Ref{Chiu:2012ir} and references therein.

Thus, the double differential cross section factorizes in the boundary regions of phase space.  By the arguments of the previous section the double differential cross section reduces to a single differential cross section of the appropriate angularity, depending on the boundary.  Because the double differential cross section must be independent on the choice of partitioning defined by $\kappa$, this will provide powerful constraints on the double differential cross section and will allow us to define an interpolating function from the boundaries into the bulk of the phase space.  This will be studied in detail in \Sec{sec:int}.

\subsection{Limit of Soft-Collinear Factorization}

We are now in a position to understand why only two factorization theorems can be written down for the double differential cross section, and using the traditional ingredients of soft-collinear factorization, no universal factorization formula could be presented.\footnote{We thank Daekyoung Kang, Iain Stewart, and Jesse Thaler for extensive discussions on this point.} Two separate arguments apply, leading to this conclusion. First, the fact that there exist two distinct soft modes as defined in \Tab{tab:scaling} implies
that there is no unique singular fixed-order cross section.  Rather, there are two different singular cross sections that depend on the scaling of the soft mode. No one soft mode covers all of phase space. 

Alternatively, one can be wholly ignorant of the power-counting and still come to the same conclusion. Formally, the SCET Lagrangians for the soft and collinear sectors at leading power are equivalent to full QCD \cite{Bauer:2008qu}. Thus one can forget about the \emph{relative} power counting of the low-scale components of the momenta between the soft and jet modes, and simply write down all possible jet and soft functions that could contribute. As long as the number of jets in the process is fixed, and hence also both the hard function and the number of eikonal lines in the soft function, the set of jet and soft functions is finite, and is controlled only by how many angularity measurements are imposed on a sector. So for a two-jet process, the only on-shell functions that can be written down are the single and double differential jet and soft functions from \Eqs{sing_diff_jet_soft_func}{double_diff_jet_soft_func}. Fixing the phase space fixes the form of the divergences, regardless of how one power-counts the modes in the sector relative to each other.\footnote{This is why the equivalence between the SCET and QCD Langrangians is important: to calculate the function once the operators are fixed, one does not need to know the power counting.} Given these functions, a simple one-loop calculation is sufficient to show which combinations are RG consistent with each other.

 As can be seen from the results of \App{app:jsfunc}, the only RG consistent combinations are those in \Eq{eq:factordd}. In particular, using only soft or collinear modes, there is no sense to the factorization theorem \Eq{eq:toy_factorization}, independent of any argument about power counting. This constitutes a remarkable test of the consistency and power of the effective theory approach: fixing the scaling of the soft modes and appropriately expanding the phase space according to the power counting automatically generates RG-consistent combinations of on-shell functions. The precise power counting must be taken seriously to have consistent factorization. Of course one must eventually consider the power counting to know where in the phase space a given factorization formula holds, and this then shows that there is no universal factorization formula using traditional ingredients of soft-collinear factorization.

The non-uniqueness of the low-scale theory has an important consequence new to multi-differential cross sections. Namely, there is no operator product expansion (OPE) from one region to the other that allows a tower of effective theories that one could construct that covers all of phase space. Thus no RG scheme can connect the different regions of phase space. This is in distinction to the single differential cross section, where the singular distribution is unique. Indeed, this uniqueness of the singular terms is what allows the various regions of the differential cross section to be connected by controlling the RG evolution of the sectors. Even when one is in the tail of the distribution, the factorization theorem is correctly reproducing a unique set of terms in the fixed-order cross section, so one only needs to add the non-singular terms in the cross section to achieve the full result.

\section{Double Differential Jet and Soft Functions}\label{sec:ddiffunc}

The factorized form of the double differential cross section from \Eq{eq:factordd} contains single as well as double differential jet and soft functions.  Soft and jet functions for individual angularities measured on a jet have been computed in \Refs{Ellis:2010rwa,broadening}, but the double differential objects are new.  As discussed in the previous section, the divergences of the double differential jet and soft functions can only have non-trivial dependence on one of the angularities, for consistency of the factorization.  However, the finite terms will have non-trivial dependence on both angularities and these contributions are necessary for improved accuracy of the double differential cross section.    Here, we use general arguments to determine the form of the double differential jet and soft functions to all orders.  The explicit calculation of the jet and soft functions is presented in \App{app:jsfunc}.

\subsection{Jet Function}\label{sec:ddjetfunc}

Much of the structure of the double differential jet function can be determined by power counting and the form of the factorization theorem.  
From the power counting of the factorization theorem, the jet function must scale as
\begin{align}
J(e_{\alpha},e_{\beta})\sim \frac{1}{\lambda^{\alpha+\beta}} \ ,
\end{align}
where the angularities scale as $e_\alpha\sim\lambda^\alpha$ and $e_\beta\sim\lambda^\beta$.  In addition, for consistency of the factorization theorem, the divergences in the double differential jet function can only have non-trivial dependence on $e_\alpha$, of exactly the same form as the single differential jet function:
\begin{equation}
\left[J(e_\alpha,e_\beta)\right]_\text{div} = \left[J(e_\alpha)\right]_\text{div} \delta(e_\beta)  \ ,
\end{equation}
where $\text{div}$ denotes the divergent parts of the jet functions.  These two observations imply that the jet function has the following general form to all orders:
\begin{equation}
J(e_{\alpha},e_{\beta})= C(\alpha_s)\,\delta(e_{\alpha})\delta(e_{\beta})+e_{\alpha}^{-1-\frac{\beta}{\alpha}}\sum_{L=1}^{\infty} D_L(\alpha_s) \left(\frac{\mu}{e_{\alpha}^{\frac{1}{\alpha}}Q}\right)^{2L\epsilon } F_L\left(\frac{e_\beta}{e_\alpha^{\frac{\beta}{\alpha}}}\right) \ .
\end{equation}
The sum runs over all loop orders $L$ with $C(\alpha_s)=1+{\cal O}(\alpha_s)$ and $D_L(\alpha_s)={\cal O}(\alpha_s^L)$.   $\epsilon$ is the dimensional regularization parameter and the jet scale $\mu$ must appear in the combination $$\frac{\mu}{e_\alpha^{1/\alpha} Q}$$ to be consistent with the anomalous dimension.  $F_L$ is a function that depends on the loop order but scales like $\lambda^0$ to all orders.  The only such combination of $e_\alpha$ and $e_\beta$ with this scaling is is $e_\alpha^{\frac{\beta}{\alpha}}/e_\beta$.

This last quality is critical so that all of the divergences can be localized at $e_\beta=0$ (as required by the factorization theorem) with the $+$-prescription \cite{Ligeti:2008ac}.  For a function $f$ with support on $[0,b]$, where $b>0$, the function can be expressed as
\begin{equation}\label{eq:plusfunc}
\Theta(x)\Theta(b-x)\, f\left(\frac{x}{b}\right) = \delta(x) \, \int_0^b dx'\, f\left( \frac{x'}{b}  \right) + \left[ \Theta(x)\Theta(b-x)\, f\left(\frac{x}{b}\right) \right]_+^b \ .
\end{equation}
The $b$ superscript denotes that the $+$-distribution is defined on $(0,b]$.  It has the property that it integrates to zero:
\begin{equation}
\int_0^b dx'\,\left[ \Theta(x')\Theta(b-x')\, f\left(\frac{x'}{b}\right) \right]_+^b = 0 \ .
\end{equation}
For the jet function, $e_\beta$ is defined on $[0,e_\alpha^{\beta/\alpha}]$, up to an ${\cal O}(1)$ factor for the upper bound.  Therefore, order-by-order, the function $F_L$ can be regulated by the $+$-prescription:
\begin{equation}
\Theta\left( e_\beta \right) \Theta(e_\alpha^{\frac{\beta}{\alpha}}-e_\beta)\, F_L\left(\frac{e_\beta}{e_\alpha^{\frac{\beta}{\alpha}}}\right) = \delta(e_\beta) \, \int_0^{e_\alpha^{\frac{\beta}{\alpha}}} de'_\beta\, F_L\left(\frac{e'_\beta}{e_\alpha^{\frac{\beta}{\alpha}}}\right)+ \left[ \Theta\left( e_\beta \right) \Theta(e_\alpha^{\frac{\beta}{\alpha}}-e_\beta)\,  F_L\left(\frac{e_\beta}{e_\alpha^{\frac{\beta}{\alpha}}}\right) \right]_+^{e_\alpha^{\frac{\beta}{\alpha}}} \ .
\end{equation}
The power counting guarantees that all of the dependence of the jet function on $e_\beta$ can be regulated by the $+$-prescription as it only appears in jet function in the combination $e_\alpha^{\frac{\beta}{\alpha}}/e_\beta$.  This is a powerful test of the consistency of the factorization theorem, since the power counting forced the particular form of the factorization in \Eq{eq:factordd}.  The explicit calculation of the double differential jet function at one-loop is presented in \App{app:jfunc}.

\subsection{Soft Function}\label{sec:softfunc}

We apply similar arguments to the the double differential soft function, $S(e_\alpha,e_\beta)$.  The scaling of the soft function is different than the jet function, because it exists near the boundary where $e_\alpha = e_\beta\sim \lambda^\beta$.  Then, from the factorization theorem the soft function scales like
\begin{equation}
S(e_\alpha,e_\beta) \sim \frac{1}{\lambda^{2\beta}} \ .
\end{equation}
For consistency of the factorization theorem, the divergences in the double differential soft function can only have non-trivial dependence on $e_\beta$ and must be of the same form as the single differential soft function:
\begin{equation}
\left[  S(e_\alpha,e_\beta) \right]_\text{div}=\left[  S(e_\beta) \right]_\text{div} \delta(e_\alpha) \ .
\end{equation}
As with the jet function, these observations imply that the soft function has the following general form to all orders:\footnote{Of course, the functions $C$, $D_L$, and $F_L$ will be different for the double differential jet and soft functions.}
\begin{equation}\label{eq:softfunc_exp}
S(e_{\alpha},e_{\beta})= C(\alpha_s)\,\delta(e_{\alpha})\delta(e_{\beta})+e_{\beta}^{-2}\sum_{L=1}^{\infty} D_L(\alpha_s) \left(\frac{\mu}{e_{\beta}Q}\right)^{2L\epsilon } F_L\left(\frac{e_\alpha}{e_\beta}\right) \ .
\end{equation}
Again, the sum runs over all loop orders $L$ with $C(\alpha_s)=1+{\cal O}(\alpha_s)$ and $D_L(\alpha_s)={\cal O}(\alpha_s^L)$. 
 The soft scale $\mu$ must appear in the combination $$\frac{\mu}{e_\beta Q}$$ to be consistent with the anomalous dimension.  $F_L$ is a function that depends on the loop order but scales like $\lambda^0$ to all orders.  For the scaling of the double differential soft function, the only such combination of $e_\alpha$ and $e_\beta$ that scales like $\lambda^0$ is $e_\alpha/e_\beta$. 

The singularities of the soft function can be localized at $e_\alpha = 0$ by the $+$-prescription.  As discussed with the jet function, because $e_\alpha$ is defined on $[0,e_\beta]$ in the soft function, the function $F_L$ can be written as a $+$-distribution:
\begin{equation}
\Theta\left( e_\alpha \right) \Theta(e_\beta-e_\alpha)\, F_L\left(\frac{e_\alpha}{e_\beta}\right) = \delta(e_\alpha) \, \int_0^{e_\beta} de'_\alpha\, F_L\left(\frac{e'_\alpha}{e_\beta}\right)+ \left[ \Theta\left( e_\alpha \right) \Theta(e_\beta-e_\alpha)\,  F_L\left(\frac{e_\alpha}{e_\beta}\right) \right]_+^{e_\beta} \ .
\end{equation}
 The calculation of the double differential soft function at one-loop is presented in \App{app:sfunc}.

\section{Interpolating between Boundary Regions}\label{sec:int}

With the boundary factorization theorem, we would like to determine the double differential cross section throughout the allowed phase space for the two angularities.  Because the factorization theorem only holds near the boundaries, we cannot claim any formal accuracy in the bulk of the phase space.  Nevertheless, because the double differential cross section must satisfy several non-trivial constraints, these can be used to determine an interpolation from one boundary of the phase space to the other.  In this section, we will present the interpolation to NLL accuracy in the boundary factorization theorem.

First, we will define what we mean by ``NLL accuracy'' for the double differential cross section.  Typically, for a single observable $e$, NLL is defined to capture the leading terms in the exponent of the cumulative distribution with the scaling that $\alpha_s \log e\sim 1$.  That is, NLL accuracy is
\begin{equation}
\log \Sigma^\text{NLL}(e) \supset \alpha_s^n \log^{n+1} e,\ \alpha_s^n \log^{n} e \ ,
\end{equation}
for all $n>0$.  For the double cumulative distribution of angularities $e_\alpha$ and $e_\beta$, we define NLL similarly, but include all possible logarithms of $e_\alpha$ and $e_\beta$:
\begin{equation}
\log \Sigma^\text{NLL}(e_\alpha,e_\beta) \supset \alpha_s^n \log^{n+1-m} e_\alpha \log^m e_\beta,\ \alpha_s^n \log^{n-l} e_\alpha\log^{l} e_\beta \ ,
\end{equation}
for all $n>0$, $0\leq m\leq n+1$ and $0\leq l\leq n$.  This definition assumes that the logarithms of the double cumulative distribution exponentiates, which we believe is a reasonable expectation.\footnote{This subtlety will be discussed further in \Sec{sec:uniint}.}  Also, as we measure the angularities on a jet, there will be non-global logarithms that arise at NLL; however, we will ignore them here.

Now, we collect the constraints that were discussed in \Sec{sec:ft} on the double differential cross section of two angularities and its factorization theorem.  With $\Sigma(e_\alpha,e_\beta)$ the double cumulative distribution of the angularities $e_\alpha$ and $e_\beta$, it must reduce on the boundaries to:
\begin{equation}\label{eq:cumconsts}
\Sigma(e_\alpha,e_\beta=e_\alpha^{\beta/\alpha})  = \Sigma(e_\alpha) \ , \qquad \Sigma(e_\alpha=e_\beta,e_\beta)  = \Sigma(e_\beta) \ .
\end{equation}
The derivatives of the cumulative distribution are also constrained:
\begin{align}\label{eq:diffconsts}
\left.\frac{\partial}{\partial e_\alpha}\Sigma(e_\alpha,e_\beta)\right|_{e_\beta=e_\alpha^{\beta/\alpha}} &= \frac{d\sigma}{de_\alpha} \ , \qquad\qquad\qquad  \left.\frac{\partial}{\partial e_\beta}\Sigma(e_\alpha,e_\beta)\right|_{e_\alpha=e_\beta} = \frac{d\sigma}{de_\beta} \ , \nonumber\\
\left.\frac{\partial}{\partial e_\alpha}\Sigma(e_\alpha,e_\beta)\right|_{e_\beta=e_\alpha} &= 0 \ , \ \, \qquad\qquad\qquad  \left.\frac{\partial}{\partial e_\beta}\Sigma(e_\alpha,e_\beta)\right|_{e_\alpha=e_\beta^{\alpha/\beta}} = 0 \ .
\end{align}
The fact that these constraints are satisfied only for the total cross section implies that the factorization into soft and collinear modes cannot occur throughout the allowed phase space.  The form of the boundary factorization theorem from \Eq{eq:factordd} is 
\begin{equation}
\frac{1}{\sigma_0}\frac{d^2\sigma}{de_\alpha de_\beta}= \Theta(e_\beta^\kappa-e_\alpha)H\times J(e_{\alpha},e_{\beta})\otimes_\alpha S(e_\alpha)+ \Theta(e_\alpha-e_\beta^\kappa)H\times J(e_{\beta})\otimes_\beta S(e_\alpha,e_\beta) \ .
\end{equation}
This must be independent of $\kappa$ for the factorization theorems at the two boundaries to be consistent with one another.

To determine a conjecture for the double differential cross section that interpolates between the boundaries of phase space subject to the above constraints, we will do the simplest thing possible.  We will set the scales in the logarithms that appear in the boundary factorization theorem appropriately so that the total cross section constraints in \Eqs{eq:cumconsts}{eq:diffconsts} are satisfied and the two boundary factorization theorems match onto one another continuously. However, because the RG evolution in the double differential cross section can only ever generate a non-zero value for one of the two angularities at the boundaries, this interpolation must be done at the level of the double cumulative cross section.

With this approach, we can then set scales in the logarithms of the double cumulative distribution on the boundary where $e_\beta=e_\alpha^{\beta/\alpha}$ (where it reduces to the cumulative distribution for $e_\alpha$ alone) so that when continued to the boundary where $e_\alpha=e_\beta$, it reduces to the cumulative distribution for $e_\beta$ and satisfies the other constraints.  As we observed in the fixed-order calculation of \Sec{sec:fo}, to satisfy the derivative constraints on the double cumulative distribution to single logarithmic accuracy required including na\"ively power-suppressed terms in the cumulative distribution.  Similar power-suppressed terms will need to be included in the resummed double cumulative distribution, in addition to setting scales, to satisfy all constraints.  Because these power-suppressed terms are not exponentiated in the boundary factorization theorem, they are otherwise arbitrary and correspond to an uncertainty in the calculation.

To illustrate our procedure for interpolation, we will study in detail the double cumulative distribution to NLL accuracy.  This will allow us to use known results for the cumulative distributions for individual recoil-free angularities at NLL.\footnote{At NLL, recoil-free angularities are identical to two-point energy correlation functions for the same value of the angular exponent $\beta$.}  Higher accuracy can be achieved by matching to fixed-order double differential cross sections, profiling the jet and soft scales in the factorization theorem \cite{Abbate:2010xh} or resumming the individual angularities to higher logarithmic order.  Here, we will only consider NLL order and will address improved accuracy in future work.

\subsection{NLL Interpolation}\label{sec:nllint}

At the boundaries of the phase space, the double cumulative distribution of two angularities $e_\alpha$ and $e_\beta$ must reduce to the cumulative distribution of a single angularity, so we start by considering the form of the cumulative distribution for a single recoil-free angularity.  To NLL accuracy,\footnote{Of course, we are ignoring non-global logarithms that first arise at NLL.} the normalized cumulative distribution of a single recoil-free angularity $e_\beta$ measured on a jet can be expressed as \cite{Banfi:2004yd,Ellis:2010rwa}
\begin{equation}\label{eq:nll_sing}
\Sigma(e_\beta) =\frac{e^{-\gamma_E R'(e_\beta)}}{\Gamma(1+R'(e_\beta))} e^{-R(e_\beta)-\gamma_i T(e_\beta)} \ .
\end{equation}
$R(e_\beta)$ is often referred to as the radiator and consists of the cusp pieces of the anomalous dimensions of the jet and soft function and to NLL accuracy, is evaluated at two-loop order.  The second term in this exponent, $\gamma_i T(e_\beta)$, is the non-cusp piece of the anomalous dimensions which result from hard collinear splittings.  For NLL accuracy, it is evaluated at one-loop order.  The prefactor accounts for the effects of multiple emissions adding together to produce a given value of the angularity $e_\beta$.  $R'(e_\beta)$ is the logarithmic derivative of the radiator:
\begin{equation}
R'(e_\beta)\equiv -\frac{\partial}{\partial \log e_\beta}R(e_\beta) \ .
\end{equation}
$\gamma_E$ is the Euler-Mascheroni constant.  The explicit expression for the cumulative distribution at NLL is given in \App{app:one}.

Because our strategy for achieving the interpolation is to only change the scale of the logarithms appearing in the single cumulative distribution, the normalized double cumulative distribution must be of the same functional form:
\begin{equation}\label{eq:nll_dub}
\Sigma(e_\alpha,e_\beta) =\frac{e^{-\gamma_E \tilde{R}(e_\alpha,e_\beta)}}{\Gamma(1+\tilde{R}(e_\alpha,e_\beta))} e^{-R(e_\alpha,e_\beta)-\gamma_i T(e_\alpha,e_\beta)} \ ,
\end{equation}
for some functions $R(e_\alpha,e_\beta)$, $T(e_\alpha,e_\beta)$ and $\tilde{R}(e_\alpha,e_\beta)$.  This then enforces the boundary conditions on the double cumulative distribution onto its constituent functions:
\begin{align}
R(e_\alpha,e_\beta =e_\alpha^{\beta/\alpha}) &= R(e_\alpha) \ , \qquad  R(e_\alpha=e_\beta,e_\beta) = R(e_\beta) \ , \nonumber \\
T(e_\alpha,e_\beta =e_\alpha^{\beta/\alpha}) &= T(e_\alpha) \ , \qquad  T(e_\alpha=e_\beta,e_\beta) = T(e_\beta) \ , \nonumber  \\
\tilde{R}(e_\alpha,e_\beta =e_\alpha^{\beta/\alpha}) &= R'(e_\alpha) \ , \quad \ \, \tilde{R}(e_\alpha=e_\beta,e_\beta) = R'(e_\beta) \ , \nonumber  
\end{align}
up to terms suppressed by positive powers of $e_\alpha,e_\beta$.  In addition, the derivatives of each constituent function must satisfy the boundary conditions so as to correctly reproduce the differential cross sections of individual angularities at the boundaries.  For example, for the derivative with respect to $e_\alpha$, we have the following derivative boundary conditions:
\begin{align}
\left.\frac{\partial}{\partial e_\alpha}R(e_\alpha,e_\beta)\right|_{e_\beta = e_\alpha^{\beta/\alpha}} &= \frac{\partial}{\partial e_\alpha}R(e_\alpha) \ , \qquad  \left.\frac{\partial}{\partial e_\alpha}R(e_\alpha,e_\beta)\right|_{e_\beta = e_\alpha} = 0 \ , \nonumber \\
\left.\frac{\partial}{\partial e_\alpha}T(e_\alpha,e_\beta)\right|_{e_\beta = e_\alpha^{\beta/\alpha}} &= \frac{\partial}{\partial e_\alpha}T(e_\alpha) \ , \qquad  \left.\frac{\partial}{\partial e_\alpha}T(e_\alpha,e_\beta)\right|_{e_\beta = e_\alpha} = 0 \ , \nonumber  \\
\left.\frac{\partial}{\partial e_\alpha}\tilde{R}(e_\alpha,e_\beta)\right|_{e_\beta = e_\alpha^{\beta/\alpha}} &= \frac{\partial}{\partial e_\alpha}R'(e_\alpha) \ , \quad \ \, \left.\frac{\partial}{\partial e_\alpha}\tilde{R}(e_\alpha,e_\beta)\right|_{e_\beta = e_\alpha} = 0 \ . \nonumber  
\end{align}
Similar constraints exist for derivatives with respect to $e_\beta$.  With these results, we can consider each function separately and determine how it can be defined so as to interpolate between the boundary regions.  As illustration of the interpolation, we will analyze the one-loop cusp component of the radiator $R(e_\alpha,e_\beta)$ and the non-cusp function $T(e_\alpha,e_\beta)$.  The complete expression for the double cumulative distribution that satisfies all constraints is given in \App{app:two}.

An important point to note is that, because they are defined by one-gluon emission, $R(e_\alpha,e_\beta)$ and $T(e_\alpha,e_\beta)$ can be directly computed in QCD.  Here, we choose to compute them via the interpolation to illustrate the procedure.  Also, we expect that the logarithmic structures generated by the interpolation are generic, and could be tested by computing anomalous dimensions at higher orders directly.  On the other hand, the multiple emissions factor $\tilde{R}(e_\alpha,e_\beta)$ can not be interpreted as the logarithmic derivative of the radiator $R(e_\alpha,e_\beta)$ and so the method for computing it directly is not clear.  However, its logarithmic structure can be determined by  matching to the boundary conditions.  This is an illustration of the power of the interpolation.

\subsubsection{One-Loop Cusp/Radiator Interpolation}\label{sec:cuspint}

Consider first the one-loop radiator function for the angularity $e_\alpha$:
\begin{align}
\label{eq:radcusp}
R^{(1)}(e_\alpha) &= \frac{C_i}{2\pi \alpha_s \beta_0^2}\left[
\frac{1}{\alpha-1}
\left( 1+2\alpha_s \beta_0 \log e_\alpha  \right) \log(1+2\alpha_s \beta_0 \log e_\alpha)\right.\nonumber \\
&\left. \qquad\qquad\qquad
-\frac{\alpha}{\alpha-1}\left(1+2\alpha_s \beta_0 \frac{\log e_\alpha}{\alpha}\right)\log\left(1+2\alpha_s \beta_0 \frac{\log e_\alpha}{\alpha}\right)
\right] \ ,
\end{align}
where $\beta_0$ is the coefficient of the one-loop $\beta$-function.  Equivalently, this can be written as an integral over the jet and soft function cusp anomalous dimensions:
\begin{equation}\label{eq:radintcusp}
R^{(1)}(e_\alpha)=-2\int_{\alpha_s(\mu_J)}^{\alpha_s(\mu)}\frac{d\alpha'}{\beta[\alpha']}\Gamma_J[\alpha']\int_{\alpha_s(\mu_J)}^{\alpha'} \frac{d\alpha''}{\beta[\alpha'']}-2\int_{\alpha_s(\mu_S)}^{\alpha_s(\mu)}\frac{d\alpha'}{\beta[\alpha']}\Gamma_S[\alpha']\int_{\alpha_s(\mu_S)}^{\alpha'} \frac{d\alpha''}{\beta[\alpha'']} \ ,
\end{equation}
where $\beta[\alpha_s]$ is the $\beta$-function and the cusp anomalous dimensions of the jet and soft function to one-loop are
\begin{equation}
\Gamma_J[\alpha_s] = \frac{\alpha_s}{\pi} C_i \frac{\alpha}{\alpha-1} \ , \qquad \Gamma_S[\alpha_s] = -\frac{\alpha_s}{\pi} C_i \frac{1}{\alpha-1} \ .
\end{equation}
$\mu$ is the renormalization scale and $\mu_J$ and $\mu_S$ are the jet and soft scales, which we take to be their canonical values:
\begin{equation}
\mu_J = e_\alpha^{1/\alpha} Q \ , \qquad \mu_S = e_\alpha Q \ ,
\end{equation}
where $Q$ is the energy of the jet.  Making this identification, we will refer to the terms in \Eq{eq:radcusp} with $\log e_\alpha$ as soft logarithms and those with $\log e_\alpha^{1/\alpha}$ as collinear.\footnote{Note that the anomalous dimensions are singular at $\alpha  = 1$.  However, as shown explicitly in \Ref{broadening}, the cross section is continuous through $\alpha = 1$ and at $\alpha = 1$ the relevant divergences transform into ultraviolet and rapidity divergences.}  In this section, we will use the expression for the radiator in \Eq{eq:radcusp} because we are only working to one-loop order.  However, the expression \Eq{eq:radintcusp} is true to all orders, and so the interpolation obtained in this section could be tested at higher orders, given the cusp anomalous dimensions of the jet and soft functions at higher orders.

This expression in \Eq{eq:radcusp} is the one-loop component of the radiator $R(e_\alpha,e_\beta)$ on the boundary $e_\beta = e_\alpha^{\beta/\alpha}$.  To this accuracy, we are free to change the argument of the soft and collinear logarithms by an order-1 number near this boundary of the phase space.  The natural such number is $e_\beta^\alpha / e_\alpha^{\beta}$, which will enable the radiator to be continued into the bulk of the phase space, away from the boundary $e_\beta = e_\alpha^{\beta/\alpha}$.

When $e_\alpha = e_\beta$, the radiator must be a function of $e_\beta$ alone.  For example, starting from the soft logarithms at the $e_\beta = e_\alpha^{\beta/\alpha}$ boundary, this means that we must choose an exponent $c$ such that 
\begin{equation}
\left.\log e_\alpha \left( \frac{e_\beta^\alpha}{e_\alpha^\beta}  \right)^{c}\right|_{e_\alpha=e_\beta} = \log e_\beta,\frac{\log e_\beta}{\beta} \ .
\end{equation}
Note that soft logarithms on one boundary can mix and become soft or collinear logarithms on the other boundary of phase space (and similarly for collinear logarithms).  Therefore, there are four possible terms that we must consider:
\begin{align}
&\log e_\alpha \to \log e_\beta \ , \qquad \log e_\alpha \to \frac{\log e_\beta}{\beta} \ , \nonumber \\
&\frac{\log e_\alpha}{\alpha} \to \log e_\beta \ , \qquad \frac{\log e_\alpha}{\alpha} \to \frac{\log e_\beta}{\beta} \ ,
\end{align}
where the arrow indicates the interpolation from boundary $e_\beta=e_\alpha^{\beta/\alpha}$ to the boundary $e_\alpha = e_\beta.$  For example, consider the interpolation $\log e_\alpha \to \log e_\beta$.  We multiply the argument of the soft logarithm on the  $e_\beta=e_\alpha^{\beta/\alpha}$ boundary by 1 on that boundary and then continue to the other boundary:
\begin{equation}
\left.\log e_\alpha\left( \frac{e_\beta^\alpha}{e_\alpha^\beta}  \right)^{c}\right|_{e_\alpha=e_\beta} = \log e_\beta \ .
\end{equation}
The exponent $c$ that satisfies this equation is $c=0$.  The three other logarithmic interpolations can be determined similarly. 

With this prescription for scale setting, the one-loop radiator is
\begin{align}\label{eq:ddrad1}
R^{(1)}(e_\alpha,e_\beta) =&\ \frac{C_i}{2\pi \alpha_s \beta_0^2}\left[
 x\,U\left(  2\alpha_s \beta_0 \log e_\alpha\right)
 +\left(\frac{1}{\alpha-1}-x\right)U\left(  2\alpha_s \beta_0 \frac{\log e_\alpha^{\alpha-1}e_\beta^{\frac{\alpha}{\beta}(1-\beta)}}{\alpha-\beta}  \right)\right. \nonumber \\
&\qquad\qquad+\left. \left(-\frac{\alpha}{\alpha - 1}-y\right) U\left(  2\alpha_s \beta_0 \frac{\log e_\beta}{\beta} \right)+  
y\, U\left(  2\alpha_s \beta_0 \frac{\log e_\alpha^{1-\beta} e_\beta^{\alpha-1}}{\alpha-\beta}\right)
  \right] \ ,
\end{align}
for some constants $x,y$.  We have used the short-hand
\begin{equation}
U(z) = (1+z)\log (1+z) \ .
\end{equation}
When $e_\beta = e_\alpha^{\beta/\alpha}$, this reduces correctly to $R^{(1)}(e_\alpha)$, and when $e_\alpha = e_\beta$, only soft and collinear logarithms of $e_\beta$ are produced.

We now enforce the boundary conditions on $R^{(1)}(e_\alpha,e_\beta)$ to determine the constants $x$ and $y$.  When $e_\alpha = e_\beta$, \Eq{eq:ddrad1} becomes
\begin{align}
\left.R^{(1)}(e_\alpha,e_\beta)\right|_{e_\alpha = e_\beta} =&\ \frac{C_i}{2\pi \alpha_s \beta_0^2}\left[
\left( 
x+y
\right)U(2\alpha_s\beta_0 \log e_\beta) \right.\nonumber \\
&\qquad\qquad\left.+ \left(
-1-x-y
\right)U\left( 2\alpha_s\beta_0 \frac{\log e_\beta}{\beta} \right)
\right] \ .
\end{align}
For this to reproduce $R^{(1)}(e_\beta)$, we must have
\begin{equation}
x+y = \frac{1}{\beta - 1} \ .
\end{equation}
To fix the remaining coefficient, we consider the derivative boundary conditions.  Taking the derivative of the radiator with respect to $e_\alpha$ and evaluating it on the boundary $e_\beta = e_\alpha$ it must vanish:
\begin{align}
\left.  \frac{\partial}{\partial e_\alpha }R^{(1)}(e_\alpha,e_\beta)\right|_{e_\beta= e_\alpha} &= 0\nonumber \\
&=\frac{C_i}{\pi e_\alpha} \left[
\left(  
x+\frac{1-\beta}{\alpha-\beta}y
\right)U'(2\alpha_s \beta_0 \log e_\beta) \right. \nonumber \\
&\qquad\qquad\left.
+\left(
\frac{1}{\alpha-1}-x
\right)U'\left(2\alpha_s\beta_0 \frac{\log e_\beta}{\beta}\right)
\right] \ ,
\end{align}
which then requires
\begin{equation}
x+\frac{1-\beta}{\alpha-\beta}y =0 \ ,\qquad \frac{1}{\alpha-1}-x=0 \ .
\end{equation}
The other derivative boundary conditions produce the same constraints on $x$ and $y$.  It then follows that 
\begin{equation}
x=\frac{1}{\alpha-1} \ ,\qquad y = \frac{\alpha-\beta}{(\alpha-1)(\beta-1)} \ ,
\end{equation}
and so the radiator function at one-loop is
\begin{align}\label{eq:rad1loop}
R^{(1)}(e_\alpha,e_\beta) =&\ \frac{C_i}{2\pi \alpha_s \beta_0^2}\left[
 \frac{1}{\alpha-1}U\left(  2\alpha_s \beta_0 \log e_\alpha\right)-\frac{\beta}{\beta-1} U\left(  2\alpha_s \beta_0 \frac{\log e_\beta}{\beta} \right)
 \right. \nonumber \\
&\qquad\qquad+\left.
\frac{\alpha-\beta}{(\alpha-1)(\beta-1)} U\left(  2\alpha_s \beta_0 \frac{\log e_\alpha^{1-\beta} e_\beta^{\alpha-1}}{\alpha-\beta}\right)
  \right] \ ,
\end{align}
which satisfies all boundary conditions.

The form of this expression is interesting and we will discuss it in more detail in \Sec{sec:mixing}.  For the radiator of a single angularity, there were only two logarithmic structures corresponding to soft or collinear logarithms.  However, the interpolating radiator for two angularities has three logarithmic structures: soft ($\log e_\alpha$), collinear ($\log e_\beta^{1/\beta}$) and what we will call ``$k_T$'' logarithms:
\begin{equation}
k_T\text{ logarithms} = \log e_\alpha^\frac{1-\beta}{\alpha-\beta} e_\beta^\frac{\alpha-1}{\alpha-\beta} \ .
\end{equation}
We use the term $k_T$ because this combination of $e_\alpha$ and $e_\beta$ reduces to $k_T = z\theta/R_0$ for one emission:
\begin{equation}
e_\alpha^\frac{1-\beta}{\alpha-\beta} e_\beta^\frac{\alpha-1}{\alpha-\beta} = \left(z\frac{\theta^\alpha}{R_0^\alpha}  \right)^\frac{1-\beta}{\alpha-\beta}\left( z\frac{\theta^\beta}{R_0^\beta} \right)^\frac{\alpha-1}{\alpha-\beta} = z\frac{\theta}{R_0} \ .
\end{equation} 
Near the boundaries of the phase space the $k_T$ logarithms appropriately reduce to either soft or collinear logarithms.
This would seem to suggest that to fully describe the bulk of the phase space requires introducing an additional mode into the effective theory.  However, because there are only two types of singularities in QCD, we do not know how this would be done.  The existence of a possible meta-effective theory that is well-defined over the entire phase space would be intriguing and deserves further study.

\subsubsection{Non-Cusp Interpolation}

As a second example of interpolation from the boundary into the bulk of the phase space, we will study the non-cusp piece, $T(e_\alpha,e_\beta)$.  To NLL accuracy, the non-cusp piece for a single angularity $e_\beta$ is
\begin{equation}
T(e_\beta) = \frac{1}{\pi \beta_0}\log\left( 1+2\alpha_s \beta_0 \frac{\log e_\beta}{\beta} \right)  \ .
\end{equation}
This expression itself satisfies the non-derivative boundary conditions on $T(e_\alpha,e_\beta)$ when continued to the boundary where $e_\beta = e_\alpha^{\beta/\alpha}$.  Then, we have
\begin{equation}
T(e_\alpha,e_\beta) = \frac{1}{\pi \beta_0}\log\left( 1+2\alpha_s \beta_0 \frac{\log e_\beta}{\beta} \right)  \ .
\end{equation}
Nevertheless, this expression does not satisfy the derivative boundary conditions.  For example, the derivative with respect to $e_\alpha$ vanishes, which satisfies the boundary condition when $e_\beta = e_\alpha$.  However, it clearly does not reproduce the correct term when $e_\beta = e_\alpha^{\beta/\alpha}$ so as to reproduce the differential cross section of $e_\alpha$.  Other terms will need to be added to $T(e_\alpha,e_\beta)$ to accomplish this.

The terms that must be added cannot spoil the logarithmic accuracy of $T(e_\alpha,e_\beta)$ and must produce the correct single logarithmic expressions when differentiated.  Therefore, we must add a term to $T(e_\alpha,e_\beta)$ that is power suppressed, but when differentiated produces singular terms.  This was anticipated in \Sec{sec:fo} where it was observed that na\"ively power-suppressed terms in the cumulative cross section were necessary to reproduce the correct single logarithms of the differential cross section.  Motivated by the expressions there, we add to $T(e_\alpha,e_\beta)$ a term that is suppressed by powers of $e_\alpha$ and $e_\beta$:
\begin{equation}
T(e_\alpha,e_\beta) =  \frac{1}{\pi \beta_0}\log\left( 1+2\alpha_s \beta_0 \frac{\log e_\beta}{\beta} \right)+2\frac{\alpha_s}{\pi}c\frac{e_\alpha^{a}e_\beta^{b}}{\beta+2\alpha_s\beta_0 \log e_\beta} \ ,
\end{equation}
for exponents $a,b$ and coefficient $c$.  For the added term to be truly power suppressed in the small $e_\alpha,e_\beta$ limit, we require $a+b>0$.  The derivative boundary conditions will constrain $a,b,c$ further.

Taking the derivative with respect to $e_\alpha$, we find
\begin{equation}
\frac{\partial}{\partial e_\alpha}T(e_\alpha,e_\beta) = 2\frac{\alpha_s}{\pi}ac\frac{e_\alpha^{a-1}e_\beta^{b}}{\beta+2\alpha_s\beta_0 \log e_\beta} \ .
\end{equation}
When $e_\alpha = e_\beta$ this must vanish.  Clearly, this is only possible if either $a$ or $c$ is zero; therefore, we only require this term to be power suppressed or beyond NLL accuracy.  For it to be power suppressed when $e_\beta=e_\alpha$ requires $a+b-1>-1$, which is the same constraint as being power-suppressed in $T(e_\alpha,e_\beta)$ itself.  When $e_\beta=e_\alpha^{\beta/\alpha}$, it must reproduce the derivative of $T(e_\alpha)$:
\begin{align}
\left.\frac{\partial}{\partial e_\alpha}T(e_\alpha,e_\beta)\right|_{e_\beta = e_\alpha^{\beta / \alpha}} &= \frac{\partial}{\partial e_\alpha}T(e_\alpha) =\frac{2}{e_\alpha}\frac{\alpha_s}{\pi} \frac{1}{\alpha+2\alpha_s\beta_0 \log e_\alpha} \nonumber \\
&= 2\frac{\alpha_s}{\pi}\frac{\alpha}{\beta}ac\frac{e_\alpha^{a-1+\frac{\beta}{\alpha}b}}{\alpha+2\alpha_s\beta_0 \log e_\alpha} \ .
\end{align}
This then requires
\begin{equation}
\frac{\alpha}{\beta}ac=1 \ ,\qquad a-1+\frac{\beta}{\alpha}b = -1 \ .
\end{equation}
There are no constraints beyond these from taking the derivative with respect to $e_\beta$.

The constraints on $T(e_\alpha,e_\beta)$ do not fully specify the parameters $a,b,c$ so we must impose an additional, arbitrary condition.  This should be interpreted as an uncertainty in the calculation that can be formally corrected by matching to the fixed-order cross section.  Note that in the cumulative distribution these power-suppressed terms are beyond NLL accuracy anyway, so are only required to satisfy the boundary conditions and not to obtain formal NLL accuracy.  Here, we will fix the parameters by considering
\begin{equation}
a+b=1 \ ,
\end{equation}
but any positive value for the sum of $a$ and $b$ would work.  One could also consider adding several power-suppressed terms to satisfy the boundary conditions.  With our choice on the sum of $a$ and $b$, the non-cusp piece becomes
\begin{equation}
T(e_\alpha,e_\beta) = \frac{1}{\pi \beta_0} \log\left( 1+2\alpha_s \beta_0 \frac{\log e_\beta}{\beta} \right) 
-2\frac{\alpha_s}{\pi} \frac{\alpha-\beta}{\alpha} \frac{e_\alpha^{-\frac{\beta}{\alpha-\beta}}e_\beta^{\frac{\alpha}{\alpha-\beta}}}{\beta+2\alpha_s \beta_0 \log e_\beta} \ ,
\end{equation}
which satisfies all of the boundary conditions to leading power at NLL accuracy.  Using the procedures developed here, all other pieces of the double cumulative cross section can be determined that satisfy the boundary conditions.  We present the full expression in \App{app:two}.

\subsection{Mixing Structure of Collinear and Soft Logarithms }
\label{sec:mixing}

We now discuss in more detail the mixing of the collinear and soft logarithms found in the radiator interpolation. When performing the interpolation we allowed for the mixing of the soft and collinear logarithms on one boundary into either of the soft or collinear logarithms on the other boundary. This allows for the presence of four possible logarithmic structures in the radiator for the double cumulative distribution. However, the consistency conditions at the phase space boundaries enforced that only three appear: the soft, collinear and $k_T$ logarithms, discussed in the previous sections. The fourth possible logarithmic structure arising from the interpolation between a soft logarithm on the $e_\beta=e_\alpha^{\beta / \alpha}$ boundary and a collinear logarithm on the $e_\alpha=e_\beta$ boundary, which has the form
\begin{equation}
\log e_\alpha^{\frac{\alpha-1}{\alpha-\beta}}e_\beta^{\frac{\alpha}{\beta}\left (\frac{1-\beta}{\alpha-\beta} \right )} \ ,
\end{equation}
does not appear. For a single emission, this combination of $e_\alpha$ and $e_\beta$ corresponds to
\begin{equation}\label{eq:badlog}
e_\alpha^{\frac{\alpha-1}{\alpha-\beta}}e_\beta^{\frac{\alpha}{\beta}\left (\frac{1-\beta}{\alpha-\beta} \right )}=\left(   z\frac{\theta^\alpha}{R_0^\alpha}     \right )^{\frac{\alpha-1}{\alpha-\beta}}         \left (      z\frac{\theta^\beta}{R_0^\beta}  \right )^{\frac{\alpha}{\beta}\left (\frac{1-\beta}{\alpha-\beta} \right )}=z^{\frac{1}{\beta}}\frac{\theta^\alpha}{R_0^\alpha}
\end{equation}
which combines collinear energy scaling and soft angle scaling. The form of the factorization theorems on the boundary guarantees that such a structure does not appear in the bulk of the phase space.

The non-appearance of this logarithmic structure implies that as we transition from one boundary to the other, either the collinear or soft logarithms (but not both) split to become a sum of collinear and soft logarithms on the other boundary. In particular, consider the transition from the $e_\beta=e_\alpha^{\beta / \alpha}$ boundary through the bulk to the $e_\alpha=e_\beta$ boundary. From \Eq{eq:rad1loop} for the one loop radiator, we see that the logarithms which reduce to soft logarithms near the $e_\beta=e_\alpha^{\beta / \alpha}$ boundary remain as soft logarithms in the bulk, and on the $e_\alpha=e_\beta$ boundary. However, the collinear logarithms near the $e_\beta=e_\alpha^{\beta / \alpha}$ boundary split into a sum of collinear and $k_T$ logarithms in the bulk, and the $k_T$ logarithms then reduce to soft logarithms near the $e_\alpha=e_\beta$ boundary. 

\begin{figure}
\begin{center}
\includegraphics[width=8cm]{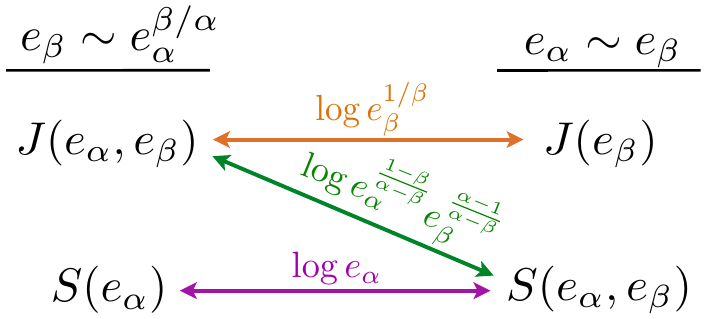}
\end{center}
\caption{
Illustration of the interpolation of the logarithmic structure between the boundaries of the phase space.  Collinear logarithms ($\log e_\beta^{1/\beta}$) always interpolate between the jet functions defined on the boundaries and soft logarithms ($\log e_\alpha$) always interpolate between the soft functions.  $k_T$ logarithms ($\log e_\alpha^\frac{1-\beta}{\alpha-\beta} e_\beta^\frac{\alpha-1}{\alpha-\beta}$) interpolate between the double differential jet and soft functions.
}
\label{fig:ddiff_flow}
\end{figure}

This can also be phrased in terms of the mixing of the anomalous dimensions of the jet and soft functions appearing in the factorization theorems on the different boundaries. Recall from \Eq{eq:radintcusp} that the one loop radiator for a single angularity can be written in terms of integrals of the jet and soft function cusp anomalous dimensions:
\begin{equation}
R^{(1)}(e_\alpha)=-2K_J(\mu_J)-2K_S(\mu_S)
\end{equation}
with
\begin{equation}
K_{J,S}(\mu_{J,S})=\int_{\alpha_s(\mu_{J,S})}^{\alpha_s(\mu)}\frac{d\alpha'}{\beta[\alpha']}\Gamma_{J,S}[\alpha']\int_{\alpha_s(\mu_{J,S})}^{\alpha'} \frac{d\alpha''}{\beta[\alpha'']}
\end{equation}
The mixing of the logarithms described in the previous paragraph is equivalent to the following mixing of the jet and soft function anomalous dimensions:
\begin{align}
&K_S^\beta(\mu_{S,\beta}) = z K_J^\alpha (\mu_{J\rightarrow S}) +K_S^\alpha (\mu_{S\rightarrow S}) |_{e_\alpha ~e_\beta}\\
&K_J^\beta(\mu_{J,\beta})=(1-z)K_J^\alpha (\mu_{J\rightarrow J})|_{e_\alpha ~e_\beta}
\end{align}
In this expression, $K_S^\beta(\mu_{S,\beta}), K_J^\beta(\mu_{J,\beta})$ are the integrals of the cusp anomalous dimensions for the factorization theorem near the $e_\alpha=e_\beta$ boundary, evaluated at their canonical scales, and the $K^\alpha$ are the similar integrals of the anomalous dimensions for the factorization theorem near the $e_\beta=e_\alpha^{\beta / \alpha}$ boundary, but evaluated at the appropriately modified scales, and $z=\frac{\beta-\alpha}{\alpha(\beta-1)}$. From this expression, we can see that as we move from the region where $e_\beta\sim e_\alpha^{\beta / \alpha}$ to $e_\alpha\sim e_\beta$, the anomalous dimension of the jet function near the $e_\beta=e_\alpha^{\beta / \alpha}$ boundary splits into two pieces, one of which contributes to the anomalous dimension of the soft function near the $e_\alpha=e_\beta$ boundary and the other to the anomalous dimension of the jet function. This mixing structure is illustrated in \Fig{fig:ddiff_flow}. \Fig{fig:ddiff_flow} also makes it transparent how the form of the factorization theorems on the two boundaries dictates that there can only be mixing between the jet functions near the $e_\beta=e_\alpha^{\beta / \alpha}$ boundary and the soft functions near the $e_\alpha=e_\beta$ boundary, as only these depend on both angularities. This clarifies why the only new logarithms in the bulk are the $k_T$ logarithms, and not the logarithms of \Eq{eq:badlog}.

To summarize, the structure of the radiator logarithms found via the interpolation procedure gives rise to a single new logarithmic structure in the bulk of the phase space, the $k_T$ logarithms. Unlike the soft $( \log e_\alpha)$ and collinear $(\log e_\beta^\frac{1}{\beta})$ logarithms, which reduce to, respectively, the soft and collinear logarithms of the two different factorization theorems near the boundaries, the $k_T$ logarithms reduce to the soft logarithms near the $e_\alpha=e_\beta$  boundary and the collinear logarithms near the $e_\beta=e_\alpha^{\beta / \alpha}$ boundary. This further clarifies the impediment to writing down a factorization theorem valid in the entire bulk region.

\subsection{Evidence for Uniqueness of Interpolation}\label{sec:uniint}

While the interpolation between the boundary regions presented above satisfies all constraints on the cross section from \Eqs{eq:cumconsts}{eq:diffconsts}, there is no guarantee that this interpolation is in any way unique.  If this is the case, then there is no sense in which the interpolation captures the logarithms to any formal accuracy in the bulk of the phase space and so matching the resummed double differential cross section to the fixed-order cross section would be meaningless.  However, making some reasonable assumptions about the structure of the logarithms in the bulk of the phase space, we will argue that the boundary conditions on the cumulative cross section are sufficiently strong to enforce the uniqueness of the interpolation up to ${\cal O}(\alpha_s^4)$.

To prove this, we assume that the logarithms in the bulk of the phase space exponentiate.  Then, the true double cumulative cross section to logarithmic accuracy can be written as
\begin{equation}\label{eq:cumintvio}
\log \Sigma(e_\alpha,e_\beta) = \log \Sigma_\text{int}(e_\alpha,e_\beta)+\sum_{n=4}^\infty  f_n\left( \log e_\alpha,\log e_\beta  \right)  \sum_{i=2}^{n-2} c_{n i} \log^i\frac{e_\alpha}{e_\beta}\log^{n-i} \frac{e_\beta^\alpha}{e_\alpha^\beta} \ ,
\end{equation}
where $\Sigma_\text{int}(e_\alpha,e_\beta)$ is the interpolation cross section that satisfies all of the boundary conditions.  To NLL accuracy, the function $f_n$ is
\begin{equation}\label{eq:ffunc}
f_n\left( \log e_\alpha,\log e_\beta  \right) = \sum_{m=0}^\infty \sum_{j=0}^m \left(d^{1n}_{mj}\alpha_s^{n+m-1}+ d^{2n}_{mj}\alpha_s^{n+m}\right)\log^{j} e_\alpha \log^{m-j} e_\beta \ ,
\end{equation}
where $d^{1n}_{mj}$ and $d^{2n}_{mj}$ are coefficients,  independent of $\alpha_s$.  We assume that the logarithms in $f_n$ cannot be rewritten in such a way that factors of $$\log \frac{e_\alpha}{e_\beta}\ , \qquad \log \frac{e_\beta^\alpha}{e_\alpha^\beta}$$exist.  That is, all dependence on these logarithms has been explicitly factored out in \Eq{eq:cumintvio}.

Because we assume that the interpolation cross section $\Sigma_\text{int}(e_\alpha,e_\beta)$ satisfies all boundary conditions, the second term, corresponding to interpolation-violating contributions, must vanish when either $e_\alpha=e_\beta$ or $e_\alpha^\beta=e_\beta^\alpha$ so that the double cumulative cross section reduces appropriately at the boundaries.  In addition, the derivatives of this term must also vanish when evaluated at the boundaries of the phase space to correctly reproduce the single differential cross sections.  These boundary conditions are automatically satisfied for the the sum over $n$ in \Eq{eq:cumintvio} to start at $n=4$ and for the sum over $i$ to range from $i=2$ to $i=n-2$.
From \Eq{eq:ffunc}, this shows that the lowest order at which interpolation-violating terms can arise is $\alpha_s^3$ in the exponent.  However, the only possible interpolation-violating (IV) contribution at ${\cal O}(\alpha_s^3)$ is leading logarithmic, which has the form
\begin{equation}
\log \Sigma_\text{IV}^{(3)}(e_\alpha,e_\beta) \sim \alpha_s^3 \log^2\frac{e_\alpha}{e_\beta}\log^{2} \frac{e_\beta^\alpha}{e_\alpha^\beta} \ ,
\end{equation}
 which should be fully captured by one-loop running of $\alpha_s$.  If this expectation is true, then the lowest order at which interpolation-violating terms can arise is $\alpha_s^4$ in the exponent of the cumulative distribution.

If exponentiation of the logarithms in the double cumulative cross section does not occur, then the lowest order at which interpolation-violating logarithms could arise is ${\cal O}(\alpha_s^2)$, with a term of the form
\begin{equation}
 \Sigma_\text{IV}^{(2)}(e_\alpha,e_\beta) \sim \alpha_s^2 \log^2\frac{e_\alpha}{e_\beta}\log^{2} \frac{e_\beta^\alpha}{e_\alpha^\beta} \ .
\end{equation}
The existence of these logarithms in the double cumulative cross section could be checked explicitly.  However, it seems very unlikely that such a term could exist because it is double logarithmic and so should be totally captured by the resummation of the double cumulative cross section presented in \Ref{Larkoski:2013paa}.  While the lack of existence of this term would not necessarily prove exponentiation, it would demonstrate that the interpolation for the double differential cross section is significantly robust.

Therefore, assuming exponentiation of logarithms in the double cumulative cross section and one-loop running capturing all leading logarithms, the lowest order at which the interpolation-violating contributions can exist is ${\cal O}(\alpha_s^4)$ in the exponent.  
If exponentiation of the logarithms of the double cumulative cross section does not occur, then the bulk of the phase space would only be described at fixed-order.  
Thus, this is strong evidence that, at least to NLL accuracy, the interpolation captures the dominant logarithmic structure of the double cumulative cross section.  We further conjecture that the interpolation presented in \Sec{sec:nllint} correctly resums all logarithms to the accuracy of the single cumulative cross sections at the boundaries.  Testing this requires at least an ${\cal O}(\alpha_s^3)$ calculation, which is well beyond the state-of-the-art for fixed-order distributions of jet observables.

\section{Comparison to Monte Carlo}\label{sec:mc}

With analytic results for the double differential cross section to NLL accuracy as defined by interpolation between the boundaries of phase space, we present a numerical analysis and compare to Monte Carlo simulation.  Because we are interested in comparing the logarithmic structure of the analytic and Monte Carlo double differential cross section of angularities $e_\alpha$ and $e_\beta$, we will plot it in the plane $\left( \log e_\beta,\log e_\alpha \right)$, with $\alpha > \beta$.  In this plane, the upper and lower boundaries of phase space are straight lines with slopes equal to 1 and $\alpha/\beta$, respectively.  We will plot the double differential cross section weighted by the two angularities:
\begin{equation}
e_\alpha e_\beta \frac{d^2\sigma}{de_\alpha \, de_\beta} = e_\alpha e_\beta \frac{\partial^2}{\partial e_\alpha \, \partial e_\beta} \Sigma(e_\alpha, e_\beta) \ .
\end{equation}
The Sudakov double logarithms manifest themselves as a concave down paraboloid in the $\left( \log e_\beta,\log e_\alpha \right)$ plane.

We generate $e^+e^- \to q \bar{q}$ events simulated with \pythia{8.165} \cite{Sjostrand:2006za,Sjostrand:2007gs} at a center-of-mass energy of 1 TeV with hadronization turned off, two-loop running of $\alpha_s$, and $\alpha_s(m_Z) = 0.118$.\footnote{The quarks that are produced are only $u$, $d$, or $s$, so mass effects should be minimal.}  To analyze the jets, we cluster jets with the $e^+e^-$ anti-$k_T$ algorithm \cite{Cacciari:2008gp} with \fastjet{3.0.3} \cite{Cacciari:2011ma} with a fat jet radius $R_0 = 1.5$.  We analyze only the hardest jet in the event, requiring that the cosine of the angle between the jet momentum axis and the initial hard parton be greater than $0.9$.  We only include particles that lie within an angle $R_0=0.4$ from the broadening axis of the hardest jet.  The energy of the jets is required to be in the range of $Q\in [450,550]$ GeV.  We then measure the recoil-free angularities for various values of the angular exponents $\alpha$ and $\beta$ of the jets in the sample.

\begin{figure}[]
\begin{center}
\subfloat[]{\label{fig:anal_2015}
\includegraphics[width=7.0cm]{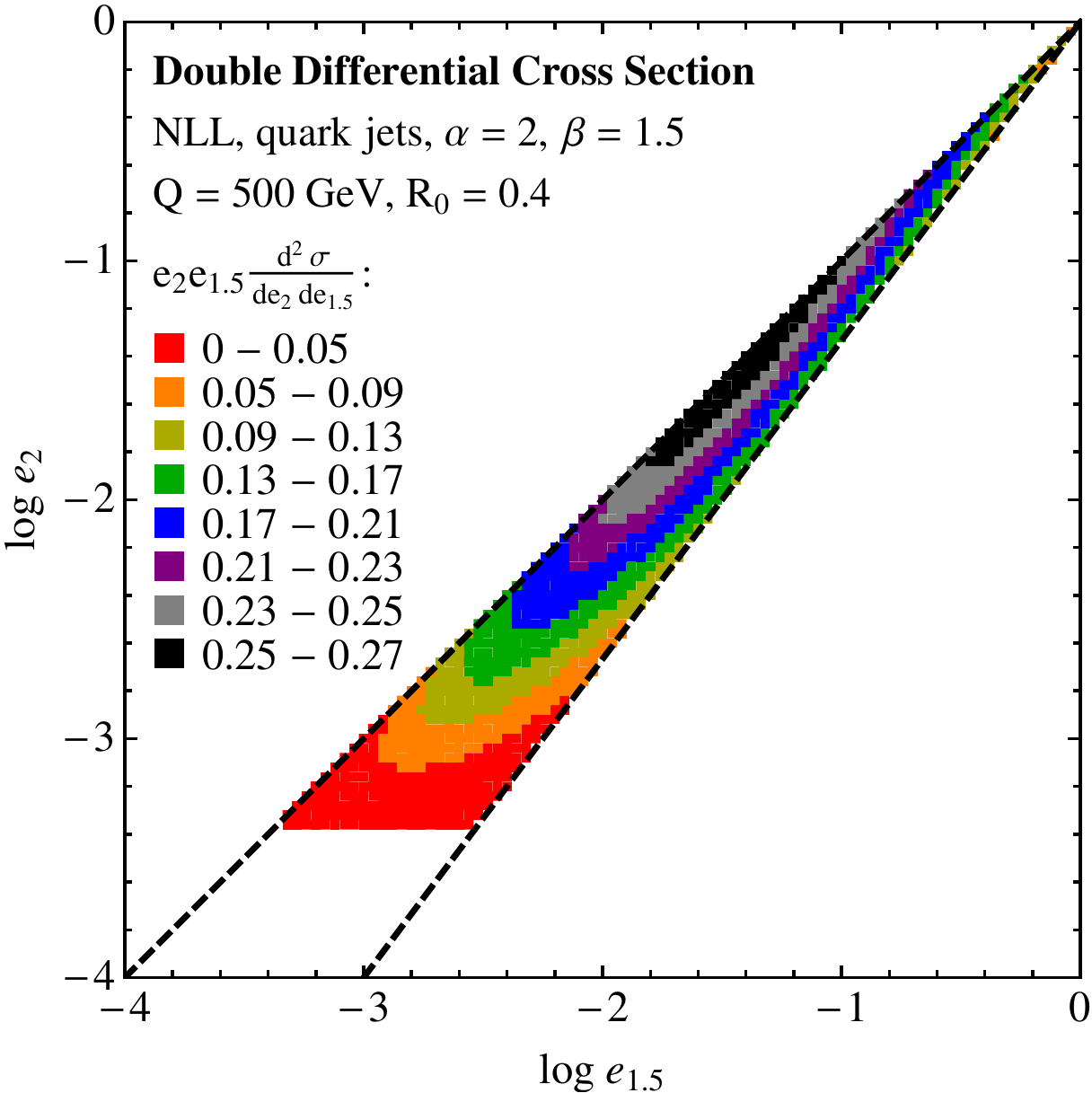}
}
$\quad$
\subfloat[]{\label{fig:anal_2010} 
\includegraphics[width=7.0cm]{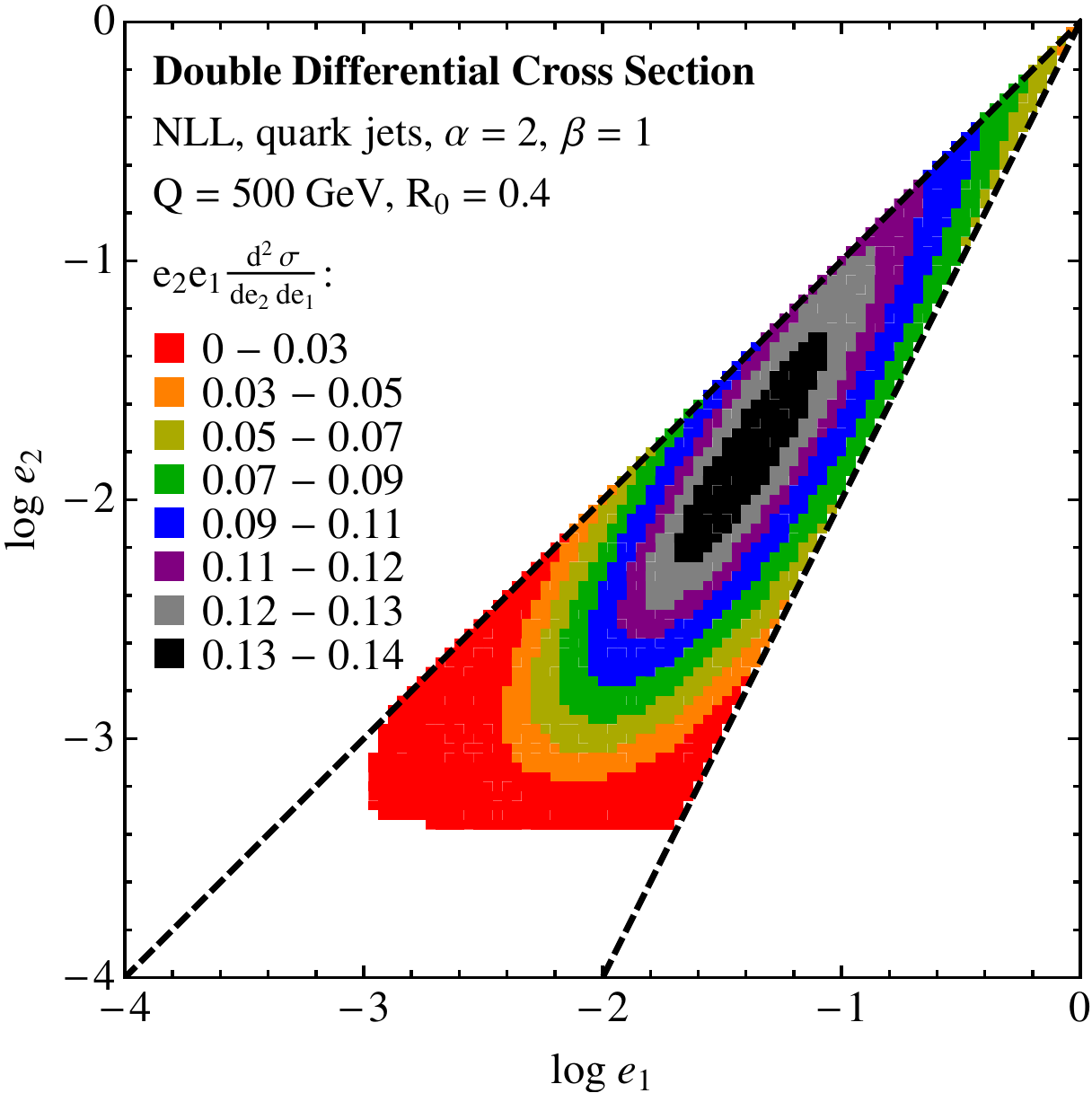}
}\\
\subfloat[]{\label{fig:anal_2005} 
\includegraphics[width=7.0cm]{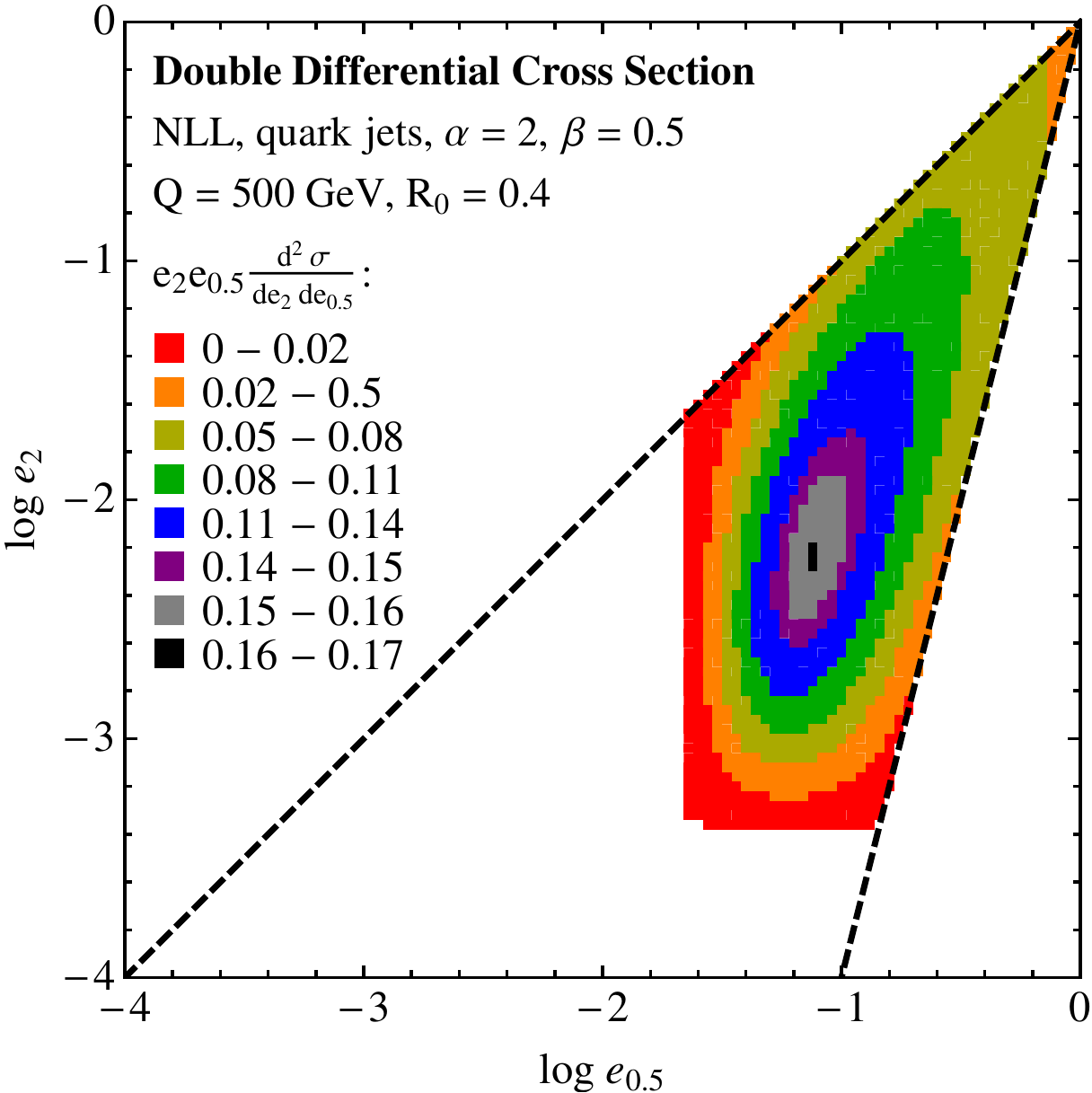}
}
$\quad$
\subfloat[]{\label{fig:anal_2002} 
\includegraphics[width=7.0cm]{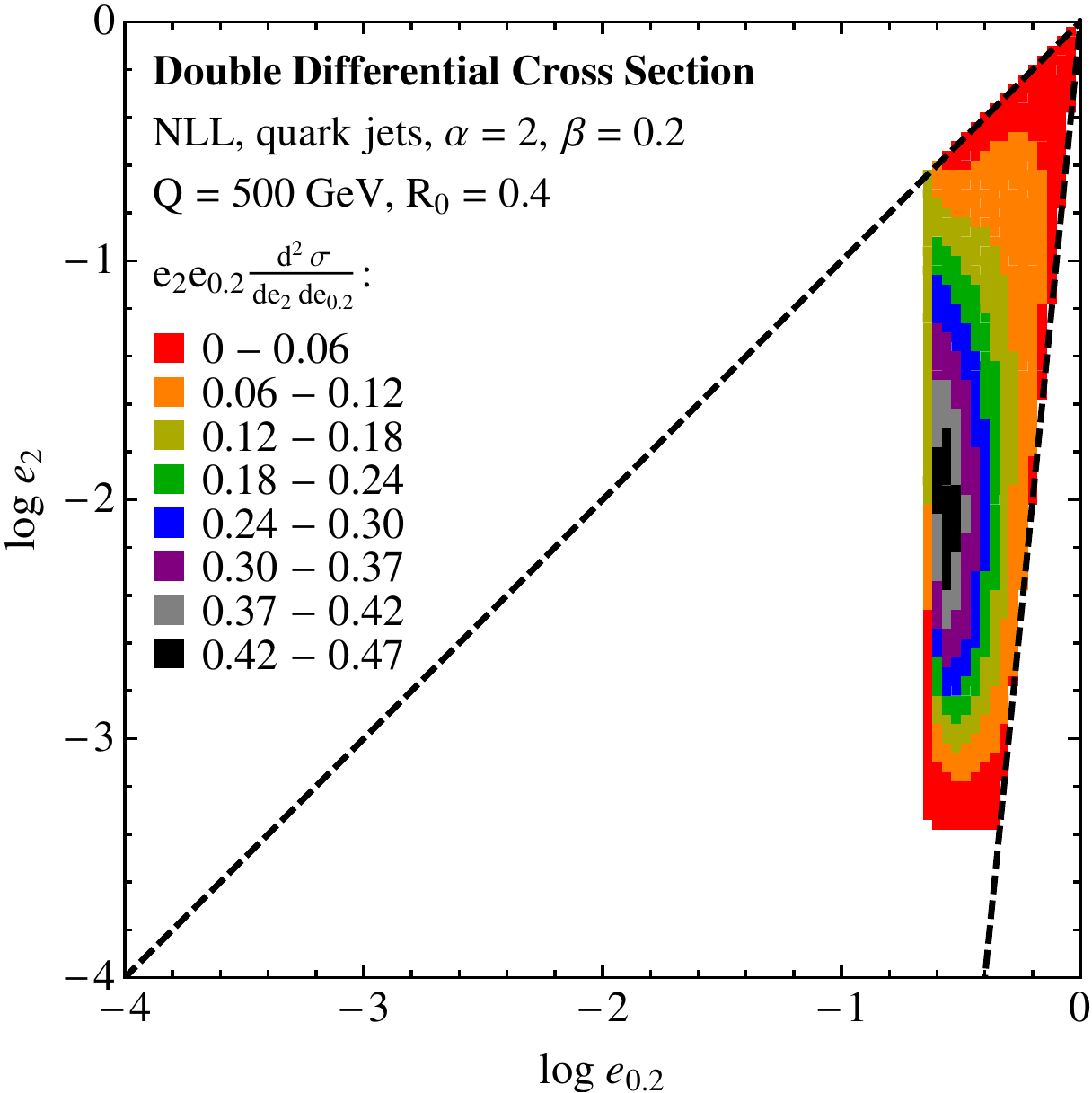}
}
\end{center}
\caption{
Plots of the double differential cross section defined from the analytic NLL interpolation of \Sec{sec:int} measured on quark jets with one angularity fixed to be thrust ($\alpha=2$) and scanning over the other angularity: $\beta = 1.5,1,0.5,0.2$.  The energy of the jets is $Q=500$ GeV and the jet radius is $R_0=0.4$.  The dashed lines on the plot correspond to the expected phase space boundary.
}
\label{fig:anal_BAng2_DD_plot}
\end{figure}

In \Figs{fig:anal_BAng2_DD_plot}{fig:PY_BAng2_DD_plot} we plot the distributions, fixing one angularity to be thrust, $e_2$, and scanning over the other angularity: $\beta = 1.5,1,0.5,0.2$.  In the NLL interpolation plots, \Fig{fig:anal_BAng2_DD_plot}, the double differential distribution has been set to zero at very small values corresponding to scales near the Landau pole of $\alpha_s$.  While the scale of the contours in the corresponding NLL interpolation and \pythia{} plots differ by up to a factor of 2, there is good qualitative agreement between the distributions.  Both exhibit a peak in the distribution in the bulk of phase space at approximately the same location.  This suggests that the correlations between angularities with different angular exponents are well-modeled in Monte Carlo.

\begin{figure}[]
\begin{center}
\subfloat[]{\label{fig:PY_2015}
\includegraphics[width=7.0cm]{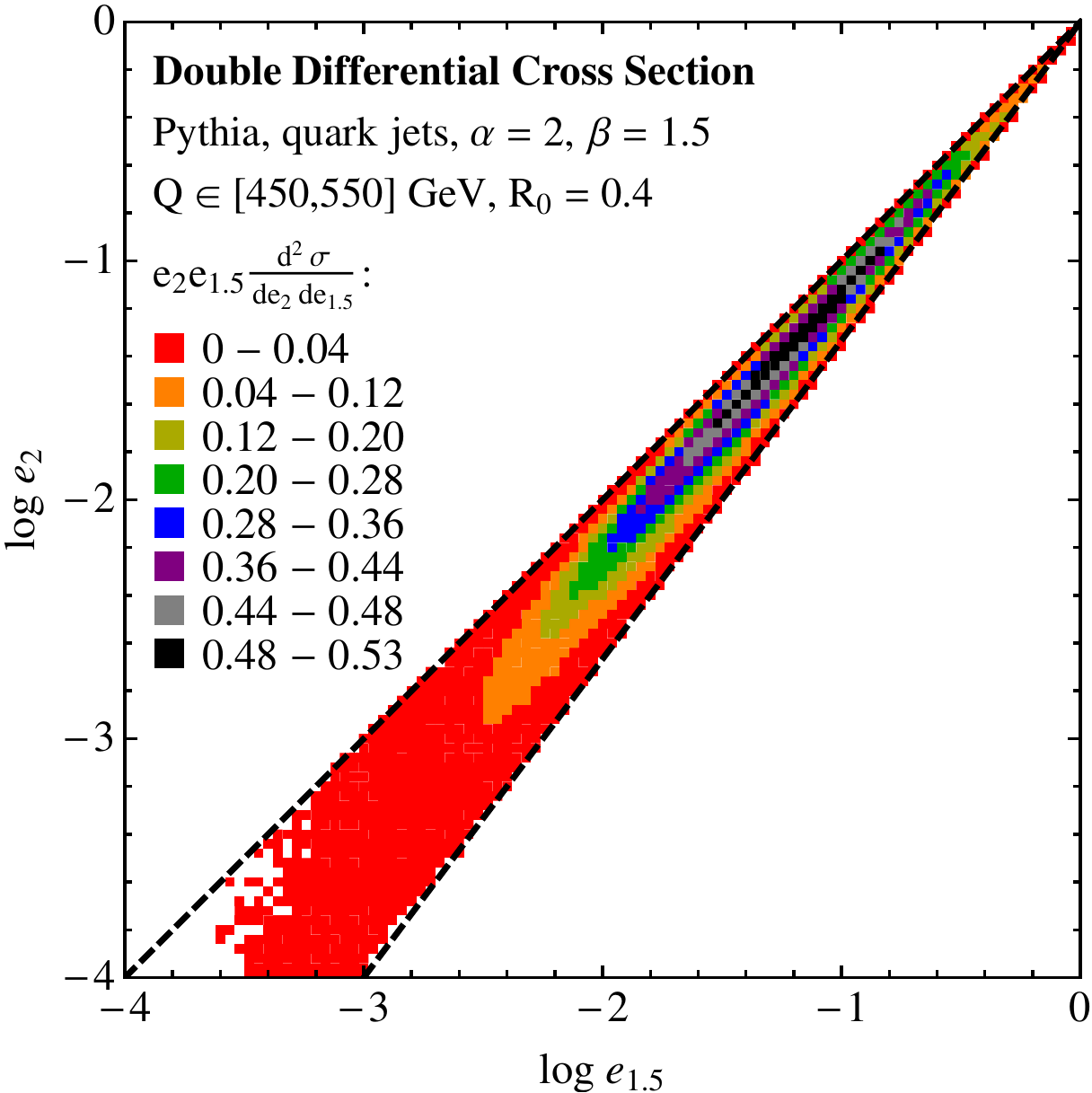}
}
$\quad$
\subfloat[]{\label{fig:PY_2010} 
\includegraphics[width=7.0cm]{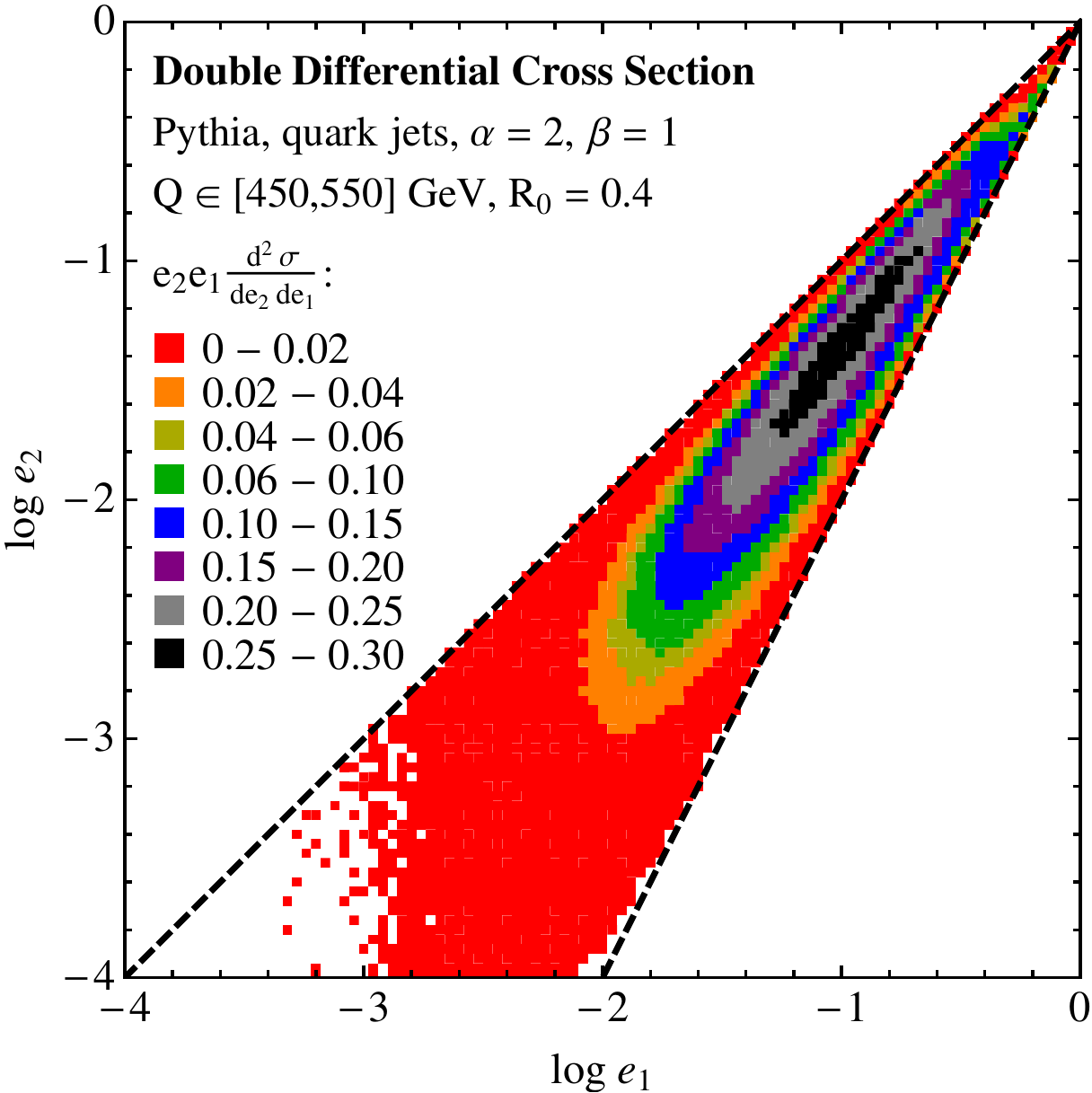}
}\\
\subfloat[]{\label{fig:PY_2005} 
\includegraphics[width=7.0cm]{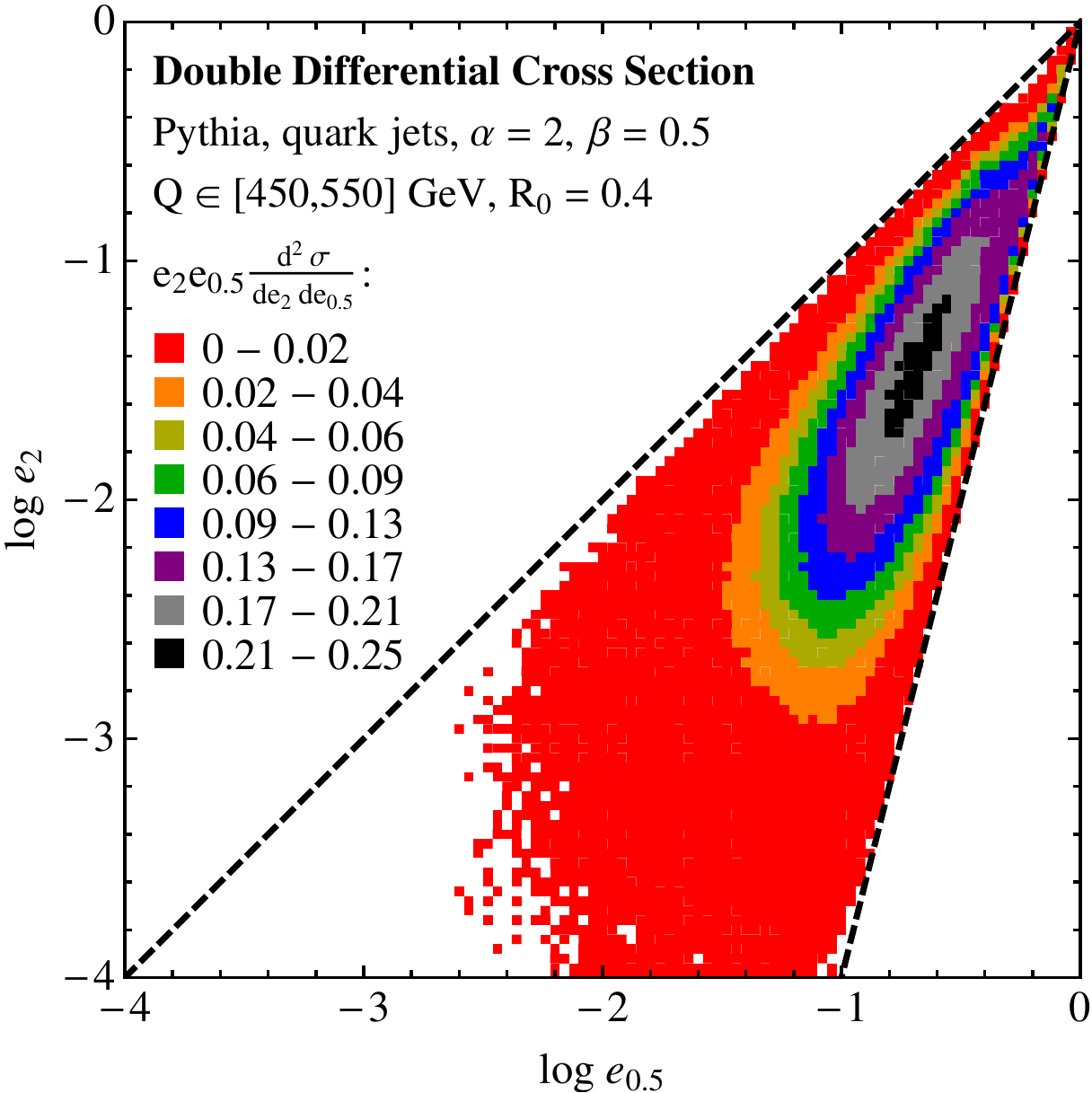}
}
$\quad$
\subfloat[]{\label{fig:PY_2002} 
\includegraphics[width=7.0cm]{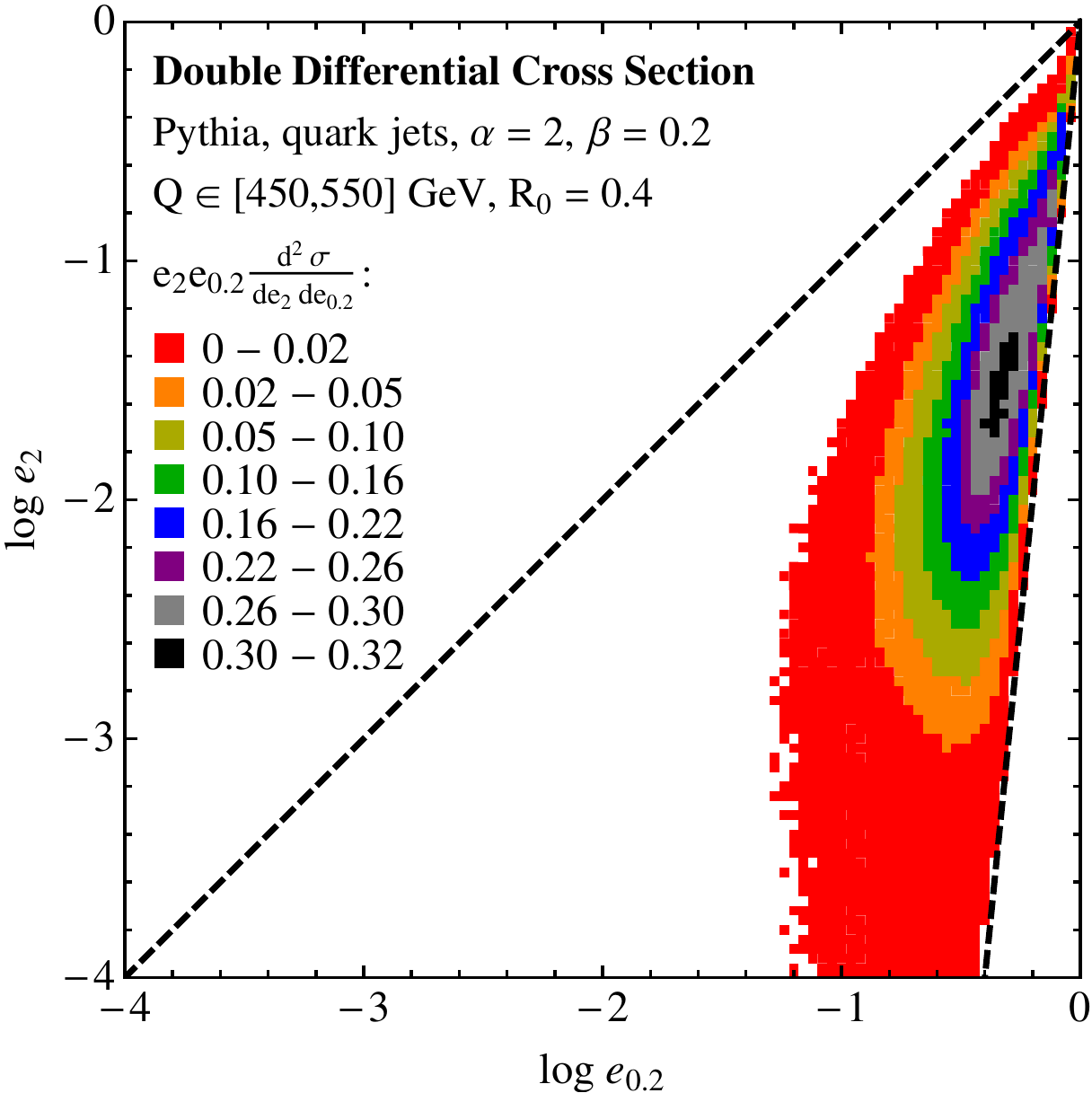}
}
\end{center}
\caption{
Plots of the double differential cross section from \pythia{} with one angularity fixed to be thrust ($\alpha=2$) and scanning over the other angularity: $\beta = 1.5,1,0.5,0.2$.  The energy of the jets is $Q\in[450,550]$ GeV and the jet radius is $R_0=0.4$.  Hadronization has been turned off.  The dashed lines on the plot correspond to the expected phase space boundary.
}
\label{fig:PY_BAng2_DD_plot}
\end{figure}

\begin{table}
\begin{center}
\begin{tabular}{c||ccc||ccc}
 & $\log e_2$ & $\log e_{1.5}$ & Peak & $\log e_2$ & $\log e_{1}$ & Peak \\
 \hline \hline
NLL Int. &  $-1.36$ & $-1.36$ & $0.27$ & $ -1.84$ & $-1.40$ & $0.14$  \\
\pythia{} & $-1.20$ & $-1.08$ & $0.53$ & $-1.24$  & $-0.92$ & $\ 0.28$  
\vspace{1em} \\
 & $\log e_2$ & $\log e_{0.5}$ & Peak & $\log e_2$ & $\log e_{0.2}$ & Peak \\
 \hline \hline
NLL Int. &  $-2.24$ & $-1.12$ & $0.16$ &  $-2.04$ & $-0.56$ & $0.46$  \\
\pythia{} & $-1.48$ & $-0.68$ & $0.22$ & $-1.56$ & $-0.36$ & $0.32$  \\
\end{tabular}
\end{center}

\caption{
Table comparing location and height of peaks of the double differential cross sections from the analytic NLL interpolation and \pythia{}.
}
\label{tab:peaks}
\end{table}

The comparison between the analytic and Monte Carlo results can be made more quantitative by comparing the location and height of the peak of the distribution.  In \Tab{tab:peaks}, we list the location and height (``Peak'') of the peak in $(\log e_\alpha,\log e_\beta)$ space for $\alpha = 2$, $\beta = 1.5,1,0.5,0.2$.  There are several features that illustrate qualitative agreement including:
\begin{itemize}
\item The location of the peak in $\log e_2$ generally becomes more negative as $\beta$ decreases.

\item The location of the peak in $\log e_\beta$ moves to less negative values as $\beta$ decreases.

\item The height of the peak is relatively large for $\beta$ near $\alpha=2$ and $\beta$ near 0 and smaller for intermediate values of $\beta$.
\end{itemize}
While this qualitative agreement is encouraging, an honest quantitative comparison between Monte Carlo and analytic results would require going to at least NLL$'$ accuracy. That is, we would include the contributions from low scale matrix elements convolved with the NLL resummation kernel.

One apparent distinction between the analytic result and \pythia{} is that the double differential cross section in \pythia{} vanishes in the region near the line $e_\alpha = e_\beta$, while it does not in the NLL cross section.  We attribute this difference to angular ordering/veto imposed in the Monte Carlo.  The line $e_\alpha = e_\beta$ requires that all emissions contributing there are located at $\theta = R_0$, the edge of the jet.  Such a configuration is exponentially suppressed in a Monte Carlo, but is allowed in our NLL expression  for the double differential cross section.  

However, one can show that at NLL$'$ in each individual factorization, the double differential cross section vanishes at the boundaries of the phase space. The argument is as follows. As noted above \Eq{eq:general_form_of_FO_result}, the general form of the fixed order singular cross section for the $e_\alpha\sim e_{\beta}^{\frac{\alpha}{\beta}}$ factorization boundary is:
\begin{align}
\frac{d^2\sigma^{\alpha}_\text{fo}}{de_\alpha\, de_\beta}&=\frac{d\sigma_\text{fo}}{de_\alpha}\delta(e_\beta)+\Theta\left(e_\alpha-c\,e_\beta^{\frac{\alpha}{\beta}}\right)\frac{1}{e_{\alpha}^{1+\beta/\alpha}}f_{+}^{\alpha}\Bigg(\frac{e_{\beta}}{e_{\alpha}^{\beta/\alpha}}\Bigg),
\end{align} 
where we have explicitly indicated we are taking the cross sections at fixed order (fo). Note that $f_{+}^{\alpha}$ encodes all non-trivial $e_\beta$ dependence, and is solely fixed by the jet function matrix element for this factorization, and we have made explicit the boundary $\Theta$-function enforced by the phase space of the jet function.\footnote{In this $\Theta$-function, $c$ is fixed number that depends on the precise definition of the angularity. For a given recoil-free observable, this boundary condition can become more complicated, but our argument remains unchanged.} If we canonically resum this distribution, the resulting cross section is:
\begin{align}
\frac{d^2\sigma^{\alpha}}{de_\alpha\, de_\beta}&=\int_{0}^{e_{\alpha}}de_\alpha'\,U(e_\alpha-e_{\alpha}')\left\{\frac{d\sigma_\text{fo}}{de_\alpha'}\delta(e_\beta)+\Theta\left(e_\alpha'-c\,e_\beta^{\frac{\alpha}{\beta}}\right)\frac{1}{e_{\alpha}'^{1+\beta/\alpha}}f_{+}^{\alpha}\Bigg(\frac{e_{\beta}}{e_{\alpha}'^{\beta/\alpha}}\Bigg)\right\}\nonumber\\ 
&=\frac{d\sigma_\text{resum}}{de_\alpha}\delta(e_\beta)+\int_{0}^{e_{\alpha}}de_\alpha'\,U(e_\alpha-e_{\alpha}')\Theta\left(e_\alpha'-c\,e_\beta^{\frac{\alpha}{\beta}}\right)\frac{1}{e_{\alpha}'^{1+\beta/\alpha}}f_{+}^{\alpha}\Bigg(\frac{e_{\beta}}{e_{\alpha}'^{\beta/\alpha}}\Bigg),
\end{align} 
where $U(e_\alpha)$ is the resummation kernel for $e_\alpha$. For non-zero $e_\beta$, only the second term contributes, thus as $e_\alpha$ approaches the boundary of phase space $c\,e_\beta^{\frac{\alpha}{\beta}}$ the cross section vanishes, since the limits of integration become squeezed to zero. Note that this argument does not depend on the particular order to which one has calculated the cross section, and thus is a robust prediction of the factorization theorem for the double-differential cross section. We leave a detailed analysis at NLL$'$, including interpolation into the bulk of the phase space, to future work.

\section{Conclusions}\label{sec:conc}

In this paper, we have used the double differential cross section of two angularities measured on a single jet as a case study  for understanding the factorization properties of double differential cross sections. We have explicitly shown that the double differential cross section for two angularities factorizes near the boundaries of the phase space, where it reduces to the single differential cross section of one of the angularities. Indeed, we have also shown the impossibility of a factorization theorem valid in the entire phase space region using only soft and collinear modes. 

We presented a conjecture for the NLL double differential cross section using an interpolation procedure, based on scale setting and the addition of subleading terms, between the two factorization theorems defined on the boundaries of phase space. This interpolation procedure has the interesting property of introducing what we termed $k_T$ logarithms in the bulk region of phase space. These logarithms reduce to soft or collinear logarithms on the boundaries of the phase space where the factorization theorem applies, but are required in the bulk to interpolate between the soft logarithms on one boundary and the collinear logarithms on the other. The conjectured double differential cross section is subject to numerous consistency constraints from the boundary factorization theorems, which guarantee that it is unique to logarithmic accuracy up to at least ${\cal O}(\alpha_s^4)$. The interpolation scale choices that we found at NLL could be tested at higher accuracy by computing the anomalous dimensions of the jet and soft functions to higher loop order. We compared our calculation for the double differential cross section of angularities with a parton shower Monte Carlo, and found qualitative agreement, evidence that Monte Carlos model the correlations between angularities well.

While we have only discussed the perturbative aspects of the double differential cross section, the effect of non-perturbative physics is also important.  Because the recoil-free angularities are additive and there exists a factorization theorem, this suggests that non-perturbative corrections can be incorporated by some kind of shape function \cite{Korchemsky:1999kt,Korchemsky:2000kp}.  In \Ref{Larkoski:2013paa}, a shape function was assumed to exist for the double differential cross section of angularities and it qualitatively agreed with the hadronization corrections in \pythia{} Monte Carlo.  Nevertheless, a rigorous definition of the non-perturbative corrections to the double differential cross section is vital for determining the effect of low energy physics on the correlations of angularities.  Angularities are additive observables and so the shape function for the double differential cross section should be similar in form to the shape function for a single differential cross section.  However, the phase space constraints can be deformed by the non-perturbative corrections, which could result in subtle, but important, effects on the differential cross section.

\subsection{Future Directions}

This paper presents the first step in a wider program with the goal of understanding the factorization and resummation properties of double differential cross sections of IRC safe observables. We therefore conclude with a number of future directions and possible applications of these techniques.

\subsubsection*{Extension to Other Observables}
Although this paper has focused specifically on the example of angularities, the conclusions and techniques should be applicable to the double differential cross sections of more phenomenologically relevant observables. In particular, the argument for the reduction of the double cumulative distribution to a single cumulative distribution on the boundary of phase space presented in \Sec{sec:ps_fact} is geometric in nature and does not rely on the detailed form of the boundaries. The only requirement is that the boundary is described by a monotonically increasing function. We therefore believe this reduction to be generic. This simplifies the problem of proving factorization theorems for double differential cross sections to that of proving factorization theorems for single differential cross sections, several of which are already known.

Armed with factorization theorems near the boundaries of phase space one can attempt an interpolation procedure by shifting scales and adding subleading terms in the cumulative distribution, as was done explicitly for the case of angularities in \Sec{sec:int}. For the relevant case when the two observables define boundary factorization theorems with different scalings for the soft modes, this interpolation procedure necessarily introduces a new logarithmic structure in the bulk of the phase space, which reduces appropriately on the boundaries to either a soft or collinear logarithm. For the case of two angularities, this was the $k_T$ logarithm discussed in \Sec{sec:cuspint}. Furthermore, with certain assumptions on the logarithmic structure, a proof similar to that given in \Sec{sec:uniint} could be used to argue for the uniqueness of the interpolation procedure. We believe that if this procedure is indeed possible for a particular pair of observables, then it gives a strong conjecture for the NLL resummed double differential cross section. This further allows for the computation of the ratio observable through marginalization. It is also interesting to speculate on the existence of a super-SCET formalism allowing for the incorporation of the additional modes required in the bulk, however, we leave this to future study. 

One observable of particular phenomenological interest is $N$-subjettiness, which merits a more detailed discussion due to the interesting structure of its phase space. The $N$-subjettiness observable $\tau_{N}^{(\beta)}$ is defined as \cite{Thaler:2010tr,Thaler:2011gf}
\begin{equation}
\tau_{N}^{(\beta)} =\frac{1}{\sum_{i\in J} p_{Ti}R_0^\beta}\sum_{i\in J} p_{T i}\min\left\{  R_{1,i}^\beta, R_{2,i}^\beta,\dotsc, R_{N,i}^\beta   \right\} \ ,
\end{equation}
where $R_0$ is the jet radius and the sums run over all particles in the jet.  $R_{n,i}$ is the angle between axis $n$ and particle $i$ and $\beta>0$ for IRC safety.  The axes in the jet can be chosen in several ways; the most elegant being to choose the axes so as to minimize the value of $\tau_N^{(\beta)}$.  The ratio of $\tau_2/\tau_1$ has proven very powerful for discriminating boosted $W$ jets from massive QCD jets \cite{Altheimer:2012mn,TheATLAScollaboration:2013tia,CMS:2013uea}.\footnote{When clear from context, we will drop the superscript $\beta$ for brevity.}  Progress has been made in computing the distribution of $\tau_2/\tau_1$ for signal jets by relating it to the event-wide thrust distribution in $e^+e^-$ collisions \cite{Feige:2012vc}.  Understanding the background distribution is a formidable challenge that has not been studied for arbitrary values of the ratio of jet mass to jet energy.

\begin{figure}
\begin{center}
\subfloat[]{\label{fig:nsub_cum_a}
\includegraphics[width=7cm]{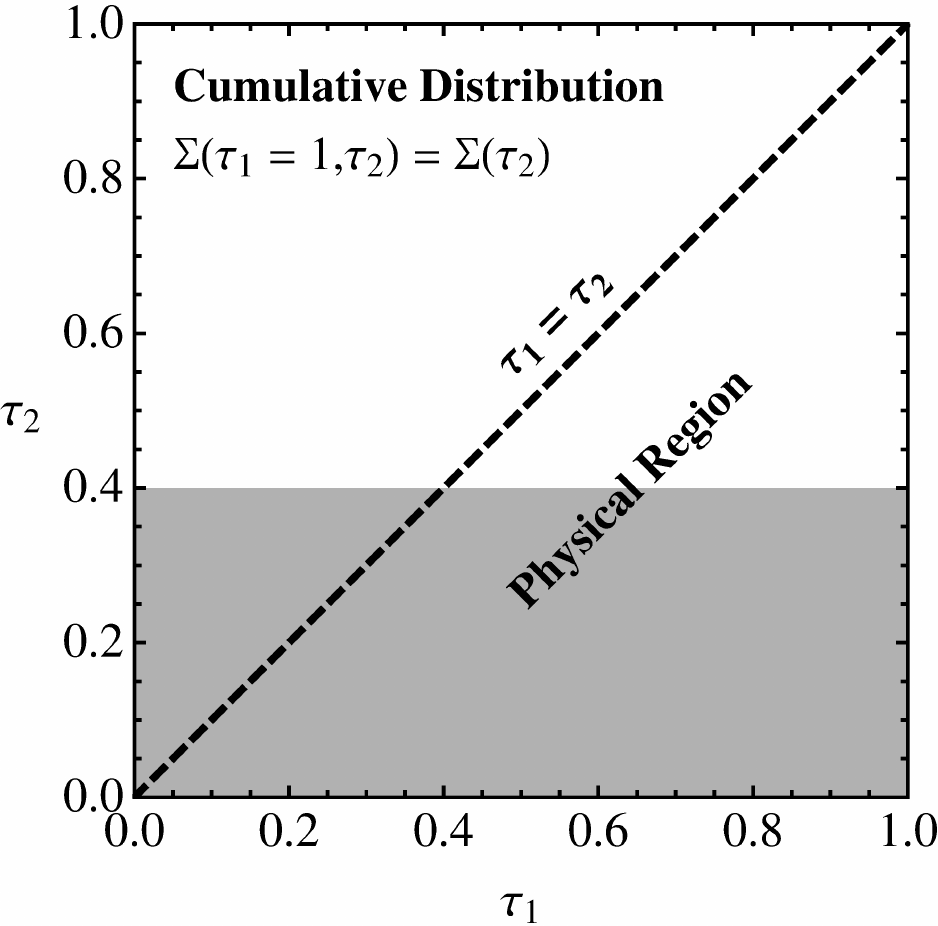}
}
$\qquad$
\subfloat[]{\label{fig:nsub_cum_b} 
\includegraphics[width=7cm]{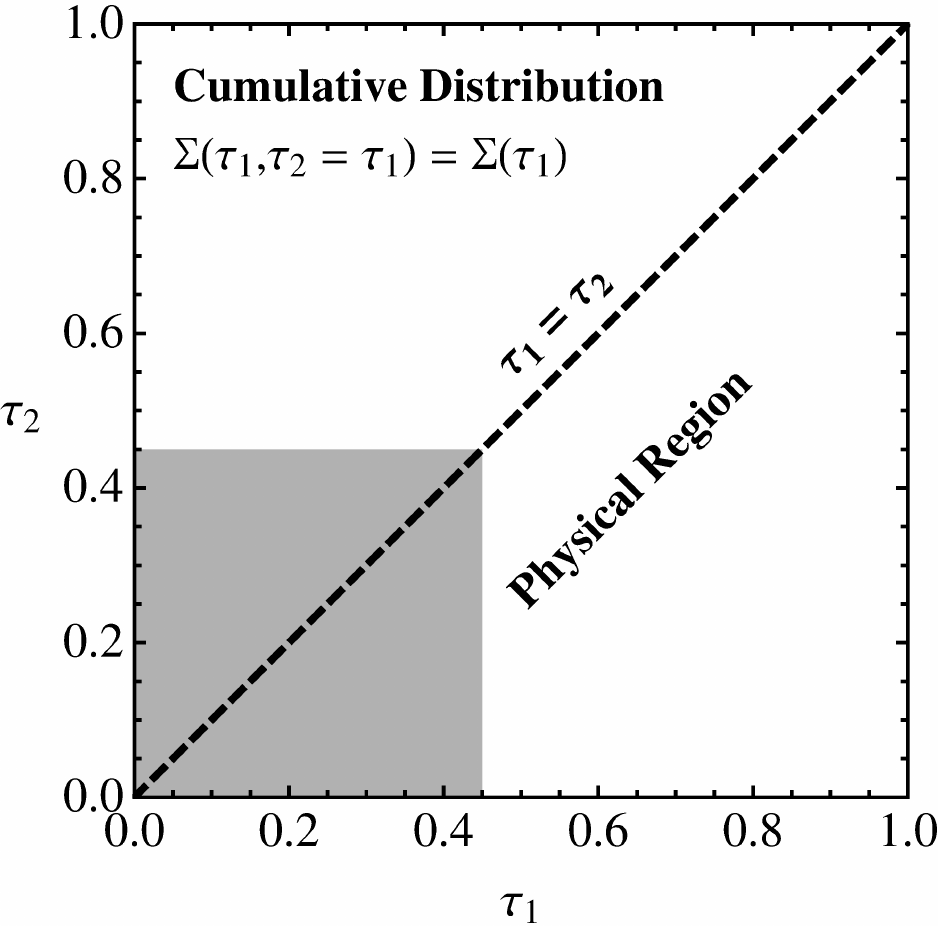}
}
\end{center}
\caption{
Illustration of the double cumulative distribution of 1- and 2-subjettiness evaluated on the boundaries of phase space.  The physical region of the phase space is indicated.  Left: Evaluated on the boundary $\tau_1 = 1$  which reduces the double cumulative distribution to $\Sigma(\tau_2)$.  Right: Evaluated on the boundary $\tau_2 = \tau_1$  which reduces the double cumulative distribution to $\Sigma(\tau_1)$.
}
\label{fig:nsub_cum_bound}
\end{figure}

Nevertheless, we suspect that a boundary factorization theorem exists for the double differential cross section of $N$-subjettiness observables $\tau_2$ and $\tau_1$.  Because $\tau_2$ is defined about two axes in the jet while $\tau_1$ is only defined about one axis, $\tau_2 < \tau_1$, with no non-trivial lower bound on the phase space.  That is, $\tau_2$ can be zero and $\tau_1$ be non-zero, which is different from the angularities considered in this paper.  However, as we illustrate in \Fig{fig:nsub_cum_bound}, the double cumulative distribution should still reduce to a single cumulative distribution on the appropriate boundaries.  For example, evaluating the double cumulative distribution on the boundary $\tau_2=\tau_1$ should reduce to the cumulative distribution for $\tau_1$ alone as $\tau_2$ has been integrated over its entire range.  From the arguments in \Sec{sec:ps_fact} this then implies that the double differential cross section reduces near this boundary:
\begin{equation}
\left.\frac{d^2\sigma}{d\tau_1\, d\tau_2} \right|_{\tau_2\sim \tau_1} \simeq \frac{d\sigma}{d\tau_1}\delta(\tau_2) +...\ ,
\end{equation}
up to terms that integrate to zero on $\tau_2\in[0,\tau_1]$.  A similar relationship holds near the boundary $\tau_1 = 1$ where the cross section reduces as
\begin{equation}
\left.\frac{d^2\sigma}{d\tau_1\, d\tau_2} \right|_{\tau_1\sim1} \simeq \frac{d\sigma}{d\tau_2}\delta(\tau_1-1) +...\ ,
\end{equation}
again, up to terms that integrate to zero on $\tau_1\in[0,1]$.\footnote{We have assumed that the maximum value of $\tau_1$ is 1.  More generally, for $\tau_1$ near its maximum value, the double differential cross section should reduce to the single differential cross section for $\tau_2$.}

Therefore, to prove that the double differential cross section of $\tau_2$ and $\tau_1$ factorizes on the boundaries of phase space only requires proving that the single differential cross sections factorize.  $\tau_1$ is just the jet angularities, for which there exists a factorization theorem and so the double differential cross section should factorize on the $\tau_2=\tau_1$ boundary.  While there is up to now no factorization theorem for the differential cross section of $\tau_2$ for QCD jets, we expect that there is a factorization theorem for $\tau_2$ when $\tau_1=1$.  On this boundary, when $\tau_1=1$, the structure that dominates $\tau_1$ is a hard, perturbative emission.  This essentially defines the jet to have two, well-separated subjets.  In this configuration, $\tau_2$ is dominated by soft and collinear radiation about those subjets, suggesting that its differential cross section is factorizable.\footnote{Because the jet has two hard subjets, the factorization theorem would probably follow from SCET$_+$ of \Ref{Bauer:2011uc}.}  A proof of the factorization of $\tau_2$ and subsequently the calculation of the double differential cross section of $\tau_1$ and $\tau_2$ would provide deep insights into the power of $N$-subjettiness as a discrimination observable as well as substantial information about the structure of QCD jets.

\subsubsection*{Exclusive PDFs}

Our results may also have consequences for the recent program of fully unintegrated or fully exclusive parton distribution functions (PDFs) \cite{Watt:2003mx,Watt:2003vf,Collins:2004vq,Collins:2005uv,Collins:2007ph,Rogers:2008jk,Mantry:2009qz,Mantry:2010mk,Jain:2011iu} that depend on all components of colliding parton momentum, not only the longitudinal component.  As we have shown in the context of angularities, the measurement of multiple observables on a single parton defines a region of the allowed phase space that is determined by the scaling of the observables with respect to one another.  Typically, analyses that resum the logarithms of the unintegrated PDFs are constrained to a particular region of phase space; in \Ref{Jain:2011iu}, they essentially take the beam broadening comparable to the square-root of the beam thrust.\footnote{This is analogous to the lower bound resulting from energy conservation in the phase space of thrust, $e_2$, and broadening, $e_1$, where $e_2\sim e_1^2$.}  Though the PDF is more exclusive and should contain more information about the colliding parton, much of that information is lost because one is forced into a small region of the phase space.  By studying the boundaries of the PDF phase space, it may be possible to interpolate between the boundary regions, producing a description of the unintegrated PDFs throughout the phase space. 

\subsubsection*{Monte Carlos}

Beyond its purely theoretic applications, the double differential cross section of two angularities can be used to tune Monte Carlos.  Typically, tuning involves adjusting the arbitrary parameters in a Monte Carlo so as to match the measured differential cross section of several observables that are sensitive to the parameters.  Tuning is not a precise science and involves significant art to choose parameters consistently so as to match many different distributions.  However, if instead Monte Carlos were tuned to joint differential cross sections of observables, correlations would be naturally incorporated.  With theoretical input for the double differential cross sections of angularities, parameters in the Monte Carlo could be adjusted appropriately to correctly model the higher order perturbative effects and separately, non-perturbative physics.  This tuning program would also require the measurement of the double differential cross sections from the experiments, something that has not yet been published in studies at the Large Hadron Collider.\footnote{However, \Ref{TheATLAScollaboration:2013tia} does contain plots of double differential cross sections from simulation comparing Qjet volitility \cite{Ellis:2012sn} and $N$-subjettiness ratio $\tau_2/\tau_1$ for QCD jets and boosted $W$ bosons.}  Because of theoretical progress, its potential application for Monte Carlo tuning, and the information it contains regarding correlations of observables, we strongly advise the ATLAS and CMS experiments to provide the measurements of double differential cross sections of jet observables.\\

\begin{acknowledgments}
We thank Jesse Thaler, Iain Stewart, Einan Gardi, Jonathan Walsh, Christopher Lee, Ilya Feige, and Matthew Schwartz for very helpful conversations.  We also thank Daekyoung Kang for collaboration in the early stages of this work.  We thank Marat Freytsis for helpful comments on the manuscript.  This work is supported by the U.S. Department of Energy (DOE) under cooperative research agreement DE-FG02-05ER-41360. I.M. is supported by NSERC of Canada and the U.S. Department of Energy (DOE) under cooperative research agreement DE-SC0011090. D.N. is also supported by an MIT Pappalardo Fellowship.  We thank the Erwin Schr\"odinger Institute and the oragnizers of the ``Jets and Quantum Fields for LHC and Future Colliders'' workshop for hospitality and support where this work was initiated.

\end{acknowledgments}

\appendix

\section{The Double Differential Jet and Soft functions}\label{app:jsfunc}

In this appendix we provide explicit calculations of the double differential jet and soft functions that appear in the factorization theorem for the double differential cross section of broadening-axis jet angularities, \Eq{eq:factordd}.

\subsection{Jet Function}\label{app:jfunc}

At one-loop, the jet consists of two particles whose momentum can be written as
\begin{equation}
(q^-,q^+,\vec{q}_\perp) \ , \qquad (Q-q^-,l^+-q^+,-\vec{q}_\perp) \ ,
\end{equation}
for a total jet momentum of $(Q,l^+,\vec{0})$ in light-cone coordinates.  In this frame, the double differential jet function of a quark jet is
\begin{align}
J^{(1)}(e_\alpha,e_\beta)&=g^2  \mu^{2\epsilon}C_F \int \frac{dl^+}{2\pi} \frac{1}{(l^+)^2}\int \frac{d^d q}{(2\pi)^d} \, \left( 4 \frac{l^+}{q^-}+(d-2)\frac{l^+-q^-}{Q-q^-}  \right)\, 2\pi\delta(q^+q^--q_\perp^2) \nonumber \\
&\qquad\qquad \times  \Theta(q^+)\Theta(q^-) \Theta(Q-q^-) \Theta(l^+-q^+)  2\pi\delta\left( l^+-q^+ -\frac{q_\perp^2}{Q-q^-}  \right) \nonumber \\
&\qquad\qquad \times \left\{   
\Theta\left( \frac{Q}{2}-q^-  \right)\delta\left( e_\alpha - Q^{\alpha-1}(Q-q^-)^{-\alpha} (q^-)^{1-\alpha}q_\perp^\alpha  \right)\right.\nonumber\\
&\qquad\qquad\quad \left.\times \ \delta\left( e_\beta - Q^{\beta-1}(Q-q^-)^{-\beta} (q^-)^{1-\beta}q_\perp^\beta   \right) \right. \nonumber\\
&\left. \qquad\qquad+\ \Theta\left( q^- -\frac{Q}{2} \right)\delta\left( e_\alpha -Q^{\alpha-1}(Q-q^-)^{1-\alpha} (q^-)^{-\alpha}q_\perp^\alpha  \right)\right.\nonumber\\
&\qquad\qquad\quad \left.\times \ \delta\left( e_\beta - Q^{\beta-1}(Q-q^-)^{1-\beta} (q^-)^{-\beta}q_\perp^\beta   \right)
\right\} \ .
\end{align} 
We have assumed that the jet radius $R_0$ is ${\cal O}(1)$ and so the jet algorithm constraint is trivial to leading power in $\lambda \ll 1$.  Evaluating this in $d=4-2\epsilon$ dimensions, we find 
\begin{align}\label{eq:jetfunccalc}
J^{(1)}(e_\alpha,e_\beta)&=\frac{\alpha_s}{\pi}\left(\frac{4\pi\mu^{2}}{Q^2}\right)^\epsilon \frac{C_F}{\Gamma(1-\epsilon)}\frac{e_\alpha^{-1+2\epsilon\frac{\beta-1}{\alpha-\beta}}e_\beta^{-1-2\epsilon\frac{\alpha-1}{\alpha-\beta}}}{\alpha-\beta}
\Theta\left(e_\alpha^\beta-2^{\alpha-\beta}e_{\beta}^\alpha\right)\Theta(e_\beta)\nonumber\\
&\qquad\times\left(1-e_\alpha^{-\frac{\beta}{\alpha-\beta}}e_\beta^{\frac{\alpha}{\alpha-\beta}}\right)^{-1-2\epsilon}\left(
2-(3+\epsilon)e_\alpha^{-\frac{\beta}{\alpha-\beta}}e_\beta^{\frac{\alpha}{\alpha-\beta}} + (3+\epsilon) \,e_\alpha^{-\frac{2\beta}{\alpha-\beta}}e_\beta^{\frac{2\alpha}{\alpha-\beta}}
\right) \ ,
\end{align}
The boundary condition enforced by the $\Theta$-functions is $$e_\beta \in \left[  0,2^{-\frac{\alpha-\beta}{\alpha}}e_\alpha^{\beta/\alpha} \right] \ ,$$ and so, by the general arguments of \Sec{sec:ddjetfunc}, all $e_\beta$ dependence in the jet function must appear in the combination
$$2^{\frac{\alpha-\beta}{\alpha}}\frac{e_\beta}{e_\alpha^{\beta/\alpha}} \  .$$

With this substitution in \Eq{eq:jetfunccalc}, we find
\begin{align}\label{eq:jetfuncexp}
J^{(1)}(e_\alpha,e_\beta)&=\frac{\alpha_s}{\pi}\left(\frac{4\pi\mu^{2}}{e_\alpha^{2/\alpha} Q^2}\right)^\epsilon \frac{C_F}{\Gamma(1-\epsilon)}\frac{e_\alpha^{-1-\frac{\beta}{\alpha}}x^{-1-2\epsilon\frac{\alpha-1}{\alpha-\beta}}}{\alpha-\beta}
\Theta\left(2^{-\frac{\alpha-\beta}{\alpha}}-x\right)\Theta(x)\nonumber\\
&\qquad\times\left(1-x^{\frac{\alpha}{\alpha-\beta}}\right)^{-1-2\epsilon}\left(
2-(3+\epsilon)x^{\frac{\alpha}{\alpha-\beta}} + (3+\epsilon) \,x^{\frac{2\alpha}{\alpha-\beta}}
\right) \ ,
\end{align}
where
\begin{equation}
x=\frac{e_\beta}{e_\alpha^{\beta/\alpha}} \ .
\end{equation}
All $x$ dependence (or equivalently $e_\beta$) can be treated with the $+$-prescription as defined in \Eq{eq:plusfunc}.  To the lowest orders in the dimensional regularization parameter $\epsilon$, all $e_\beta$ dependence can be expressed as\footnote{This expansion has been performed with the \mathematica package \texttt{HypExp} \cite{Huber:2005yg,Huber:2007dx}.}
\begin{align}
&x^{-1-2\epsilon\frac{\alpha-1}{\alpha-\beta}}\left(1-x^{\frac{\alpha}{\alpha-\beta}}\right)^{-1-2\epsilon}\left(
2-(3+\epsilon)x^{\frac{\alpha}{\alpha-\beta}} + 3 \,x^{\frac{2\alpha}{\alpha-\beta}}
\right)\Theta\left(2^{-\frac{\alpha-\beta}{\alpha}}-x\right)\Theta(x)\nonumber \\
&=\left[-\frac{\alpha-\beta}{\alpha-1}\frac{1}{\epsilon} -\frac{3}{2}\frac{\alpha-\beta}{\alpha}-\epsilon\frac{\alpha-\beta}{3\alpha^2} \left( -9+\pi^2+18\alpha-2\alpha\pi^2-9\log 2+3\alpha \log 2  \right) \right]e_\alpha^{\beta/\alpha}\delta(e_\beta) \nonumber \\
&\qquad +\left[ \frac{\Theta(e_\beta)\Theta(e_\alpha^\beta - 2^{\alpha-\beta}e_\beta^\alpha)}{1-e_\alpha^{-\frac{\beta}{\alpha-\beta}}e_\beta^{\frac{\alpha}{\alpha-\beta}}} \left(2\frac{e_\alpha^{\beta/\alpha}}{e_\beta} 
-3 \left(\frac{e_\beta}{e_\alpha^{\beta/\alpha}}\right)^{\frac{\beta}{\alpha-\beta}}
+3\left(\frac{e_\beta}{e_\alpha^{\beta/\alpha}}  \right)^{\frac{\alpha+\beta}{\alpha-\beta}}
\right) \right]_+^{2^{-\frac{\alpha-\beta}{\alpha}}e_\alpha^{\beta/\alpha}} \ .
\end{align}
The remaining $\epsilon$ dependence in \Eq{eq:jetfuncexp} after this expansion can be regularized by familiar $+$-distributions in $e_\alpha$.  With the $\overline{\text{MS}}$ prescription the double differential jet function is
\begin{equation}
J^{(1)}(e_\alpha,e_\beta) = J^{(1)}(e_\alpha,e_\beta)_\text{div}+J^{(1)}(e_\alpha,e_\beta)_\text{fin} \ , 
\end{equation}
where the divergent term is
\begin{align}
J^{(1)}(e_\alpha,e_\beta)_\text{div}&=\frac{\alpha_s}{2\pi} C_F \delta(e_\beta) \left\{ \left[ \frac{\alpha}{\alpha-1}\frac{1}{\epsilon^2}+\frac{\alpha}{\alpha-1}\frac{1}{\epsilon}\log\frac{\mu^2}{Q^2}+\frac{3}{2}\frac{1}{\epsilon} \right] \delta(e_\alpha)-\frac{2}{\epsilon}\frac{1}{\alpha-1}\left[\frac{\Theta(e_\alpha)}{e_\alpha} \right]_+ \right\}
\end{align}
and the finite term is
\begin{align}
J^{(1)}(e_\alpha,e_\beta)_\text{fin}&=\frac{\alpha_s}{2\pi}C_F\delta(e_\beta)\left\{
\left[ \frac{3}{2}\alpha \log\frac{\mu^2}{Q^2}+\frac{\alpha}{\alpha-1}\frac{1}{2}\log^2\frac{\mu^2}{Q^2} - \frac{\alpha}{\alpha-1}\frac{\pi^2}{12}\right.\right.\nonumber \\
&\qquad\qquad\qquad\qquad\left. \left.
-\frac{3}{\alpha}+\frac{1}{\alpha}\frac{\pi^2}{3}+6-\frac{2\pi^2}{3}-\frac{3\log 2}{\alpha}+\log 2
 \right]\delta(e_\alpha) \right.\nonumber \\
 &\qquad\qquad\left.
 \!\!\!\!- \frac{3}{\alpha}\left[  \frac{\Theta(e_\alpha)}{e_\alpha} \right]_+-\frac{2}{\alpha-1}\log\frac{\mu^2}{Q^2}\left[  \frac{\Theta(e_\alpha)}{e_\alpha} \right]_+
 +\frac{4}{\alpha(\alpha-1)}\left[ \Theta(e_\alpha) \frac{\log e_\alpha}{e_\alpha}  \right]_+
\right\}\nonumber\\
&+\frac{\alpha_s}{\pi}\frac{C_F}{\alpha-\beta}e_\alpha^{-1-\frac{\beta}{\alpha}}\left[ \frac{\Theta(e_\beta)\Theta(e_\alpha^\beta - 2^{\alpha-\beta}e_\beta^\alpha)}{1-e_\alpha^{-\frac{\beta}{\alpha-\beta}}e_\beta^{\frac{\alpha}{\alpha-\beta}}} \left(2\frac{e_\alpha^{\beta/\alpha}}{e_\beta} 
-3 \left(\frac{e_\beta}{e_\alpha^{\beta/\alpha}}\right)^{\frac{\beta}{\alpha-\beta}}\right.\right.\nonumber \\
&\left.\left.
\qquad\qquad\qquad\qquad\qquad\qquad\qquad\qquad\qquad+\ 3\left(\frac{e_\beta}{e_\alpha^{\beta/\alpha}}  \right)^{\frac{\alpha+\beta}{\alpha-\beta}}
\right) \right]_+^{2^{-\frac{\alpha-\beta}{\alpha}}e_\alpha^{\beta/\alpha}} 
\end{align}
These expressions are consistent with the general arguments of \Sec{sec:ddjetfunc}.

\subsection{Soft Function}\label{app:sfunc}

At one-loop, the double differential soft function of a jet in the process $e^+e^-\to q\bar{q}$ has the following form:
\begin{align}
S^{(1)}(e_\alpha,e_\beta)&= 2g^2\mu^{2\epsilon}  C_F \int\frac{d^d k}{(2\pi)^d} \frac{2}{k^+ k^-} 2\pi\delta(k^+k^- - k_\perp^2)\Theta\left( 1 - \frac{k^+}{k^-}  \right) \nonumber\\
&\qquad\times \delta\left( e_\alpha - Q^{-1} (k^+)^{\frac{\alpha}{2}}(k^-)^{1-\frac{\alpha}{2}}   \right) \delta\left( e_\beta - Q^{-1} (k^+)^{\frac{\beta}{2}}(k^-)^{1-\frac{\beta}{2}}   \right)   \ .
\end{align}
The $\Theta$-function is the constraint of the jet algorithm for the definition of the angularities from \Eq{eq:ang_def}. In $d=4-2\epsilon$ dimensions, the result is
\begin{equation}
S^{(1)}(e_\alpha,e_\beta)= 2\frac{\alpha_s}{\pi}\left(  \frac{4\pi \mu^2}{Q^2}\right)^\epsilon \frac{C_F}{\Gamma(1-\epsilon)}  \frac{e_\alpha^{-1+2\epsilon\frac{\beta-1}{\alpha-\beta}}e_\beta^{-1-2\epsilon\frac{\alpha-1}{\alpha-\beta}}}{\alpha-\beta}\Theta(e_\beta-e_\alpha)\Theta(e_\alpha) \ .
\end{equation}
From the arguments of \Sec{sec:softfunc}, we expect that the expansion in $\epsilon$ produces appropriate $+$-distributions.  To show this explicitly, focus on the factor containing $e_\alpha$ and $e_\beta$ first.  This can be rewritten as
\begin{equation}
e_\alpha^{-1+2\epsilon\frac{\beta-1}{\alpha-\beta}}e_\beta^{-1-2\epsilon\frac{\alpha-1}{\alpha-\beta}}\Theta(e_\beta-e_\alpha) \Theta(e_\alpha)= e_\beta^{-2-2\epsilon}\times \left( \frac{e_\alpha}{e_\beta}  \right)^{-1+2\epsilon\frac{\beta-1}{\alpha-\beta}} \Theta(e_\beta-e_\alpha)\Theta(e_\alpha) \ ,
\end{equation}
which is very similar in form to the expected result of \Eq{eq:softfunc_exp}.  The second factor can be expanded with the $+$-prescription as
\begin{align}
&\left( \frac{e_\alpha}{e_\beta}  \right)^{-1+2\epsilon\frac{\beta-1}{\alpha-\beta}} \Theta(e_\beta-e_\alpha)\Theta(e_\alpha)
= \frac{1}{2\epsilon}\frac{\alpha-\beta}{\beta-1} e_\beta \, \delta(e_\alpha) + \left[ \frac{e_\beta}{e_\alpha} \Theta(e_\beta-e_\alpha)\Theta(e_\alpha)  \right]_+^{e_\beta} +{\cal O}(\epsilon)  \ .
\end{align}

With this expansion of $e_\alpha$, we can now expand the remaining $\epsilon$ dependence in $+$-distributions of $e_\beta$.  With the $\overline{\text{MS}}$ prescription, the soft function is
\begin{equation}
S^{(1)}(e_\alpha,e_\beta) = S^{(1)}(e_\alpha,e_\beta)_\text{div}+S^{(1)}(e_\alpha,e_\beta)_\text{fin} \ ,
\end{equation}
where the divergent term is
\begin{align}
S^{(1)}(e_\alpha,e_\beta)_\text{div}&=\frac{\alpha_s}{2\pi} \frac{C_F}{\beta-1} \delta(e_\alpha)\left\{\left[ -\frac{1}{\epsilon^2}  - \frac{1}{\epsilon} \log\frac{\mu^2}{Q^2} \right]\delta(e_\beta) +\frac{2}{\epsilon}\left[ \frac{\Theta(e_\beta)}{e_\beta}  \right]_+\right\}
\end{align}
and the finite term is
\begin{align}
S^{(1)}(e_\alpha,e_\beta)_\text{fin}&=\frac{\alpha_s}{2\pi} \frac{C_F}{\beta-1} \delta(e_\alpha)\left\{\left[ -\frac{1}{2}\log^2\frac{\mu^2}{Q^2}  + \frac{\pi^2}{12}  \right]\delta(e_\beta) +2\left[ \frac{\Theta(e_\beta)}{e_\beta}  \right]_+\!\!-4\left[ \Theta(e_\beta)\frac{\log e_\beta}{e_\beta}  \right]_+\right\}\nonumber \\
&\qquad+2\frac{\alpha_s}{\pi}\frac{C_F}{\alpha-\beta}e_\beta^{-2}\left[ \frac{e_\beta}{e_\alpha}\Theta(e_\alpha)\Theta(e_\beta-e_\alpha)  \right]_+^{e_\beta} \ ,
\end{align}
where higher-order terms in $\epsilon$ have been ignored.  This form of the soft function is in agreement with the general arguments made in \Sec{sec:softfunc}.

\section{The Cumulative Distribution of a Single Angularity}\label{app:one}

To NLL order, the cumulative distribution of a recoil-free angularity $e_\beta$ can be expressed as
\begin{equation}\label{eq:resummed_x_sec_single}
\Sigma(e_\beta) =\frac{e^{-\gamma_E R'(e_\beta)}}{\Gamma(1+R'(e_\beta))} e^{-R(e_\beta)-\gamma_i T(e_\beta)} \ .
\end{equation}
$R(e_\beta)$ is the radiator and consists of the cusp pieces of the jet and soft function anomalous dimensions.  To NLL accuracy, the cusp anomalous dimensions are evaluated at two loop order and the radiator is
\begin{align}\label{eq:resummed_x_sec_single_rad}
R(e_\beta) &= \frac{C_i}{2\pi \alpha_s \beta_0^2}\frac{1}{\beta-1}\left[
\left( 1+\lambda  \right) \log(1+\lambda)-(\beta+\lambda)\log\left(1+\frac{\lambda}{\beta}\right)
\right] \nonumber \\
&\qquad\qquad+\frac{C_i}{4\pi^2 \beta_0^2}\frac{1}{\beta-1}\left[
\left( \frac{\Gamma^1_{\text{cusp}}}{\Gamma^0_{\text{cusp}}} -2\pi \frac{\beta_1}{\beta_0} \right)\left(
\beta \log\left( 1+\frac{\lambda}{\beta} \right)-\log(1+\lambda)
\right)\right.\nonumber \\
&\qquad\qquad\qquad\left.+\, \pi\frac{\beta_1}{\beta_0}\left(   
\log^2(1+\lambda) - \beta \log^2\left( 1+\frac{\lambda}{\beta}  \right)
\right)
\right] \ .
\end{align}
Here, $C_i$ is the color of the jet, $\lambda = 2\alpha_s \beta_0 \log e_\beta$, $\beta_0$ and $\beta_1$ are the one- and two-loop $\beta$-functions:
\begin{equation}
\beta_0 = \frac{11}{12\pi} C_A - \frac{n_f}{6\pi}  \ , \qquad \beta_1 = \frac{17}{24\pi^2}C_A^2-\frac{5}{24\pi^2} C_A n_f -\frac{C_F n_f}{8\pi^2} \ ,
\end{equation}
and the ratio of the two- to the one-loop cusp anomalous dimensions is
\begin{equation}
\frac{\Gamma^1_{\text{cusp}}}{\Gamma^0_{\text{cusp}}} = \left( \frac{67}{18}-\frac{\pi^2}{6}  \right) C_A - \frac{5}{9}n_f \ .
\end{equation}

For the non-cusp term in the exponent, the function $T(e_\beta)$ is
\begin{equation}\label{eq:resummed_x_sec_single_non_cusp}
T(e_\beta) = \frac{1}{\pi \beta_0}\log\left(  1+2\alpha_s\beta_0 \frac{\log e_\beta}{\beta}  \right) \ ,
\end{equation}
and $\gamma_i$ is the non-cusp anomalous dimension to one loop.  For quarks and gluons, it is
\begin{equation}
\gamma_q = \frac{3}{4}C_F \ , \qquad \gamma_g = \frac{11 C_A -2 n_f}{12} \ .
\end{equation}
Finally, $R'(e_\beta)$ is just the logarithmic derivative of the radiator:
\begin{equation}
R'(e_\beta) \equiv -\frac{\partial}{\partial \log e_\beta} R(e_\beta) \ .
\end{equation}
For NLL accuracy, this only needs to be evaluated at one-loop:
\begin{equation}\label{eq:resummed_x_sec_single_rad_der}
R'(e_\beta)_\text{NLL} = \frac{C_i}{\pi \beta_0}\frac{1}{\beta-1}\left[\log\left( 1+2\alpha_s\beta_0 \frac{\log e_\beta}{\beta}  \right)-\log\left( 1+2\alpha_s\beta_0 \log e_\beta  \right) \right] \ .
\end{equation}

\section{The Double Cumulative Distribution for Two Angularities}\label{app:two}

From \Sec{sec:nllint}, the ansatz of the form of the double cumulative cross section for angularities $e_\alpha$ and $e_\beta$ to NLL is
\begin{equation}
\Sigma(e_\alpha,e_\beta) =\frac{e^{-\gamma_E \tilde{R}(e_\alpha,e_\beta)}}{\Gamma(1+\tilde{R}(e_\alpha,e_\beta))} e^{-R(e_\alpha,e_\beta)-\gamma_i T(e_\alpha,e_\beta)} \ .
\end{equation}
The functions $R(e_\alpha,e_\beta)$, $T(e_\alpha,e_\beta)$ and $\tilde{R}(e_\alpha,e_\beta)$ can be found by setting scales in the logarithms so as to satisfy the boundary conditions.  We find, for the radiator $R(e_\alpha,e_\beta)$
\begin{align}
R(e_\alpha,e_\beta) =&\ \frac{C_i}{2\pi \alpha_s \beta_0^2}\left[
 \frac{1}{\alpha-1}U\left(  2\alpha_s \beta_0 \log e_\alpha\right)-\frac{\beta}{\beta-1} U\left(  2\alpha_s \beta_0 \frac{\log e_\beta}{\beta} \right)
 \right. \nonumber \\
&\qquad\qquad+\left.
\frac{\alpha-\beta}{(\alpha-1)(\beta-1)} U\left(  2\alpha_s \beta_0 \frac{\log e_\alpha^{1-\beta} e_\beta^{\alpha-1}}{\alpha-\beta}\right)
  \right] \nonumber \\
  &+\frac{C_i}{4\pi^2 \beta_0^2}\left[
\left( \frac{\Gamma^1_{\text{cusp}}}{\Gamma^0_{\text{cusp}}} -2\pi \frac{\beta_1}{\beta_0} \right)\left(
\frac{\beta}{\beta-1} \log\left( 1+2\alpha_s \beta_0\frac{\log e_\beta}{\beta} \right)\right.\right.\nonumber \\
&\left. \quad -\frac{1}{\alpha-1}\log(1+2\alpha_s \beta_0 \log e_\alpha)
-\frac{\alpha-\beta}{(\alpha-1)(\beta-1)}\log\left(
1+2\alpha_s\beta_0 \frac{\log e_\alpha^{1-\beta}e_\beta^{\alpha-1}}{\alpha-\beta}
\right)
\right)\nonumber \\
&\quad+\, \pi\frac{\beta_1}{\beta_0}\left(   
\frac{1}{\alpha-1}\log^2(1+2 \alpha_s \beta_0 \log e_\alpha) 
- \frac{\beta}{\beta-1} \log^2\left( 1+2\alpha_s\beta_0\frac{\log e_\beta}{\beta}  \right)\right.\nonumber \\
&\left.\left.
\quad+\, \frac{\alpha-\beta}{(\alpha-1)(\beta-1)}\log^2\left(
1+2\alpha_s\beta_0 \frac{\log e_\alpha^{1-\beta}e_\beta^{\alpha-1}}{\alpha-\beta}
\right)
\right)
\right] 
  \ ,
\end{align}
where $U(z) = (1+z)\log(1+z)$.
For the non-cusp piece $T(e_\alpha,e_\beta)$ we find.
\begin{equation}
T(e_\alpha,e_\beta) = \frac{1}{\pi \beta_0} \log\left( 1+2\alpha_s \beta_0 \frac{\log e_\beta}{\beta} \right) 
-2\frac{\alpha_s}{\pi} \frac{\alpha-\beta}{\alpha} \frac{e_\alpha^{-\frac{\beta}{\alpha-\beta}}e_\beta^{\frac{\alpha}{\alpha-\beta}}}{\beta+2\alpha_s \beta_0 \log e_\beta} \ .
\end{equation}
Finally, for the multiple emissions piece $\tilde{R}(e_\alpha,e_\beta)$ we find
\begin{align}
\tilde{R}(e_\alpha,e_\beta) &= \frac{C_i}{\pi \beta_0} \left\{
\frac{1}{\beta-1}\log\left( 1+2\alpha_s \beta_0 \frac{\log e_\beta}{\beta}  \right) - \frac{1}{\alpha-1} \log\left(  1+2\alpha_s\beta_0 \log e_\alpha \right)
\right. \nonumber \\
&\qquad\qquad\left.
-\frac{\alpha-\beta}{(\alpha-1)(\beta-1)}\log\left(
1+2\alpha_s\beta_0 \frac{\log e_\alpha^{1-\beta}e_\beta^{\alpha-1}}{\alpha-\beta} 
\right)\right.\nonumber \\
&\left.\qquad\qquad
+\ 2\alpha_s\beta_0 \frac{\alpha-\beta}{\alpha}\frac{e_\alpha^{-\frac{\beta}{\alpha-\beta}}e_\beta^{\frac{\alpha}{\alpha-\beta}}}{\beta+2\alpha_s \beta_0 \log e_\beta}
\right\}  \ .
\end{align}
The power suppressed terms have been chosen so that the sum of the exponents of $e_\alpha$ and $e_\beta$ is 1. Explicitly evaluating these functions on the appropriate boundaries reproduces \Eq{eq:resummed_x_sec_single} with the appropriate values for \eqref{eq:resummed_x_sec_single_rad}, \eqref{eq:resummed_x_sec_single_non_cusp}, and \eqref{eq:resummed_x_sec_single_rad_der}.

\bibliography{ddiff_paper}

\providecommand{\href}[2]{#2}\begingroup\raggedright\begin{thebibliography}{10}

\bibitem{Catani:1991kz}
S.~Catani, G.~Turnock, B.~Webber, and L.~Trentadue, {\it {Thrust distribution
  in e+ e- annihilation}},  {\em Phys.Lett.} {\bf B263} (1991) 491--497.

\bibitem{Catani:1991bd}
S.~Catani, G.~Turnock, and B.~Webber, {\it {Heavy jet mass distribution in e+
  e- annihilation}},  {\em Phys.Lett.} {\bf B272} (1991) 368--372.

\bibitem{Catani:1992jc}
S.~Catani, G.~Turnock, and B.~Webber, {\it {Jet broadening measures in $e^{+}
  e^{-}$ annihilation}},  {\em Phys.Lett.} {\bf B295} (1992) 269--276.

\bibitem{Seymour:1997kj}
M.~Seymour, {\it {Jet shapes in hadron collisions: Higher orders, resummation
  and hadronization}},  {\em Nucl.Phys.} {\bf B513} (1998) 269--300,
  [\href{http://xxx.lanl.gov/abs/hep-ph/9707338}{{\tt hep-ph/9707338}}].

\bibitem{Dokshitzer:1998kz}
Y.~L. Dokshitzer, A.~Lucenti, G.~Marchesini, and G.~Salam, {\it {On the QCD
  analysis of jet broadening}},  {\em JHEP} {\bf 9801} (1998) 011,
  [\href{http://xxx.lanl.gov/abs/hep-ph/9801324}{{\tt hep-ph/9801324}}].

\bibitem{Berger:2003iw}
C.~F. Berger, T.~Kucs, and G.~F. Sterman, {\it {Event shape / energy flow
  correlations}},  {\em Phys.Rev.} {\bf D68} (2003) 014012,
  [\href{http://xxx.lanl.gov/abs/hep-ph/0303051}{{\tt hep-ph/0303051}}].

\bibitem{Banfi:2004yd}
A.~Banfi, G.~P. Salam, and G.~Zanderighi, {\it {Principles of general
  final-state resummation and automated implementation}},  {\em JHEP} {\bf
  0503} (2005) 073, [\href{http://xxx.lanl.gov/abs/hep-ph/0407286}{{\tt
  hep-ph/0407286}}].

\bibitem{Schwartz:2007ib}
M.~D. Schwartz, {\it {Resummation and NLO matching of event shapes with
  effective field theory}},  {\em Phys.Rev.} {\bf D77} (2008) 014026,
  [\href{http://xxx.lanl.gov/abs/0709.2709}{{\tt arXiv:0709.2709}}].

\bibitem{Becher:2008cf}
T.~Becher and M.~D. Schwartz, {\it {A precise determination of $\alpha_s$ from
  LEP thrust data using effective field theory}},  {\em JHEP} {\bf 0807} (2008)
  034, [\href{http://xxx.lanl.gov/abs/0803.0342}{{\tt arXiv:0803.0342}}].

\bibitem{Ellis:2010rwa}
S.~D. Ellis, C.~K. Vermilion, J.~R. Walsh, A.~Hornig, and C.~Lee, {\it {Jet
  Shapes and Jet Algorithms in SCET}},  {\em JHEP} {\bf 1011} (2010) 101,
  [\href{http://xxx.lanl.gov/abs/1001.0014}{{\tt arXiv:1001.0014}}].

\bibitem{Abbate:2010xh}
R.~Abbate, M.~Fickinger, A.~H. Hoang, V.~Mateu, and I.~W. Stewart, {\it {Thrust
  at N$^3$LL with Power Corrections and a Precision Global Fit for
  alphas(mZ)}},  {\em Phys.Rev.} {\bf D83} (2011) 074021,
  [\href{http://xxx.lanl.gov/abs/1006.3080}{{\tt arXiv:1006.3080}}].

\bibitem{Chiu:2012ir}
J.-Y. Chiu, A.~Jain, D.~Neill, and I.~Z. Rothstein, {\it {A Formalism for the
  Systematic Treatment of Rapidity Logarithms in Quantum Field Theory}},  {\em
  JHEP} {\bf 1205} (2012) 084, [\href{http://xxx.lanl.gov/abs/1202.0814}{{\tt
  arXiv:1202.0814}}].

\bibitem{Feige:2012vc}
I.~Feige, M.~D. Schwartz, I.~W. Stewart, and J.~Thaler, {\it {Precision Jet
  Substructure from Boosted Event Shapes}},  {\em Phys.Rev.Lett.} {\bf 109}
  (2012) 092001, [\href{http://xxx.lanl.gov/abs/1204.3898}{{\tt
  arXiv:1204.3898}}].

\bibitem{Becher:2012qc}
T.~Becher and G.~Bell, {\it {NNLL Resummation for Jet Broadening}},  {\em JHEP}
  {\bf 1211} (2012) 126, [\href{http://xxx.lanl.gov/abs/1210.0580}{{\tt
  arXiv:1210.0580}}].

\bibitem{Chien:2012ur}
Y.-T. Chien, R.~Kelley, M.~D. Schwartz, and H.~X. Zhu, {\it {Resummation of Jet
  Mass at Hadron Colliders}},  {\em Phys.Rev.} {\bf D87} (2013) 014010,
  [\href{http://xxx.lanl.gov/abs/1208.0010}{{\tt arXiv:1208.0010}}].

\bibitem{Larkoski:2012eh}
A.~J. Larkoski, {\it {QCD Analysis of the Scale-Invariance of Jets}},  {\em
  Phys.Rev.} {\bf D86} (2012) 054004,
  [\href{http://xxx.lanl.gov/abs/1207.1437}{{\tt arXiv:1207.1437}}].

\bibitem{Dasgupta:2012hg}
M.~Dasgupta, K.~Khelifa-Kerfa, S.~Marzani, and M.~Spannowsky, {\it {On jet mass
  distributions in Z+jet and dijet processes at the LHC}},  {\em JHEP} {\bf
  1210} (2012) 126, [\href{http://xxx.lanl.gov/abs/1207.1640}{{\tt
  arXiv:1207.1640}}].

\bibitem{Jouttenus:2013hs}
T.~T. Jouttenus, I.~W. Stewart, F.~J. Tackmann, and W.~J. Waalewijn, {\it {Jet
  Mass Spectra in Higgs $+$ One Jet at NNLL}},  {\em Phys.Rev.} {\bf D88}
  (2013) 054031, [\href{http://xxx.lanl.gov/abs/1302.0846}{{\tt
  arXiv:1302.0846}}].

\bibitem{Dasgupta:2013ihk}
M.~Dasgupta, A.~Fregoso, S.~Marzani, and G.~P. Salam, {\it {Towards an
  understanding of jet substructure}},  {\em JHEP} {\bf 1309} (2013) 029,
  [\href{http://xxx.lanl.gov/abs/1307.0007}{{\tt arXiv:1307.0007}}].

\bibitem{broadening}
A.~J. Larkoski, D.~Neill, and J.~Thaler, {\it {Jet Shapes with the Broadening
  Axis}},  \href{http://xxx.lanl.gov/abs/1401.2158}{{\tt arXiv:1401.2158}}.

\bibitem{Davison:2008vx}
R.~Davison and B.~Webber, {\it {Non-Perturbative Contribution to the Thrust
  Distribution in e+ e- Annihilation}},  {\em Eur.Phys.J.} {\bf C59} (2009)
  13--25, [\href{http://xxx.lanl.gov/abs/0809.3326}{{\tt arXiv:0809.3326}}].

\bibitem{Laenen:2000ij}
E.~Laenen, G.~F. Sterman, and W.~Vogelsang, {\it {Recoil and threshold
  corrections in short distance cross-sections}},  {\em Phys.Rev.} {\bf D63}
  (2001) 114018, [\href{http://xxx.lanl.gov/abs/hep-ph/0010080}{{\tt
  hep-ph/0010080}}].

\bibitem{Kelley:2011ng}
R.~Kelley, M.~D. Schwartz, R.~M. Schabinger, and H.~X. Zhu, {\it {The two-loop
  hemisphere soft function}},  {\em Phys.Rev.} {\bf D84} (2011) 045022,
  [\href{http://xxx.lanl.gov/abs/1105.3676}{{\tt arXiv:1105.3676}}].

\bibitem{Jouttenus:2011wh}
T.~T. Jouttenus, I.~W. Stewart, F.~J. Tackmann, and W.~J. Waalewijn, {\it {The
  Soft Function for Exclusive N-Jet Production at Hadron Colliders}},  {\em
  Phys.Rev.} {\bf D83} (2011) 114030,
  [\href{http://xxx.lanl.gov/abs/1102.4344}{{\tt arXiv:1102.4344}}].

\bibitem{Hornig:2011iu}
A.~Hornig, C.~Lee, I.~W. Stewart, J.~R. Walsh, and S.~Zuberi, {\it {Non-global
  Structure of the $O({\alpha}_s^2)$ Dijet Soft Function}},  {\em JHEP} {\bf
  1108} (2011) 054, [\href{http://xxx.lanl.gov/abs/1105.4628}{{\tt
  arXiv:1105.4628}}].

\bibitem{Bauer:2011uc}
C.~W. Bauer, F.~J. Tackmann, J.~R. Walsh, and S.~Zuberi, {\it {Factorization
  and Resummation for Dijet Invariant Mass Spectra}},  {\em Phys.Rev.} {\bf
  D85} (2012) 074006, [\href{http://xxx.lanl.gov/abs/1106.6047}{{\tt
  arXiv:1106.6047}}].

\bibitem{Abdesselam:2010pt}
A.~Abdesselam, E.~B. Kuutmann, U.~Bitenc, G.~Brooijmans, J.~Butterworth,
  et~al., {\it {Boosted objects: A Probe of beyond the Standard Model
  physics}},  {\em Eur.Phys.J.} {\bf C71} (2011) 1661,
  [\href{http://xxx.lanl.gov/abs/1012.5412}{{\tt arXiv:1012.5412}}].

\bibitem{Altheimer:2012mn}
A.~Altheimer, S.~Arora, L.~Asquith, G.~Brooijmans, J.~Butterworth, et~al., {\it
  {Jet Substructure at the Tevatron and LHC: New results, new tools, new
  benchmarks}},  {\em J.Phys.} {\bf G39} (2012) 063001,
  [\href{http://xxx.lanl.gov/abs/1201.0008}{{\tt arXiv:1201.0008}}].

\bibitem{Altheimer:2013yza}
A.~Altheimer, A.~Arce, L.~Asquith, J.~Backus~Mayes, E.~Bergeaas~Kuutmann,
  et~al., {\it {Boosted objects and jet substructure at the LHC}},
  \href{http://xxx.lanl.gov/abs/1311.2708}{{\tt arXiv:1311.2708}}.

\bibitem{ATLAS-CONF-2011-073}
{\bf ATLAS Collaboration} Collaboration, {\it Measurement of jet mass and
  substructure for inclusive jets in Ãs = 7 tev pp collisions with the atlas
  experiment},  Tech. Rep. ATLAS-CONF-2011-073, CERN, Geneva, May, 2011.

\bibitem{Miller:2011qg}
{\bf ATLAS Collaboration} Collaboration, D.~W. Miller, {\it {Jet substructure
  in ATLAS}},  Tech. Rep. ATL-PHYS-PROC-2011-142, 2011.

\bibitem{ATLAS-CONF-2011-053}
{\bf ATLAS Collaboration} Collaboration, {\it Light-quark and gluon jets in
  atlas},  Tech. Rep. ATLAS-CONF-2011-053, CERN, Geneva, Apr, 2011.

\bibitem{ATLAS:2012am}
{\bf ATLAS Collaboration} Collaboration, G.~Aad et~al., {\it {Jet mass and
  substructure of inclusive jets in $\sqrt{s}=7$ TeV $pp$ collisions with the
  ATLAS experiment}},  {\em JHEP} {\bf 1205} (2012) 128,
  [\href{http://xxx.lanl.gov/abs/1203.4606}{{\tt arXiv:1203.4606}}].

\bibitem{ATLAS:2012xna}
{\bf ATLAS Collaboration} Collaboration, {\it {Identification and Tagging of
  Double b-hadron jets with the ATLAS Detector}}, .

\bibitem{Aad:2012meb}
{\bf ATLAS Collaboration} Collaboration, G.~Aad et~al., {\it {ATLAS
  measurements of the properties of jets for boosted particle searches}},  {\em
  Phys.Rev.} {\bf D86} (2012) 072006,
  [\href{http://xxx.lanl.gov/abs/1206.5369}{{\tt arXiv:1206.5369}}].

\bibitem{Aad:2012raa}
{\bf ATLAS Collaboration} Collaboration, G.~Aad et~al., {\it {Search for
  resonances decaying into top-quark pairs using fully hadronic decays in $pp$
  collisions with ATLAS at $\sqrt{s}=7$ TeV}},  {\em JHEP} {\bf 1301} (2013)
  116, [\href{http://xxx.lanl.gov/abs/1211.2202}{{\tt arXiv:1211.2202}}].

\bibitem{ATLAS:2012dp}
{\bf ATLAS Collaboration} Collaboration, G.~Aad et~al., {\it {Search for pair
  production of massive particles decaying into three quarks with the ATLAS
  detector in $\sqrt{s}=7$ TeV $pp$ collisions at the LHC}},  {\em JHEP} {\bf
  1212} (2012) 086, [\href{http://xxx.lanl.gov/abs/1210.4813}{{\tt
  arXiv:1210.4813}}].

\bibitem{ATLAS-CONF-2012-066}
{\bf ATLAS Collaboration} Collaboration, {\it Studies of the impact and
  mitigation of pile-up on large-$r$ and groomed jets in atlas at $\sqrt{s}=7$
  tev},  Tech. Rep. ATLAS-CONF-2012-066, CERN, Geneva, Jul, 2012.

\bibitem{ATLAS-CONF-2012-065}
{\bf ATLAS Collaboration} Collaboration, {\it Performance of large-r jets and
  jet substructure reconstruction with the atlas detector},  Tech. Rep.
  ATLAS-CONF-2012-065, CERN, Geneva, Jul, 2012.

\bibitem{Aad:2013fba}
{\bf ATLAS Collaboration} Collaboration, G.~Aad et~al., {\it {Measurement of
  jet shapes in top pair events at sqrt(s) = 7 TeV using the ATLAS detector}},
  \href{http://xxx.lanl.gov/abs/1307.5749}{{\tt arXiv:1307.5749}}.

\bibitem{Aad:2013gja}
{\bf ATLAS Collaboration} Collaboration, G.~Aad et~al., {\it {Performance of
  jet substructure techniques for large-$R$ jets in proton-proton collisions at
  $\sqrt{s}$ = 7 TeV using the ATLAS detector}},  {\em JHEP} {\bf 1309} (2013)
  076, [\href{http://xxx.lanl.gov/abs/1306.4945}{{\tt arXiv:1306.4945}}].

\bibitem{TheATLAScollaboration:2013pia}
{\bf ATLAS Collaboration} Collaboration, {\it {Pile-up subtraction and
  suppression for jets in ATLAS}},  Tech. Rep. ATLAS-CONF-2013-083,
  ATLAS-COM-CONF-2013-097, 2013.

\bibitem{TheATLAScollaboration:2013qia}
{\bf ATLAS Collaboration} Collaboration, {\it {Performance of boosted top quark
  identification in 2012 ATLAS data}},  Tech. Rep. ATLAS-CONF-2013-084,
  ATLAS-COM-CONF-2013-074, 2013.

\bibitem{TheATLAScollaboration:2013ria}
{\bf ATLAS Collaboration} Collaboration, {\it {Performance of pile-up
  subtraction for jet shapes}},  Tech. Rep. ATLAS-CONF-2013-085,
  ATLAS-COM-CONF-2013-100, 2013.

\bibitem{TheATLAScollaboration:2013sia}
{\bf ATLAS Collaboration} Collaboration, {\it {Jet Charge Studies with the
  ATLAS Detector Using $\sqrt{s} = 8$ TeV Proton-Proton Collision Data}},
  Tech. Rep. ATLAS-CONF-2013-086, ATLAS-COM-CONF-2013-101, 2013.

\bibitem{TheATLAScollaboration:2013tia}
{\bf ATLAS Collaboration} Collaboration, {\it {Performance and Validation of
  Q-Jets at the ATLAS Detector in pp Collisions at $\sqrt{s}$=8 TeV in 2012}},
  Tech. Rep. ATLAS-CONF-2013-087, ATLAS-COM-CONF-2013-099, 2013.

\bibitem{CMS-PAS-QCD-10-041}
{\bf CMS Collaboration} Collaboration, {\it Measurement of the subjet
  multiplicity in dijet events from proton-proton collisions at sqrt(s) = 7
  tev},  Tech. Rep. CMS-PAS-QCD-10-041, CERN, Geneva, 2010.

\bibitem{CMS-PAS-JME-10-013}
{\bf CMS Collaboration} Collaboration, {\it Jet substructure algorithms},
  Tech. Rep. CMS-PAS-JME-10-013, CERN, Geneva, 2011.

\bibitem{CMS:2011bqa}
{\bf CMS Collaboration} Collaboration, {\it {Search for BSM ttbar Production in
  the Boosted All-Hadronic Final State}},  Tech. Rep. CMS-PAS-EXO-11-006, 2011.

\bibitem{Chatrchyan:2012mec}
{\bf CMS Collaboration} Collaboration, S.~Chatrchyan et~al., {\it {Shape,
  transverse size, and charged hadron multiplicity of jets in pp collisions at
  7 TeV}},  {\em JHEP} {\bf 1206} (2012) 160,
  [\href{http://xxx.lanl.gov/abs/1204.3170}{{\tt arXiv:1204.3170}}].

\bibitem{Chatrchyan:2012tt}
{\bf CMS Collaboration} Collaboration, S.~Chatrchyan et~al., {\it {Measurement
  of the underlying event activity in $pp$ collisions at $\sqrt{s} = 0.9$ and 7
  TeV with the novel jet-area/median approach}},  {\em JHEP} {\bf 1208} (2012)
  130, [\href{http://xxx.lanl.gov/abs/1207.2392}{{\tt arXiv:1207.2392}}].

\bibitem{Chatrchyan:2012sn}
{\bf CMS Collaboration} Collaboration, S.~Chatrchyan et~al., {\it {Search for a
  Higgs boson in the decay channel $H$ to ZZ(*) to $q$ qbar $\ell^-$ l+ in $pp$
  collisions at $\sqrt{s}=7$ TeV}},  {\em JHEP} {\bf 1204} (2012) 036,
  [\href{http://xxx.lanl.gov/abs/1202.1416}{{\tt arXiv:1202.1416}}].

\bibitem{CMS:2013kfa}
{\bf CMS Collaboration} Collaboration, C.~Collaboration, {\it {Performance of
  quark/gluon discrimination in 8 TeV pp data}},  Tech. Rep.
  CMS-PAS-JME-13-002, 2013.

\bibitem{CMS:2013wea}
{\bf CMS Collaboration} Collaboration, C.~Collaboration, {\it {Pileup Jet
  Identification}},  Tech. Rep. CMS-PAS-JME-13-005, 2013.

\bibitem{CMS:2013vea}
{\bf CMS Collaboration} Collaboration, C.~Collaboration, {\it {Performance of b
  tagging at sqrt(s)=8 TeV in multijet, ttbar and boosted topology events}},
  Tech. Rep. CMS-PAS-BTV-13-001, 2013.

\bibitem{CMS:2013uea}
{\bf CMS Collaboration} Collaboration, C.~Collaboration, {\it {Identifying
  Hadronically Decaying Vector Bosons Merged into a Single Jet}},  Tech. Rep.
  CMS-PAS-JME-13-006, 2013.

\bibitem{Thaler:2010tr}
J.~Thaler and K.~Van~Tilburg, {\it {Identifying Boosted Objects with
  N-subjettiness}},  {\em JHEP} {\bf 1103} (2011) 015,
  [\href{http://xxx.lanl.gov/abs/1011.2268}{{\tt arXiv:1011.2268}}].

\bibitem{Thaler:2011gf}
J.~Thaler and K.~Van~Tilburg, {\it {Maximizing Boosted Top Identification by
  Minimizing N-subjettiness}},  {\em JHEP} {\bf 1202} (2012) 093,
  [\href{http://xxx.lanl.gov/abs/1108.2701}{{\tt arXiv:1108.2701}}].

\bibitem{Larkoski:2013eya}
A.~J. Larkoski, G.~P. Salam, and J.~Thaler, {\it {Energy Correlation Functions
  for Jet Substructure}},  {\em JHEP} {\bf 1306} (2013) 108,
  [\href{http://xxx.lanl.gov/abs/1305.0007}{{\tt arXiv:1305.0007}}].

\bibitem{Thaler:2008ju}
J.~Thaler and L.-T. Wang, {\it {Strategies to Identify Boosted Tops}},  {\em
  JHEP} {\bf 0807} (2008) 092, [\href{http://xxx.lanl.gov/abs/0806.0023}{{\tt
  arXiv:0806.0023}}].

\bibitem{Almeida:2008yp}
L.~G. Almeida, S.~J. Lee, G.~Perez, G.~F. Sterman, I.~Sung, et~al., {\it
  {Substructure of high-$p_T$ Jets at the LHC}},  {\em Phys.Rev.} {\bf D79}
  (2009) 074017, [\href{http://xxx.lanl.gov/abs/0807.0234}{{\tt
  arXiv:0807.0234}}].

\bibitem{Jankowiak:2011qa}
M.~Jankowiak and A.~J. Larkoski, {\it {Jet Substructure Without Trees}},  {\em
  JHEP} {\bf 1106} (2011) 057, [\href{http://xxx.lanl.gov/abs/1104.1646}{{\tt
  arXiv:1104.1646}}].

\bibitem{Jankowiak:2012na}
M.~Jankowiak and A.~J. Larkoski, {\it {Angular Scaling in Jets}},  {\em JHEP}
  {\bf 1204} (2012) 039, [\href{http://xxx.lanl.gov/abs/1201.2688}{{\tt
  arXiv:1201.2688}}].

\bibitem{Soyez:2012hv}
G.~Soyez, G.~P. Salam, J.~Kim, S.~Dutta, and M.~Cacciari, {\it {Pileup
  subtraction for jet shapes}},  \href{http://xxx.lanl.gov/abs/1211.2811}{{\tt
  arXiv:1211.2811}}.

\bibitem{Larkoski:2013paa}
A.~J. Larkoski and J.~Thaler, {\it {Unsafe but Calculable: Ratios of
  Angularities in Perturbative QCD}},  {\em JHEP} {\bf 1309} (2013) 137,
  [\href{http://xxx.lanl.gov/abs/1307.1699}{{\tt arXiv:1307.1699}}].

\bibitem{Dasgupta:2007wa}
M.~Dasgupta, L.~Magnea, and G.~P. Salam, {\it {Non-perturbative QCD effects in
  jets at hadron colliders}},  {\em JHEP} {\bf 0802} (2008) 055,
  [\href{http://xxx.lanl.gov/abs/0712.3014}{{\tt arXiv:0712.3014}}].

\bibitem{Bauer:2000yr}
C.~W. Bauer, S.~Fleming, D.~Pirjol, and I.~W. Stewart, {\it {An Effective field
  theory for collinear and soft gluons: Heavy to light decays}},  {\em
  Phys.Rev.} {\bf D63} (2001) 114020,
  [\href{http://xxx.lanl.gov/abs/hep-ph/0011336}{{\tt hep-ph/0011336}}].

\bibitem{Bauer:2000ew}
C.~W. Bauer, S.~Fleming, and M.~E. Luke, {\it {Summing Sudakov logarithms in $B
  \to X(s \gamma)$ in effective field theory}},  {\em Phys.Rev.} {\bf D63}
  (2000) 014006, [\href{http://xxx.lanl.gov/abs/hep-ph/0005275}{{\tt
  hep-ph/0005275}}].

\bibitem{Bauer:2001yt}
C.~W. Bauer, D.~Pirjol, and I.~W. Stewart, {\it {Soft collinear factorization
  in effective field theory}},  {\em Phys.Rev.} {\bf D65} (2002) 054022,
  [\href{http://xxx.lanl.gov/abs/hep-ph/0109045}{{\tt hep-ph/0109045}}].

\bibitem{Bauer:2001ct}
C.~W. Bauer and I.~W. Stewart, {\it {Invariant operators in collinear effective
  theory}},  {\em Phys.Lett.} {\bf B516} (2001) 134--142,
  [\href{http://xxx.lanl.gov/abs/hep-ph/0107001}{{\tt hep-ph/0107001}}].

\bibitem{Bauer:2002nz}
C.~W. Bauer, S.~Fleming, D.~Pirjol, I.~Z. Rothstein, and I.~W. Stewart, {\it
  {Hard scattering factorization from effective field theory}},  {\em
  Phys.Rev.} {\bf D66} (2002) 014017,
  [\href{http://xxx.lanl.gov/abs/hep-ph/0202088}{{\tt hep-ph/0202088}}].

\bibitem{Dasgupta:2001sh}
M.~Dasgupta and G.~Salam, {\it {Resummation of nonglobal QCD observables}},
  {\em Phys.Lett.} {\bf B512} (2001) 323--330,
  [\href{http://xxx.lanl.gov/abs/hep-ph/0104277}{{\tt hep-ph/0104277}}].

\bibitem{Banfi:2005gj}
A.~Banfi and M.~Dasgupta, {\it {Problems in resumming interjet energy flows
  with $k_t$ clustering}},  {\em Phys.Lett.} {\bf B628} (2005) 49--56,
  [\href{http://xxx.lanl.gov/abs/hep-ph/0508159}{{\tt hep-ph/0508159}}].

\bibitem{Becher:2011pf}
T.~Becher, G.~Bell, and M.~Neubert, {\it {Factorization and Resummation for Jet
  Broadening}},  {\em Phys.Lett.} {\bf B704} (2011) 276--283,
  [\href{http://xxx.lanl.gov/abs/1104.4108}{{\tt arXiv:1104.4108}}].

\bibitem{Bertolini:2013iqa}
D.~Bertolini, T.~Chan, and J.~Thaler, {\it {Jet Observables Without Jet
  Algorithms}},  \href{http://xxx.lanl.gov/abs/1310.7584}{{\tt
  arXiv:1310.7584}}.

\bibitem{Rakow:1981qn}
P.~E. Rakow and B.~Webber, {\it {Transverse Momentum Moments of Hadron
  Distributions in QCD Jets}},  {\em Nucl.Phys.} {\bf B191} (1981) 63.

\bibitem{Ellis:1986ig}
R.~K. Ellis and B.~Webber, {\it {QCD Jet Broadening in Hadron Hadron
  Collisions}},  {\em Conf.Proc.} {\bf C860623} (1986) 74.

\bibitem{Bauer:2008qu}
C.~W. Bauer, O.~Cata, and G.~Ovanesyan, {\it {On different ways to quantize
  Soft-Collinear Effective Theory}},
  \href{http://xxx.lanl.gov/abs/0809.1099}{{\tt arXiv:0809.1099}}.

\bibitem{Ligeti:2008ac}
Z.~Ligeti, I.~W. Stewart, and F.~J. Tackmann, {\it {Treating the b quark
  distribution function with reliable uncertainties}},  {\em Phys.Rev.} {\bf
  D78} (2008) 114014, [\href{http://xxx.lanl.gov/abs/0807.1926}{{\tt
  arXiv:0807.1926}}].

\bibitem{Sjostrand:2006za}
T.~Sjostrand, S.~Mrenna, and P.~Z. Skands, {\it {PYTHIA 6.4 Physics and
  Manual}},  {\em JHEP} {\bf 0605} (2006) 026,
  [\href{http://xxx.lanl.gov/abs/hep-ph/0603175}{{\tt hep-ph/0603175}}].

\bibitem{Sjostrand:2007gs}
T.~Sjostrand, S.~Mrenna, and P.~Z. Skands, {\it {A Brief Introduction to PYTHIA
  8.1}},  {\em Comput.Phys.Commun.} {\bf 178} (2008) 852--867,
  [\href{http://xxx.lanl.gov/abs/0710.3820}{{\tt arXiv:0710.3820}}].

\bibitem{Cacciari:2008gp}
M.~Cacciari, G.~P. Salam, and G.~Soyez, {\it {The Anti-k(t) jet clustering
  algorithm}},  {\em JHEP} {\bf 0804} (2008) 063,
  [\href{http://xxx.lanl.gov/abs/0802.1189}{{\tt arXiv:0802.1189}}].

\bibitem{Cacciari:2011ma}
M.~Cacciari, G.~P. Salam, and G.~Soyez, {\it {FastJet User Manual}},  {\em
  Eur.Phys.J.} {\bf C72} (2012) 1896,
  [\href{http://xxx.lanl.gov/abs/1111.6097}{{\tt arXiv:1111.6097}}].

\bibitem{Korchemsky:1999kt}
G.~P. Korchemsky and G.~F. Sterman, {\it {Power corrections to event shapes and
  factorization}},  {\em Nucl.Phys.} {\bf B555} (1999) 335--351,
  [\href{http://xxx.lanl.gov/abs/hep-ph/9902341}{{\tt hep-ph/9902341}}].

\bibitem{Korchemsky:2000kp}
G.~Korchemsky and S.~Tafat, {\it {On power corrections to the event shape
  distributions in QCD}},  {\em JHEP} {\bf 0010} (2000) 010,
  [\href{http://xxx.lanl.gov/abs/hep-ph/0007005}{{\tt hep-ph/0007005}}].

\bibitem{Watt:2003mx}
G.~Watt, A.~Martin, and M.~Ryskin, {\it {Unintegrated parton distributions and
  inclusive jet production at HERA}},  {\em Eur.Phys.J.} {\bf C31} (2003)
  73--89, [\href{http://xxx.lanl.gov/abs/hep-ph/0306169}{{\tt
  hep-ph/0306169}}].

\bibitem{Watt:2003vf}
G.~Watt, A.~Martin, and M.~Ryskin, {\it {Unintegrated parton distributions and
  electroweak boson production at hadron colliders}},  {\em Phys.Rev.} {\bf
  D70} (2004) 014012, [\href{http://xxx.lanl.gov/abs/hep-ph/0309096}{{\tt
  hep-ph/0309096}}].

\bibitem{Collins:2004vq}
J.~C. Collins and X.~Zu, {\it {Initial state parton showers beyond leading
  order}},  {\em JHEP} {\bf 0503} (2005) 059,
  [\href{http://xxx.lanl.gov/abs/hep-ph/0411332}{{\tt hep-ph/0411332}}].

\bibitem{Collins:2005uv}
J.~Collins and H.~Jung, {\it {Need for fully unintegrated parton densities}},
  \href{http://xxx.lanl.gov/abs/hep-ph/0508280}{{\tt hep-ph/0508280}}.

\bibitem{Collins:2007ph}
J.~Collins, T.~Rogers, and A.~Stasto, {\it {Fully unintegrated parton
  correlation functions and factorization in lowest-order hard scattering}},
  {\em Phys.Rev.} {\bf D77} (2008) 085009,
  [\href{http://xxx.lanl.gov/abs/0708.2833}{{\tt arXiv:0708.2833}}].

\bibitem{Rogers:2008jk}
T.~C. Rogers, {\it {Next-to-Leading Order Hard Scattering Using Fully
  Unintegrated Parton Distribution Functions}},  {\em Phys.Rev.} {\bf D78}
  (2008) 074018, [\href{http://xxx.lanl.gov/abs/0807.2430}{{\tt
  arXiv:0807.2430}}].

\bibitem{Mantry:2009qz}
S.~Mantry and F.~Petriello, {\it {Factorization and Resummation of Higgs Boson
  Differential Distributions in Soft-Collinear Effective Theory}},  {\em
  Phys.Rev.} {\bf D81} (2010) 093007,
  [\href{http://xxx.lanl.gov/abs/0911.4135}{{\tt arXiv:0911.4135}}].

\bibitem{Mantry:2010mk}
S.~Mantry and F.~Petriello, {\it {Transverse Momentum Distributions from
  Effective Field Theory with Numerical Results}},  {\em Phys.Rev.} {\bf D83}
  (2011) 053007, [\href{http://xxx.lanl.gov/abs/1007.3773}{{\tt
  arXiv:1007.3773}}].

\bibitem{Jain:2011iu}
A.~Jain, M.~Procura, and W.~J. Waalewijn, {\it {Fully-Unintegrated Parton
  Distribution and Fragmentation Functions at Perturbative $k_T$}},  {\em JHEP}
  {\bf 1204} (2012) 132, [\href{http://xxx.lanl.gov/abs/1110.0839}{{\tt
  arXiv:1110.0839}}].

\bibitem{Ellis:2012sn}
S.~D. Ellis, A.~Hornig, T.~S. Roy, D.~Krohn, and M.~D. Schwartz, {\it {Qjets: A
  Non-Deterministic Approach to Tree-Based Jet Substructure}},  {\em
  Phys.Rev.Lett.} {\bf 108} (2012) 182003,
  [\href{http://xxx.lanl.gov/abs/1201.1914}{{\tt arXiv:1201.1914}}].

\bibitem{Huber:2005yg}
T.~Huber and D.~Maitre, {\it {HypExp: A Mathematica package for expanding
  hypergeometric functions around integer-valued parameters}},  {\em
  Comput.Phys.Commun.} {\bf 175} (2006) 122--144,
  [\href{http://xxx.lanl.gov/abs/hep-ph/0507094}{{\tt hep-ph/0507094}}].

\bibitem{Huber:2007dx}
T.~Huber and D.~Maitre, {\it {HypExp 2, Expanding Hypergeometric Functions
  about Half-Integer Parameters}},  {\em Comput.Phys.Commun.} {\bf 178} (2008)
  755--776, [\href{http://xxx.lanl.gov/abs/0708.2443}{{\tt arXiv:0708.2443}}].

\end{thebibliography}\endgroup

\end{document}